\documentclass[]{JHEP3}
\pdfoutput=1

\usepackage{amssymb}
\usepackage{graphicx}
\usepackage{amsmath}
\usepackage{multirow}

\def\lsim{\mathrel{\rlap{\lower3pt\hbox{\hskip0pt$\sim$}}
   \raise1pt\hbox{$<$}}}         
\def\gsim{\mathrel{\rlap{\lower4pt\hbox{\hskip1pt$\sim$}}
   \raise1pt\hbox{$>$}}}         

\newcommand{\eps}{\epsilon}

\newcommand{\mSFOS}{m_{SF\text{-}OS}}
\newcommand{\bea}{\begin{eqnarray}}
\newcommand{\eea}{\end{eqnarray}}
\newcommand{\bit}{\begin{itemize}}
\newcommand{\eit}{\end{itemize}}

 \def    \ptj           {\mbox{$p_{T j}$}}
 \def    \etaj           {\mbox{$\eta_{ j}$}}
 \def    \ptl           {\mbox{$p_{T l}$}}
 \def    \etal           {\mbox{$\eta_{ l}$}}
 \def    \gev            {\mbox{$\mathrm{GeV}$}}
\newcommand{\etmiss}{ \not \hskip -4pt E_T}

\title{Strong Double Higgs Production at the LHC}

\author{Roberto Contino\\
	Dipartimento di Fisica, Universit\`a di Roma ``La Sapienza" \\
	INFN, Sezione di Roma, Italy\\
	CERN, Physics Department, Theory Unit, Geneva, Switzerland
	}

\author{Christophe Grojean\\
	CERN, Physics Department, Theory Unit, Geneva, Switzerland\\
	Institut de Physique Th\'eorique, CEA Saclay, France
	}

\author{Mauro Moretti\\
	Dipartimento di Fisica, Universit\`a di Ferrara, and INFN, Sezione di Ferrara, Italy
	}
	
\author{Fulvio Piccinini\\
	INFN, Sezione di Pavia, Italy
	}
	
\author{Riccardo Rattazzi\\
	Institut de Th\'eorie des Ph\'enom\`enes Physiques, EPFL, Lausanne, Switzerland
	}
		
\abstract{The hierarchy problem and the electroweak data, together, provide a plausible motivation for considering a light Higgs emerging as a 
pseudo-Goldstone boson from a strongly-coupled sector. In that scenario, the rates for Higgs production and decay  differ significantly 
from those in the Standard Model. However, one genuine strong coupling signature is the growth with energy of the scattering amplitudes 
among the  Goldstone bosons, the longitudinally polarized  vector bosons as well as the Higgs boson itself. The rate for double Higgs 
production in vector boson fusion is thus enhanced with respect to its negligible rate in the SM.
We study that reaction in $pp$ collisions, where the production of two Higgs bosons at high $p_T$ is associated with the emission of two 
forward jets.
We concentrate on the decay mode $hh \to WW^{(*)}WW^{(*)}$ and study the semi-leptonic decay chains of the $W$'s with 2, 3 or 4 leptons in the 
final states. While the 3 lepton final states are the most relevant and can lead to a 3$\sigma$ signal significance with 300 fb$^{-1}$ collected 
at a 14~TeV LHC, the two same-sign lepton final states provide complementary information. We also comment on the prospects for improving 
the detectability of  double Higgs production at the foreseen LHC energy and
luminosity upgrades.
}

\preprint{CERN-PH-TH/2009-036}
\keywords{Higgs, Electroweak symmetry breaking, Composite models}

\begin{document}

\section{Introduction
\label{sec:introduction}}

It is clear that,  in addition to the four known fundamental forces (gravity, electromagnetism, the weak and the strong interactions),
new dynamics must exist in order to account for the observed phenomenon of electroweak symmetry breaking (EWSB). Luckily the state of 
our knowledge is  about to change as 
the Large Hadron Collider (LHC) is set to directly explore, for the first time in history, the nature of this dynamics. A basic question the LHC 
will address concerns the strength of the new dynamics: is the force behind EWSB  a weak or a strong one? In most regards this question is 
equivalent to asking whether a light Higgs boson exists or not. This is because in the absence of new states (in particular the Higgs boson) 
the strength of the interaction among the longitudinally polarized vector bosons grows with energy becoming
strong at around 1 or 2 TeV's. The Standard Model (SM) Higgs boson plays instead the role of `moderator' of the strength of interactions, and 
allows the model to be extrapolated at weak coupling down to very short distances, possibly down to the Unification or Planck 
scale~\cite{HiggsMediator}. In order to achieve this amazing goal the couplings of the SM Higgs are extremely constrained and predicted 
in terms of just one new parameter, the mass of the Higgs itself. In such situation, the SM Higgs is for all practical purposes an elementary particle.
However it is also possible, and plausible in some respects, that a light and narrow Higgs-like scalar does exist, but that this particle is a bound 
state from some strong dynamics not much above the weak scale. In such a situation the couplings of the Higgs to fermions and vector bosons are 
expected to deviate in a significant way from those in the SM, thus indicating the presence of an underlying strong dynamics.
Provided such deviations are discovered, the issue will be to understand the nature of the strong dynamics.
In that perspective the importance of having a well founded, but simple, theoretical picture to
study  the Higgs couplings at the LHC cannot be overemphasized.

The hierarchy problem and electroweak data, together, provide a plausible motivation for considering a light composite Higgs. 
It is well known that the absence of an elementary Higgs scalar nullifies the hierarchy problem. Until recently the idea of Higgs compositeness 
was basically seen as coinciding  with the so called Higgsless limit, where there exists no narrow light scalar resonance. 
The standard realization of this scenario is given by Technicolor models~\cite{TC}. However, another possibility, which is now more seriously considered, 
is that the Higgs, and not just the eaten Goldstone bosons, arises as a naturally light pseudo-Goldstone boson from  strong dynamics
just above the weak scale~\cite{compositeHiggs,Agashe:2004rs,Contino:2006qr,Giudice:2007fh}. 
This possibility is preferable over standard Technicolor in view of electroweak precision constraints. The reason  is that the electroweak breaking scale $v$
is not fixed to coincide exactly with the strong dynamics scale $f$, like it was for Technicolor. Indeed $v$ is now determined by additional parameters 
(in explicit models these can be the top Yukawa and the SM gauge couplings) and it is conceivable to have a situation where there is a small separation 
of scales. As a matter of fact $v\lsim 0.3 f$ is enough to largely eliminate all tension with the data. The pseudo-Goldstone Higgs
is therefore a plausible scenario at the LHC. In that respect one should mention another possibility that was  considered recently where the role of the 
Higgs is partially played  by a composite dilaton, that is the pseudo-Goldstone boson of spontaneously broken scale invariance~\cite{dilaton}. 
This second possibility is less motivated than the previous one as regards electroweak data, in that, like in Technicolor, no parameter exists to adjust 
the size of $S$ (and $T$). However  it makes   definite predictions for the structure of the couplings, that are distinguished from the pseudo-Goldstone 
case. The existence of the dilaton example suggests that it may be useful to keep a more ample perspective on ``Higgs'' physics.

The effective Lagrangian for a composite light Higgs was characterized  in Ref.~\cite{Giudice:2007fh}, also focussing on the pseudo-Goldstone scenario. 
It was shown that the Lagrangian is described at lowest order by a very  few parameters, and, in particular, in the pseudo-Goldstone case, only two 
parameters $c_H$ and $c_y$ are relevant at the LHC.
Both parameters modify in a rather restricted way the Higgs production rate and branching ratios.
In particular, the parameter $c_H$, that corresponds to the leading non-linearity in the $\sigma$-model kinetic term,  gives a genuine ``strong coupling'' 
signature by determining a growing amplitude for the scattering among longitudinal vector bosons. As seen in the unitary gauge, because of its modified 
coupling to vectors, the Higgs fails to completely unitarize the scattering amplitude. This is the same $\sigma$-model signature one has in Technicolor. 
The novelty is that the Higgs is also composite belonging to the $\sigma$-model, and thus the same growth with energy is found in the amplitude for 
$V_LV_L\to hh$ ($V=W,Z$). One signature of this class of models at hadron collider is therefore a significant enhancement over the (negligible) SM rate 
for the production of two Higgs bosons at high $p_T$ along  with two forward jets associated with the two primary partons that radiated the $V_LV_L$ 
pair. The goal of the present paper is to study the detectability of this process at the LHC and at its foreseen energy and luminosity upgrades.

\section{General parametrization of Higgs couplings}
\label{parametrization}
 
In this section we will introduce a general parametrization of the Higgs couplings to vectors and fermions. 
The goal is to describe deviations from the SM in  Higgs production and decay. 
 
 We are interested in the general situation in which a light scalar $h$ exists in addition to the vectors and the eaten Goldstones associated to the breaking $SU(2)\times U(1)_Y\to U(1)_Q$. 
 By the request of custodial symmetry, the Goldstone bosons describe the coset $SO(4)/SO(3)$ and can be fit into the $2\times 2$ matrix
\begin{equation}
\Sigma =e^{i\sigma_a\pi^a/v}\qquad\qquad v=246 \,{\rm GeV}\, .
\end{equation}
By working at sufficiently low energy with respect to any possible strong scale, we can perform a derivative expansion. The leading effects 
growing with energy arise at the 2-derivative level, and so we truncate our Lagrangian at this order. Moreover we assume that the gauge fields 
are coupled to the strong sector via weak gauging: the operators involving the field strengths $W_{\mu\nu}$ and $B_{\mu\nu}$ will appear with 
loop suppressed coefficients, and we neglect them. Similarly, we assume that the elementary fermions are coupled to the strong sector only via 
the (proto)-Yukawa interactions, so that the leading effects
will not involve derivatives (\textit{e.g.} operators involving the product of a fermionic and $\sigma$-model current will be suppressed). 

\begin{figure}[!t]
\begin{center}
\includegraphics[height=22mm]{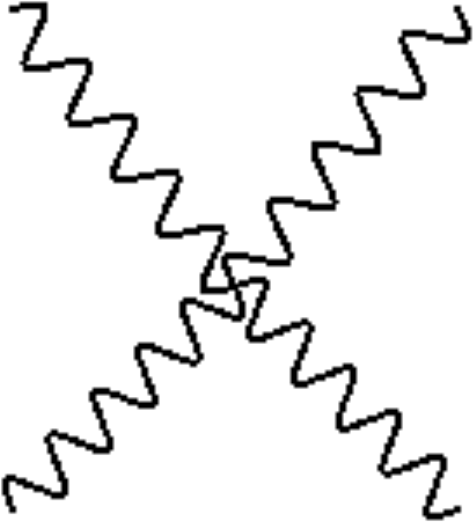} \hspace{1cm}
\includegraphics[height=22mm]{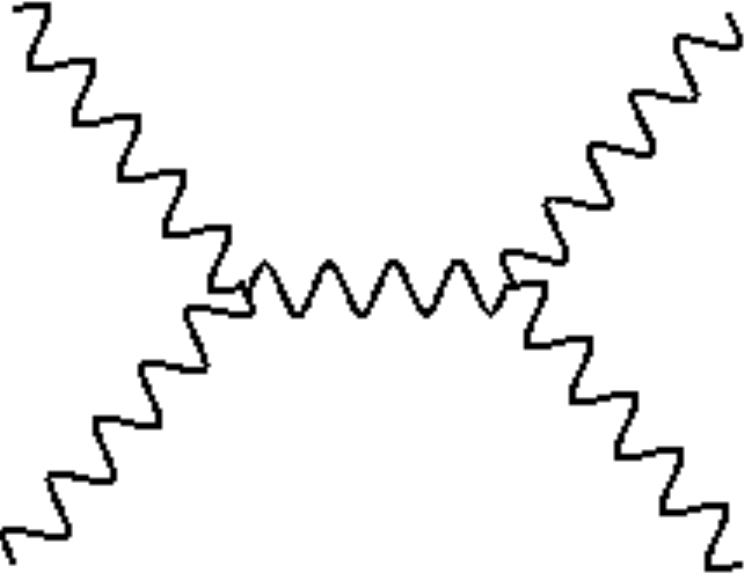} \hspace{1cm}
\includegraphics[height=22mm]{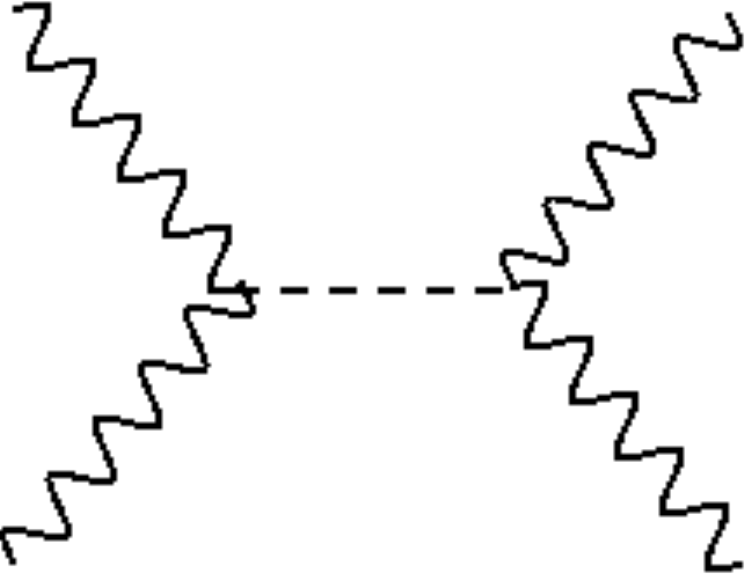}  \hspace{0.1cm}
\begin{minipage}[t]{0.12\linewidth}
\vspace{-1.3cm}
+ crossed 
\end{minipage}
\\[0.6cm]
\includegraphics[height=22mm,width=20mm]{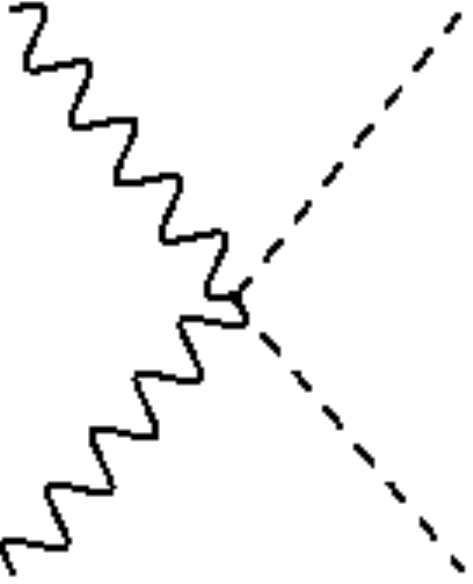} \hspace{1cm}
\includegraphics[height=22mm,width=20mm]{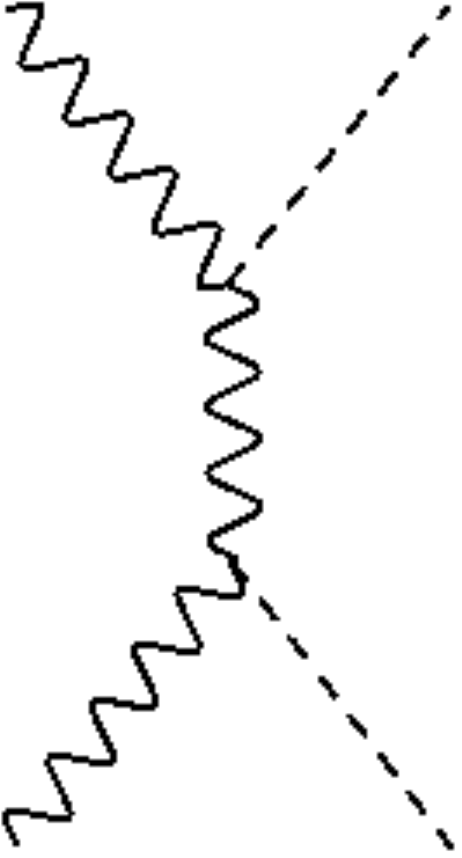}  \hspace{0.1cm}
\begin{minipage}[t]{0.12\linewidth}
\vspace{-1.3cm}
+ crossed 
\end{minipage}
\caption[]{ \label{fig:WWscat}
\small
Leading diagrams for the $V_LV_L\to V_LV_L$ (upper row) and $V_LV_L \to hh$ (lower row) scatterings at high energies.
}
\end{center}
\end{figure}

Under these assumptions the most general Lagrangian is~\footnote{In general $c$
can be a matrix in flavor space, but in the following we will assume  for simplicity that
it is proportional to unity in the basis in which the mass matrix is diagonal. In this way no
flavor-changing neutral current effects originate from the tree-level exchange of $h$.}
\begin{equation}
\begin{split}
{\cal L} =
 &\frac{1}{2}(\partial_\mu h)^2 - V(h) +\frac{v^2}{4}{\rm Tr}\left (D_\mu \Sigma^\dagger D^\mu \Sigma\right )
    \left [ 1+2a\, \frac{h}{v}+b\,\frac{h^2}{v^2}+\dots\right ] \\
 &-m_i\,\bar \psi_{Li}\, \Sigma\left (1+c\,\frac{h}{v} + \dots \right)\psi_{Ri}\,+\,{\rm h.c.}\, ,
\end{split}
\end{equation}
where $V(h)$ denotes the potential for $h$
\begin{equation}
V(h) = \frac{1}{2} m_h^2 h^2 + d_3\, \frac{1}{6} \left( \frac{3m_h^2}{v} \right) h^3 + d_4\, \frac{1}{24} \left( \frac{3m_h^2}{v^2} \right) h^4 + \dots
\end{equation}
and $a$, $b$, $c$, $d_3$, $d_4$ are arbitrary numerical parameters.
We have neglected  terms of higher order in $h$ (denoted by the dots) as they do not affect the leading 
$2\to 2$ processes.  For $a=b=c=d_3=d_4=1$ and vanishing higher order terms, the scalar $h$ can be embedded into a linear multiplet
\begin{equation}
U\equiv \left (1+\frac{h}{v}\right ) \Sigma\, ,
\end{equation}
and one obtains the SM Higgs doublet Lagrangian. The role of $a$, $b$ and $c$ in $2\to 2$ processes is easily seen 
by working in the equivalent Goldstone boson approximation~\cite{Chanowitz:1985hj}, according to which longitudinal vector bosons can be replaced
by the corresponding Goldstone bosons at high energy, $V_L^i \leftrightarrow \pi_i$. 
The parameter $a$ controls the strength of the $V_LV_L\to V_LV_L$ scattering ($V=W,Z$), see Fig.~\ref{fig:WWscat} (upper row).  
At the two derivative level the Goldstone scattering amplitude is
\begin{equation}
{\cal A}(\pi_i \pi_j\to \pi_k \pi_l)=\delta_{ij}\delta_{kl}\,\mathcal{A}(s)+\delta_{ik}\delta_{jl}\,\mathcal{A}(t)+\delta_{il}\delta_{jk}\,\mathcal{A}(u)
\end{equation}
with 
\begin{equation}
\mathcal{A}(s)\simeq \frac{s}{v^2}(1-a^2) 
\end{equation}
where subleading terms in $(M_W^2/s)$ have been omitted.
Perturbative unitarity is thus satisfied for $a=1$.
The parameter $b$ instead controls the process $V_LV_L\to hh$, 
see Fig.~\ref{fig:WWscat} (lower row), 
\begin{equation}
{\cal A}({\pi_i \pi_j\to hh}) \simeq \delta_{ij} \frac{s}{v^2}(b-a^2)\, .
\end{equation}
In this case perturbative unitarity is satisfied for $b=a^2$. 
Notice that an additional contribution from the s-channel Higgs exchange 
via the  trilinear coupling $d_3$  has been omitted
because subleading at high energy.  In fact, as it will be shown in the following sections,
in a realistic analysis of double Higgs production at the LHC
such contribution can be numerically important
and lead to a significant model dependency.
Finally the parameter $c$ controls the $V_LV_L\to \psi\bar\psi$ amplitude
\begin{equation}
{\cal A}({\pi_i \pi_j \to \psi\bar\psi}) = \delta_{ij} \frac{m_\psi\sqrt {s}}{v^2}(1-ac)\,  ,
\end{equation}
which is weak for $ac=1$. Hence, as well known,  only for the SM choice of parameters $a=b=c=1$ the theory 
is weakly coupled at all scales.
%

From the above general perspective, the study of $VV\to VV$, $VV\to hh$ and $VV\to \bar \psi\psi $ tests three different parameters. 
However, in specific models $a$, $b$  and $c$ can be related to each other.   For instance in the pseudo-Goldstone Higgs models based 
on the coset $SO(5)/SO(4)$~\cite{Agashe:2004rs, Contino:2006qr}, indicating by $f$ the decay constant of the $\sigma$-model and 
defining  $\xi\equiv v^2/f^2$, one has
\begin{equation}
a=\sqrt {1-\xi}\qquad\qquad b=1-2\xi\, .
\label{abgoldstone}
\end{equation}
The parameter $c$, on the other hand,  depends on which $SO(5)$ representation the SM fermions belong to. For examples,
fermions in spinorial and fundamental representations of $SO(5)$ imply:
\begin{align}
c =& \sqrt {1-\xi} &&  {\rm {spinorial \,representation\,\,(4\,of}}\,SO(5))
\label{c4goldstone}\\[0.3cm]
c =& \frac{1-2\xi}{\sqrt {1-\xi}} && {\rm {fundamental \,representation\,\,(5\,of}}\,SO(5))\, .
 \label{c5goldstone}
\end{align}
By expanding the above equations at small $\xi$, the result matches the general expressions obtained by using the 
Strongly Interacting Light Higgs (SILH) Lagrangian in the notation of Ref.~\cite{Giudice:2007fh}
\begin{equation}
a=1-\frac{c_H}{2}\xi \qquad\qquad b=1-2c_H\xi
\qquad\qquad c=1-\left(\frac{c_H}{2}+c_y\right)\xi\,.
\end{equation}
In particular, fermions in the spinorial (fundamental) representations of $SO(5)$ correspond to $c_y=0$ ($c_y=1$).  
Notice however that the general SILH parametrization applies more generally to a light composite $SU(2)_L$ 
Higgs doublet, regardless of whether it has a pseudo-Goldstone boson interpretation.  
The prediction for $d_3$ and $d_4$ is more model dependent, as it relies on the way the Higgs potential is
generated. As benchmark values for the trilinear coupling $d_3$ we consider  those predicted in the
$SO(5)/SO(4)$ minimal models of Ref.~\cite{Agashe:2004rs} (MCHM4) and Ref.~\cite{Contino:2006qr} (MCHM5), 
respectively with spinorial and fundamental fermion representations, where the Higgs potential is entirely generated 
by loops of SM fields:
~\footnote{The singularity  for $\xi\to 1$ in Eqs.~(\ref{c5goldstone}) and (\ref{eq:d3MCHM5})
appears because this limit is approached by keeping  the  mass of the Higgs and of the fermions fixed.}
\begin{align}
\label{eq:d3MCHM4}
d_3 =& \sqrt {1-\xi} &&   \text{MCHM4 with spinorial \,representations of $SO(5)$}\\[0.3cm]
d_3 =& \frac{1-2\xi}{\sqrt {1-\xi}} && \text{MCHM5 with vector \,representations of $SO(5)$}\, .
\label{eq:d3MCHM5}
\end{align}

Another, distinct example arises when $h$ represents the dilaton from spontaneously broken scale invariance. 
There one obtains a different relation among $a$, $b$ and $c$. Indeed the dilaton case corresponds to the 
choice $a^2=b=c^2$ with the derivative terms in the Lagrangian exactly truncated at quadratic order in $h$. For this choice one 
can define the dilaton decay constant by $v/a\equiv f_D$, the dilaton field as 
\begin{equation}
e^{\phi/f_D}=1+\frac{h}{f_D}
\end{equation}
and the Lagrangian can be rewritten as~\cite{dilaton}
\begin{equation}
{\cal L}=e^{2\phi/f_D}\left [\frac{ 1}{2}(\partial_\mu \phi)^2+\frac{v^2 }{4}{\rm Tr}
 \left (D_\mu \Sigma^\dagger D^\mu \Sigma\right )\right ]
 - \left( m_i \, e^{\phi/f_D}\bar \psi_{Li}\Sigma \psi_{Ri}\,+\,{\rm h.c.} \right)\\
\end{equation}
as dictated by invariance under dilatations
\begin{equation}
\phi(x)\to \phi(x e^\lambda)+\lambda f_D 
 \qquad\quad \pi_a(x) \to  \pi_a(x e^\lambda)
 \qquad\quad \psi(x)\to e^{3\lambda/2}\,\psi(xe^\lambda)\, .
\end{equation}
Notice that in the case of a SILH  all the amplitudes of  the three processes discussed above grow with the energy.
On the other hand, in the dilaton case the relation $a^2=b$ ensures that the amplitude for $VV\to hh$ does not feature 
the leading growth $\propto s$. 
The wildly different behaviour of the process $VV\to hh$ is what distinguishes the case of a genuine,
but otherwise composite Higgs, from a light scalar, the dilaton, which is not directly linked to the breakdown of  the electroweak symmetry.
Another difference which is worth pointing out between the specific case of a pseudo--Goldstone Higgs and a dilaton or a composite 
non-Goldstone Higgs has to do with the range of $a, b, c$. In the case of a pseudo-Goldstone Higgs one can prove  in general  that 
$a,b <1$~\cite{Low:2009di}, while all known models also satisfy $c<1$. Instead one easily sees that in the dilaton case depending on  
$f_D>v$ or $f_D< v$  one respectively has   $a,b,c<1$ or $a,b,c>1$.

\begin{figure}[!t]
\begin{center}
\includegraphics[width=0.485\textwidth,clip,angle=0]{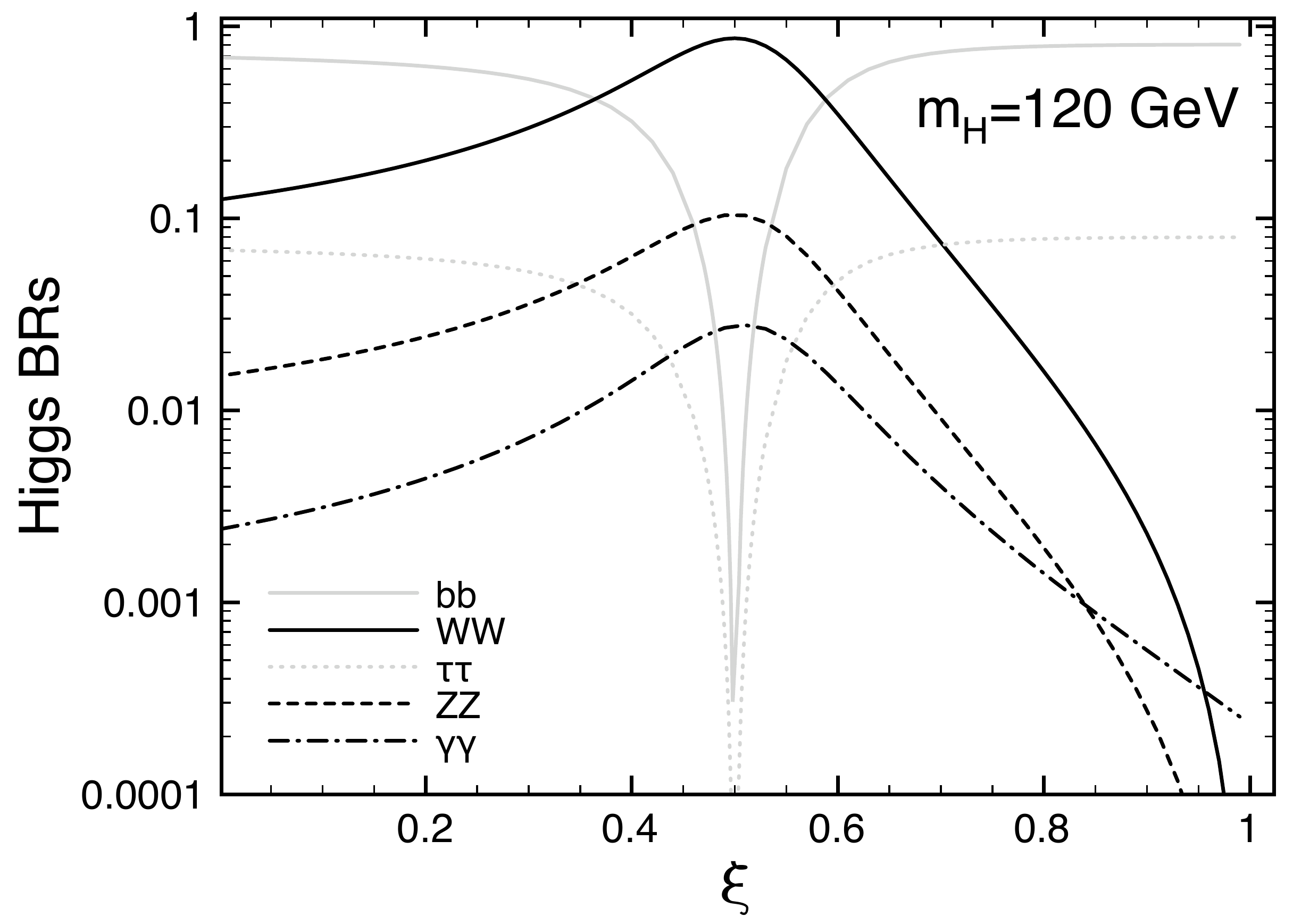}
\hspace{0.2cm}
\includegraphics[width=0.485\textwidth,clip,angle=0]{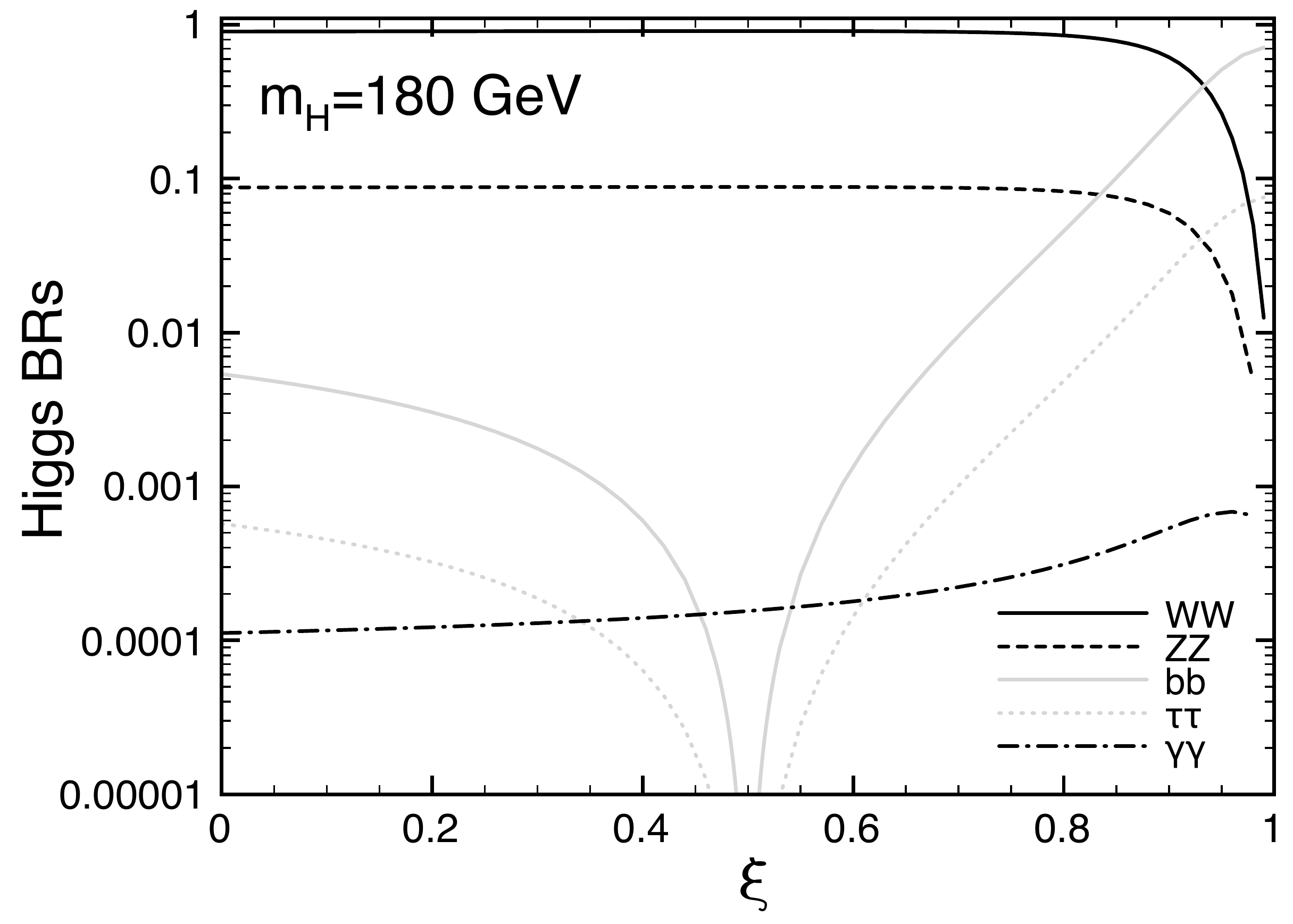}
\caption[]{
 \label{fig:BRs}
\small
Higgs decay branching ratios as a function of $\xi$ for SM fermions embedded into fundamental 
representations of $SO(5)$ for two benchmark Higgs masses: $m_h=120$~GeV (left plot) and 
$m_h=180$~GeV (right plot).  For $\xi=0.5$, the Higgs is fermiophobic, while in the Technicolor limit, 
$\xi \to 1$, the Higgs becomes gaugephobic.
}
\end{center}
\end{figure}

In general the couplings $a,b,c$ also 
parametrize deviations from the SM in the Higgs branching ratios. However, for the specific  case of 
the dilaton the relative branching ratios into vectors and fermions are  not affected. Instead, for loop induced 
processes like $h \to \gamma\gamma$ or $gg \to h$, deviations of order 1 with respect to the Standard Higgs 
occur due to the trace anomaly contribution \cite{Giudice:2000av}. Similarly, in the 
pseudo-Goldstone Higgs case with matter in the spinorial representation, the dominant branching ratios 
to fermions and vectors are not affected.  On the other hand, in the case with matter in the fundamental representation
the phenomenology can be dramatically changed when $\xi\sim O(1)$. 
From Eqs.~(\ref{abgoldstone}) and (\ref{c5goldstone}) we have
\begin{equation}
\frac{\Gamma(h\to\bar\psi\psi)}{\Gamma(h\to VV)}=\left (\frac{1-2\xi}{1-\xi}\right )^2 
 \frac{\Gamma(h\to\bar\psi\psi)}{\Gamma(h\to VV)}\Big\vert_{SM}\, ,
\end{equation}
so that around $\xi\sim 1/2$ the width into fermions is suppressed. In this case, even for  
$m_h$ significantly below the $2W$ threshold the dominant decay channel could be the one to $WW^*$. 
In Fig.~\ref{fig:BRs}  we show the Higgs branching ratios  as a function of $\xi$ 
in this particular model.
The possibility to have a moderately light Higgs decaying predominantly to 
vectors is relevant to our study of double Higgs production. It turns out that only when such decay channel 
dominates do we have a chance to spot the signal over the SM backgrounds.
%

One final remark must be made concerning the indirect constraints that exist on $a,b,c$.
As stressed by the authors of Ref.~\cite{Barbieri:2007bh}, the parameter $a$ is constrained by the LEP precision data:
modifying the Higgs coupling to the SM vectors changes the 
one-loop infrared contribution to the electroweak parameters $\epsilon_{1,3}$ by an amount 
$\Delta \epsilon_{1,3} = c_{1,3} (1-a^2) \log (\Lambda^2/m_h^2)$, 
where $c_1 = - 3\,\alpha(M_Z)/16\pi \cos^2\theta_W$,  $c_3 = \alpha(M_Z)/48\pi \sin^2\theta_W$
and $\Lambda$ denotes the mass scale  of the resonances of the strong sector.
For example, assuming no additional corrections to the precision observables and setting
$m_h = 120\,$GeV, $\Lambda = 2.5\,$TeV, one obtains $0.8 \lesssim a^2 \lesssim  1.5$ at 99$\%$~CL.
However, such constraint can become weaker (or stronger) in presence of additional contributions
to $\epsilon_{1,3}$.   For that reason in our analysis of double Higgs production we will keep an open mind on the possible values of $a$.
On the other hand,  no indirect constraint exists on the parameters $b$, $c$,  
thus leaving open the possibility of large deviations from perturbative unitarity in the $VV\to hh$ and $VV\to\psi\psi$
scatterings.

\section {Anatomy of $VV\to VV$ and $VV\to hh$ scatterings}
\label{sec:anatomy}

\subsection{$VV \to VV$ scattering}

The key feature of strong electroweak symmetry breaking is the occurrence of scattering amplitudes that grow 
with the energy above the weak scale. We thus expect them to dominate over the background at high enough energy. 
Indeed,  with no Higgs to unitarize the  amplitudes, on dimensional grounds, and by direct inspection of the relevant Feynman 
diagrams, one estimates~\cite{LET}
\begin{equation}
\mathcal{A} (V_T V_T \to V_T V_T) \sim g^2 f(t/s)
\quad \textrm{and} \quad
\mathcal{A} (V_L V_L \to V_L V_L)\sim \frac{s}{v^2} 
\label{eq:NDA}
\end{equation}
with  $f(t/s)$ a rational function which is $O(1)$, at  least formally, in the central region $-t=O(s)$. 
Then, according to the above estimates, in the central region we have
\begin{equation}
\frac{d\sigma_{LL\to LL}/dt}{d\sigma_{TT\to TT}/dt}\Big\vert_{t\sim -s/2}= N_h\frac{s^2}{M_W^4}\, ,
\label{hard}
\end{equation}
where  $N_h$  is a numerical factor expected to be of order 1. On the other hand, $f(t/s)$ has simple Coulomb poles in the forward region, 
due to $t$- and $u$-channel  vector exchange. Then,
after imposing a cut~\footnote{The offshellness of the 
$W$'s radiated by the quarks in fact provides a natural cut on $|t|$ and $|u|$ of the order of $p_{Tjet}^4/s$.
Nevertheless, the total inclusive 
cross section is dominated by soft physics and does not probe the dynamics of EW symmetry breaking.} 
$-s+Q^2_\textrm{min}<t<-Q^2_\textrm{min}$, with  $M_W^2\ll Q^2_\textrm{min}\ll s$,
the expectation  for the integrated cross sections is
\begin{equation}
\frac{\sigma_{LL\to LL}(Q_\textrm{min})}{\sigma_{TT\to TT}(Q_\textrm{min})}= N_s\frac{s \,Q^2_\textrm{min}}{M_W^4}\, .
\label{soft}
\end{equation}
Here again $N_s$ is a numerical factor expected to be of order 1.  By the above estimates, we expect 
the longitudinal cross section, both the hard one and the more inclusive one, to 
become larger than the transverse cross section right above the vector boson mass scale. 

\begin{figure}[tbp]
\begin{center}
\includegraphics[width=0.24\textwidth,clip,angle=0]{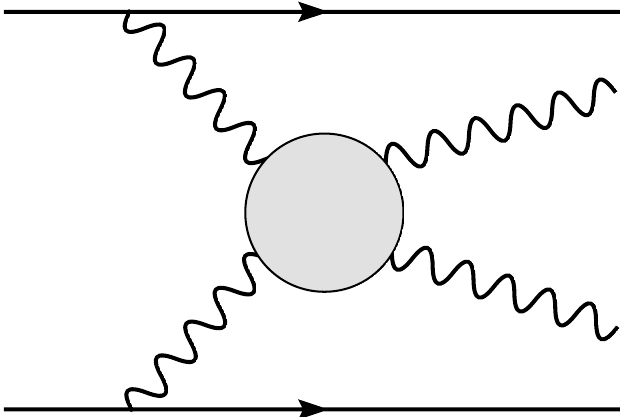}
\hspace{1cm}
\includegraphics[width=0.24\textwidth,clip,angle=0]{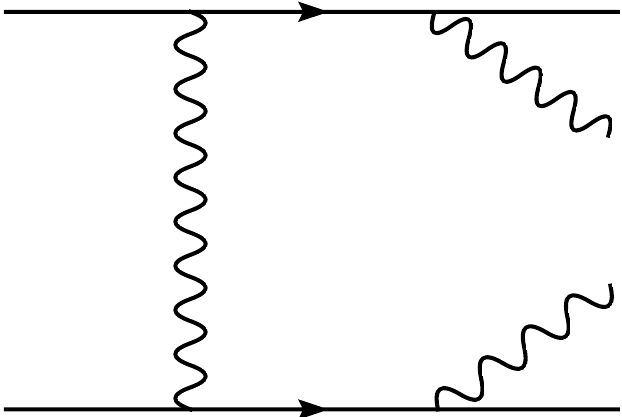}
\hspace{1cm}
\includegraphics[width=0.24\textwidth,clip,angle=0]{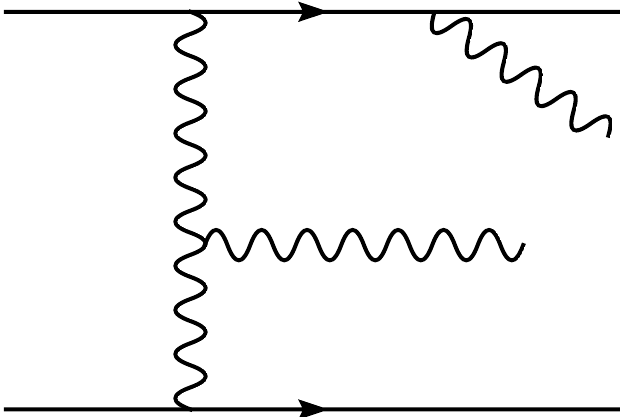}
\caption[]{
\label{fig:WWdiag}
\small
The full set of diagrams for $qq\to WWqq$ at order $g_W^4$. The blob indicates the
sum of all possible $WW\to WW$ subdiagrams. It is understood that the bremsstrahlung diagrams 
(second and third diagrams) correspond to all possible ways to attach an outgoing  $W$ to the quark lines.
}
\end{center}
\end{figure}


In reality the situation is more complicated because, since we do not posses on-shell vector boson beams, the $V$'s have first to be radiated from the colliding  protons. Then the physics of vector boson scattering is the more accurately reproduced the closer to on-shell the internal vector boson lines are, see Fig.~\ref{fig:WWdiag}.
This is the limit in which the 
process factorizes into the collinear (slow) emission of  virtual vector bosons \`a la Weizsacker--Williams and their subsequent hard (fast) scattering~\cite{EWA, Accomando:2006mc}. As evident from the collision kinematics,  the virtuality of the vector bosons is of the order of the   $p_T$ of the outgoing quarks. Thus the interesting limit is the one where the transverse momentum  of the two spectator jets is much smaller than the other relevant scales. In particular when
\begin{equation}
p_{Tjet}\ll p_{TW} \qquad\qquad M_W\ll p_{TW}
\end{equation}
where $p_{TW}$ and $p_{Tjet}$ respectively represent the transverse momenta of the outgoing vector bosons and  jets. In this kinematical region, 
the virtuality of the incoming vector bosons can be neglected with respect to the virtuality that characterizes the hard scattering subdiagrams. 
Then the  cross section can be written as a convolution of vector boson distribution functions with the hard vector cross section. It turns out that
the densities for respectively transverse and longitudinal polarizations have different sizes, and this adds an extra relative factor in the comparison
schematized above.  In particular, the emission of transverse vectors is logarithmically 
distributed in $p_T$,  like for the Weizsacker--Williams photon spectrum. 
Thus we have that the transverse parton splitting function is~\cite{EWA}
\begin{equation}
\label{eq:PTW}
P^{T}(z)=\frac{g_A^2+g_V^2}{4\pi^2}\,\frac{1+(1-z)^2}{2z}\ln \left[ \frac{\bar p_{T}^2}{(1-z)M_W^2} \right]\, ,
\end{equation}
where $z$ indicates the fraction of energy carried by the vector boson
and $\bar p_T$ is the largest value allowed for $p_{T}$.
On the other hand, the emission of longitudinal vectors is peaked at $p_{T}\sim M_W$ and shows no logarithmic enhancement 
when allowing large $p_{T}$~\cite{EWA}
\begin{equation}
\label{eq:PLW}
P^{L}(z)=\frac{g_A^2+g_V^2}{4\pi^2}\,\frac{1-z}{z}\, .
\end{equation}
Hence, by choosing a cut $p_{Tjet}<\bar p_{T}$ with $\bar p_{T}\gg M_W$ the cross section
for transverse vectors is enhanced due to  their luminosity by a  factor $(\ln \bar p_{T}/ M_W)^2$. 
For reasonable cuts this is not a very important 
effect though. At least, it is less important than the numerical factors $N_h$ and $N_s$ that come out from the explicit computation of the hard 
cross section, and which we shall analyze in a moment.

One last comment concerns the subleading corrections to the effective vector boson approximation (EWA). On general grounds, we expect the corrections 
to be controlled by the ratio $p_{Tjet}^2/p_{TW}^2$, that is the ratio between the virtuality of the incoming vector lines and the virtuality of the hard 
$VV\to VV$ subprocess~\footnote{In fact, another kinematic parameter controlling the approximation is given by the invariant mass of the 
$W+{\rm jet}$  subsystem $m^2_{JW}=(p_W+p_{jet})^2$. In the region 
$m^2_{JW}\ll s$ the bremsstrahlung diagrams are enhanced by a collinear singularity. In a realistic experimental situation
this region is practically eliminated by a cut on the relative angle between the jet and the (boosted) decay products of the $W$'s.}. In particular, both in the 
fully hard region $p_{Tjet}\sim p_{TW}\sim \sqrt s$ and in the forward region $p_{Tjet}, p_{TW}\lsim m_W $ we expect the approximation to break down. In 
these other kinematic regions, the contribution of the other diagrams in Fig.~\ref{fig:WWdiag}  is not only important but essential to obtain a physically 
meaningful gauge independent result ~\cite{Accomando:2006mc}.
 For the process $qq\to qqV_T V_T$, when the cross section is integrated over $p_{Tjet}$ up to $ \bar p_{T}$ the subleading corrections to EWA become 
only suppressed by $1/(\ln \bar p_{T}/ M_W)$.
This is the same log that appears in  $P^{T}(z)$. The process $qq\to qqV_T V_T$ is not significantly affected by a strongly coupled Higgs sector. 
On the other hand, in the presence of a strongly coupled Higgs sector, for $qq\to qqV_LV_L$ the EWA is further enhanced with respect to 
subleading effects because of the underlying $V_LV_L\to V_LV_L$  strong subprocess. Indeed, by
applying the axial gauge analysis of Ref.~\cite{Kunszt:1987tk}, one finds that, independent of the cut on $p_{Tjet}$, the subleading effects 
to the EWA are suppressed by at least $M_W^2/p_{TW}^2$. 

Having made the above comments on vector boson scattering in hadron collisions, let us now concentrate on the partonic process. 
We will illustrate our point with the example of the $W^+ W^+\to W^+W^+$ process (similar results can be 
obtained for the other processes) in the case of the composite pseudo-Goldstone Higgs, where for  $a\not =1$ the longitudinal scattering 
is dominated at large energies by the (energy-growing) contact interaction.
Let us then compare the semihard and hard cross sections for different polarizations as
prospected in Eqs.~(\ref{hard}) and (\ref{soft}). Considering first the case $s \gg M_W^2$ with fixed $t$ and $u$, for each
polarization channel  we can write the amplitude as~\footnote{Here and in the following equations
the high-energy approximation consists in neglecting terms of order~$(M_W^2/s)$.}
\begin{equation}
\mathcal{A} \simeq \frac{A_\gamma^t \, s}{t} + \frac{A_Z^t\, s}{t-M_Z^2}
  +\frac{A_\gamma^u \, s}{u} + \frac{A_Z^u\, s}{u-M_Z^2} +  A_\textrm{reg.} + A_s \,\frac{s}{v^2} \, ,
\label{eq:amplitudedecomposition}
\end{equation}
where the $A$'s are numerical constants which 
take different values for the different polarization channels (see Table~\ref{table:multiplicity}).
\begin{table}[pt]
\begin{center}
$
\begin{array}{|ccc|cccccc|}
\hline
 &\textrm{channel} & \textrm{weight} & A_\gamma^t & A_Z^t & A_\gamma^u & A_Z^u & \hspace{-.5cm} A_\textrm{reg.} & A_s \\
\hline
{\vrule height 18pt depth 10pt width 0pt}
LL\to LL & & 1/2& 2e^2 & \frac{g^2(1-2c_W^2)^2}{2c_W^2} & 2e^2 
  & \frac{g^2(1-2c_W^2)^2}{2c_W^2} & \frac{2(M_W^2-c_W^2 m_h^2)}{c_W^2\,v^2 } 
  & a^2-1 \\
\hline
 \multirow{2}{*}{{\vrule height 20pt depth 8pt width 0pt} $LL\to TT$} 
 & {\vrule height 15pt depth 8pt width 0pt} 
{\scriptstyle LL\;\to\; ++ } & 1 & 0 &0 & 
 0 & 0 & g^2 (1-a^2)/2 & 0 \\
 & {\vrule height 0pt depth 8pt width 0pt} 
{\scriptstyle LL\;\to\; +- } & 1& 0 & 0 & 
0 &0 &  g^2 (a^2-1)/2  &  0 \\
\hline
{\vrule height 15pt depth 8pt width 0pt} 
LT\to LT 
& {\scriptstyle L+\;\to\; L+} & 4
& 2 e^2 & - g^2 (1-2c_W^2) & 0 &  0
& g^2 (a^2-1)/2 & 0\\
\hline
 \multirow{2}{*}{{\vrule height 20pt depth 8pt width 0pt} $TT\to TT$} 
 & {\vrule height 15pt depth 8pt width 0pt} 
{\scriptstyle ++\;\to\; ++ } & 1& 2 e^2 & 2 g^2 c_W^2 & 
 2 e^2 & 2 g^2 c_W^2 & 0 &  0 \\
 & {\vrule height 0pt depth 8pt width 0pt} 
{\scriptstyle +-\;\to\; +- } & 2 & 2 e^2 & 2 g^2 c_W^2 & 
   0 & 0 & 2 g^2 &  0 \\
\hline
\end{array}
$
\end{center}
\caption[]{
\label{table:multiplicity}
\small
$W^+W^+\to W^+W^+$ scattering: coefficients for the decomposition of the amplitude according to Eq.~(\ref{eq:amplitudedecomposition}). 
Of the 13 independent polarization channels those not shown above can be obtained by either crossing 
or complex conjugation due to Bose symmetry. 
Cross sections can be computed by weighting each channel in the table  by the corresponding
multiplicity factor reported in the third column.
Only channels with non-vanishing coefficients in the decomposition of Eq.~(\ref{eq:amplitudedecomposition}) are shown. Terms proportional to  $(1-a^2)$ have been omitted for simplicity in the expression of $A_\textrm{reg.}$ in the $LL\to LL$ channel.}
\end{table}
The coefficients $A^{t,u}$ are easily computed in the eikonal approximation and are directly related to the 
electric- and $SU(2)_L$- charges of the $W$'s:
\begin{equation}
A_\gamma = 2 \times  (\textrm{electric charge of } W^+)^2
\quad \textrm{and} \quad 
A_Z = 2 \times (\textrm{``$SU(2)_L$ charge'' of } W^+)^2\, .
\end{equation}
Since $U(1)_{em}$ is unbroken, the longitudinal and transverse $W$'s have the same electric charge~$e$, but their 
$SU(2)_L$ charges are different: the charge of the transverse $W$'s, $g c_W$, is directly obtained from the triple 
point interaction $W^+W^-Z$, whereas the charge of the longitudinal $W$, $g(c_W^2-s_W^2)/(2c_W)$, can be deduced 
from the coupling of the $Z$ to the Goldstones $\pi^\pm$  of the Higgs doublet.

\begin{figure}[!t]
\begin{center}
\includegraphics[width=0.485\textwidth,clip,angle=0]{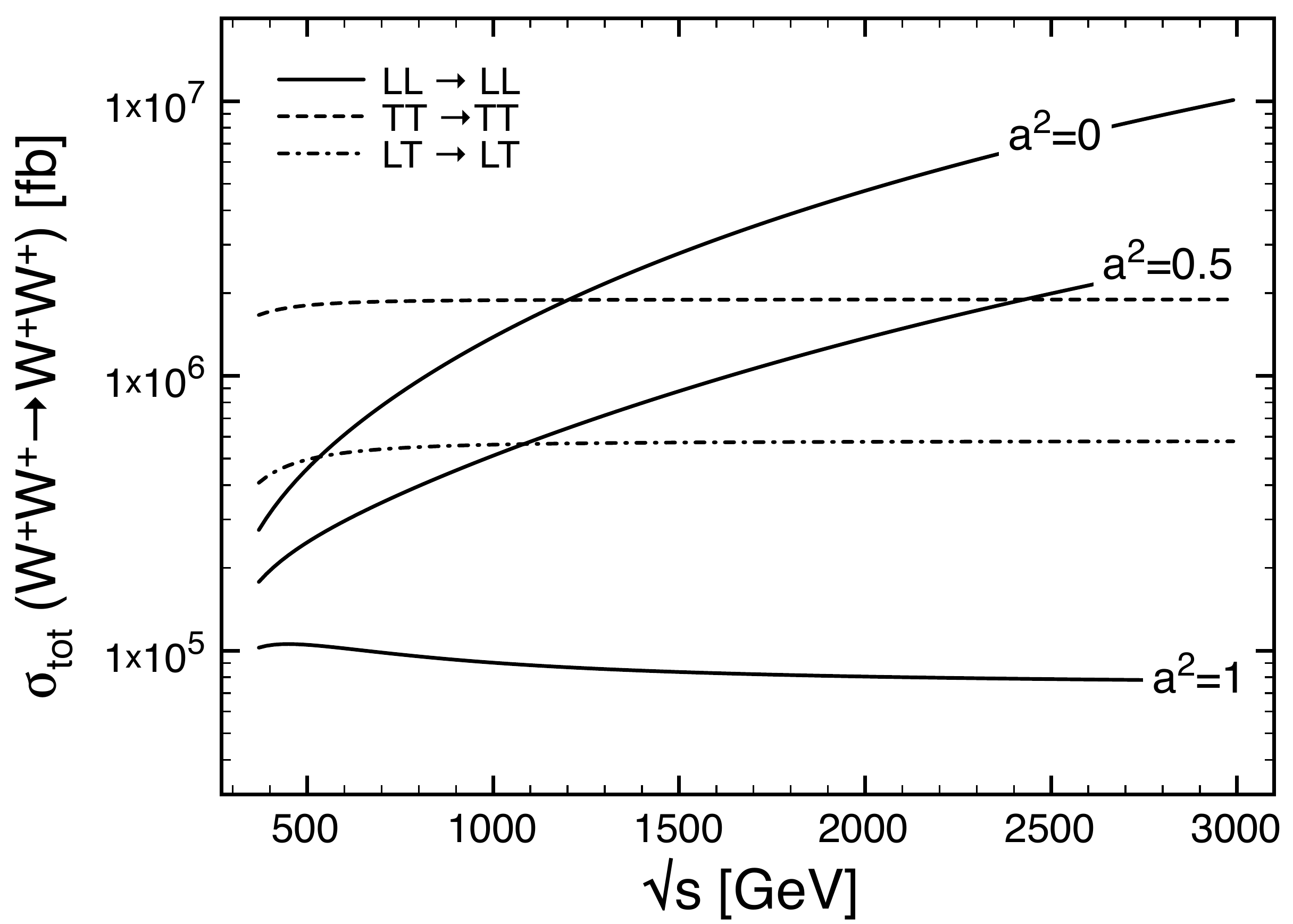}
\hspace{0.2cm}
\includegraphics[width=0.485\textwidth,clip,angle=0]{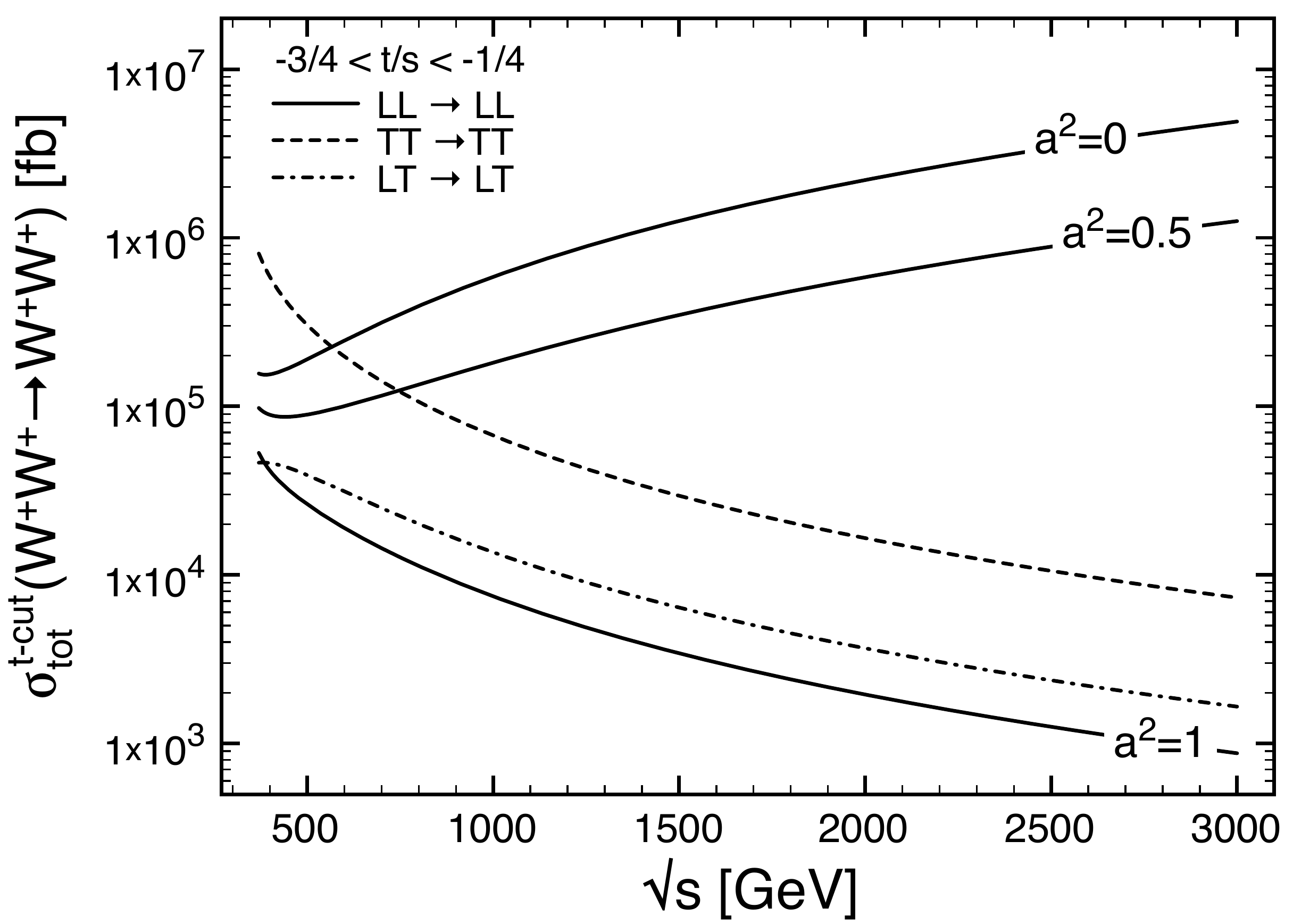}
\caption[]{
\label{fig:WpWpTOWpWp}
\small
 Cross section for the hard scattering $W^+W^+\to W^+W^+$ as a function of the center of mass energy 
for two different cuts on $t$ and $m_h = 180$~GeV. 
The left plot shows the almost inclusive cross section with $-s+ 4 M_W^2 < t < -M_W^2$.
The right plot shows the hard cross section with $-3/4 < t/s < -1/4$.
}
\end{center}
\end{figure}

The energy-growing term in Eq.~(\ref{eq:amplitudedecomposition}) has a non-vanishing coefficient $A_s$ 
only for the scattering of longitudinal modes (and $a\not =1$), in which case it dominates the differential cross section.
At large $s$ and for  $|t|,\,|u|>Q^2_\textrm{min}\gg M_W^2$ ($s\gg Q^2_\textrm{min}$) one has:
\begin{equation}
\sigma_{LL\to LL}(Q_\textrm{min})\simeq \frac{(1-a^2)^2\, s}{32 \pi\, v^4} \, .
\end{equation}
On the other hand,  the 
scattering of transverse modes is dominated by the forward $t$- and $u$-poles
\begin{equation}
\sigma_{TT\to TT}(Q_\textrm{min}) \simeq  \frac{g^4}{\pi} \left( \frac{s_W^4}{Q^2_\textrm{min}} + 
 \frac{c_W^4}{Q^2_\textrm{min}+M_Z^2} \right)\sim
 \frac{g^4}{\pi} \frac{s_W^4+c_W^4}{Q^2_\textrm{min}} \, ,
\end{equation}
and  the ratio of the longitudinal to transverse cross section is
\begin{equation}
\frac{\sigma_{LL\to LL}(Q_\textrm{min})}{\sigma_{TT\to TT}(Q_\textrm{min})} \simeq
 \frac{(1-a^2)^2}{512} \frac{Q^2_\textrm{min}}{s_W^4+c_W^4} \frac{s}{M_W^4}
\end{equation}
corresponding to a numerical factor $N_s\sim 1/500$ ! By using Table~\ref{table:multiplicity} one can directly check that this factor 
simply originates from a pile up of trivial effects (factors of 2) in the amplitudes. 
Interestingly, this numerical enhancement 
occurs for the $TT\to TT$ and $LT\to LT$ scattering channels, 
as clearly displayed by the left plot of Fig.~\ref{fig:WpWpTOWpWp},
while it is absent in $TT\to LL$ (this latter channel is not shown in Fig.~\ref{fig:WpWpTOWpWp} because its cross section 
is much smaller than the others).
%

\begin{figure}[!t]
\begin{center}
\vphantom{b}
\includegraphics[width=0.485\textwidth,clip,angle=0]{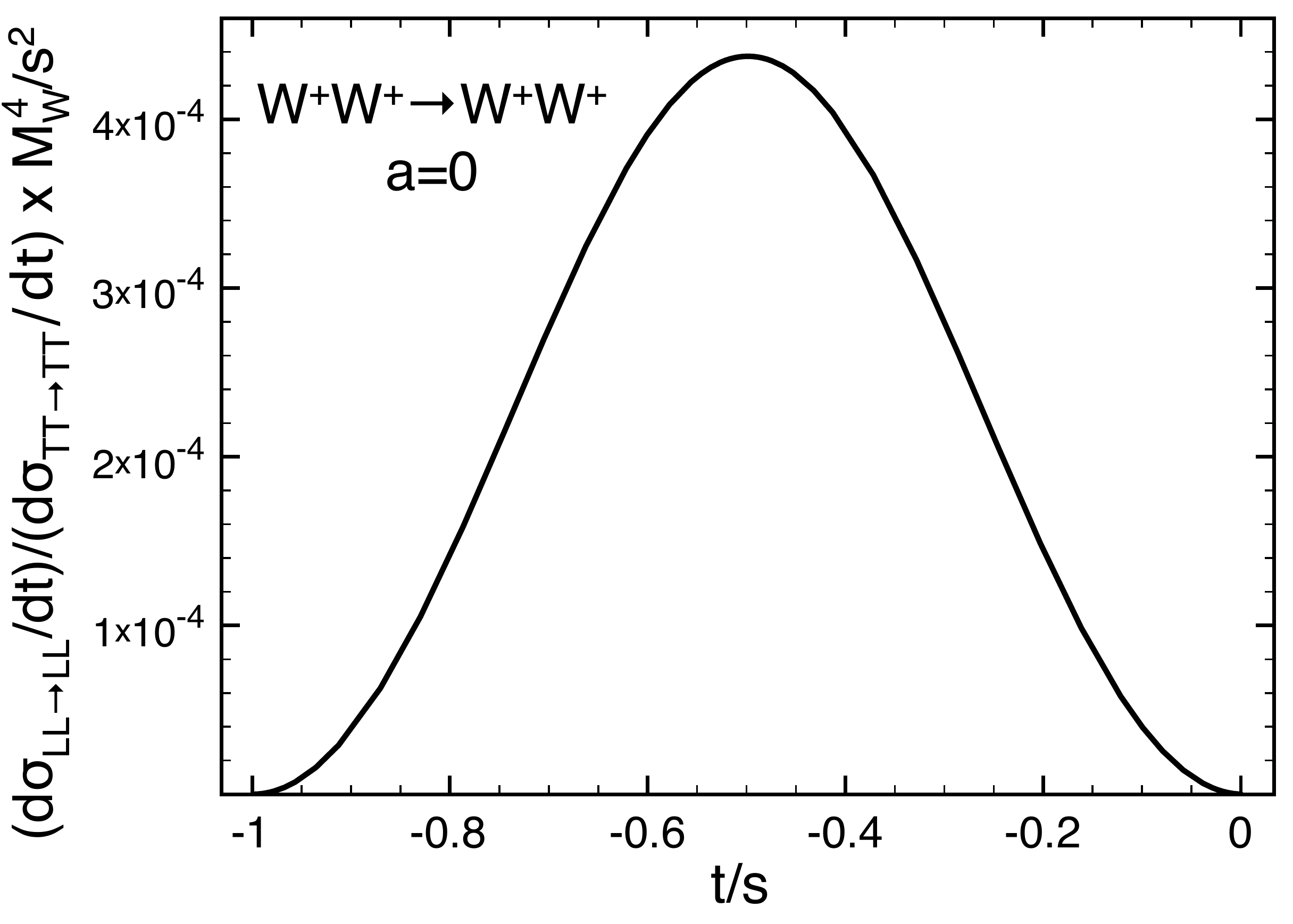}
\hspace{0.2cm}
\includegraphics[width=0.485\textwidth,clip,angle=0]{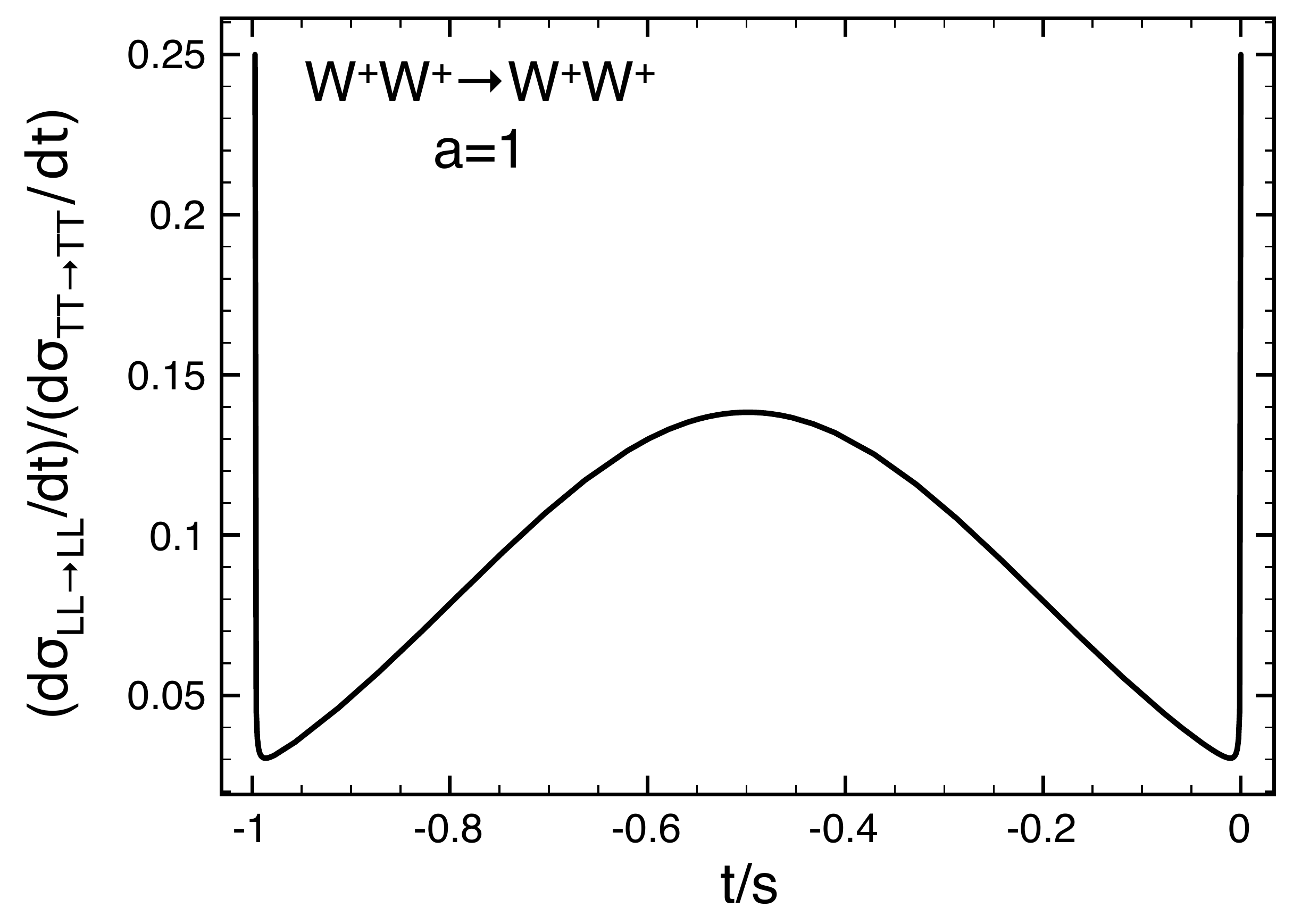}
\caption[]{
\label{fig:diffcrosssection}
\small
Differential cross section for longitudinal versus transverse polarizations for $a =0$  (left plot)
and  in the Standard Model ($a=1$, right plot). 
The different normalization reflects the different naive expectation in the two cases:
in the SM, both differential cross sections scale like $1/s^2$ at large energy, whereas for $a=0$ 
the longitudinal differential cross section stays constant, see Eq.~(\ref{eq:diffxsecs}).
}
\end{center}
\end{figure}

Of course the best way to test hard vector boson scattering is to go to the central region where the `background' from the Coulomb singularity of $Z$ and $\gamma$ exchange is absent.
Figure~\ref{fig:diffcrosssection} reports the ratio of the differential cross sections as a function of $t$
both for $a =0$ (left plot) and $a =1$ (right plot).
It is shown that even for exactly central $W$'s ($t=-s/2$) the ratio is still smaller than its naive estimate, 
the suppression factor being $N_h\sim 4\times 10^{-4}$ for $a=0$.
The origin of this numerical (as opposed to parametric) suppression is in the value of the coefficients
$A_i$ entering the various scattering channels. 
Indeed, for $t = -s/2$ Eq.~(\ref{eq:amplitudedecomposition}) simplifies to 
\begin{equation}
\mathcal{A}\simeq -2 \left( A_\gamma^t + A_Z^t + A_\gamma^u + A_Z^u \right)
+  A_\textrm{reg.} + A_s \, \frac{s}{v^2} \, ,
\end{equation}
which leads to the differential cross sections (for $t=-s/2$):
\begin{equation}
\label{eq:diffxsecs}
\begin{gathered}
\frac{d\sigma_{LL\to LL}}{dt}\Big|_{a \not = 1} \simeq \frac{(1-a^2)^2}{32 \pi \, v^4}\, , \qquad
\frac{d\sigma_{LL\to LL}}{dt}\Big|_{a = 1} \simeq 
\frac{g^4\left( c_W^2 m_h^2 + 3 M_W^2\right)^2}{128 \pi c_W^4 M_W^4\, s^2}
 \\[0.5cm]
\frac{d\sigma_{TT\to TT}}{dt} \simeq \frac{g^4\left( 64+2 \times 4 \right)}{16 \pi \, s^2}=\frac{72\, g^4}{16 \pi \, s^2}\, .
\end{gathered}
\end{equation}
Hence
\begin{equation}
\frac{d\sigma_{LL\to LL}/dt}{d\sigma_{TT\to TT}/dt}\Big\vert_{t\sim -s/2}= \frac{(1-a^2)^2}{2304}\frac{s^2}{M_W^4}
 \qquad\quad \text{for } a \not =1\, ,
\label{hard1}
\end{equation}
corresponding to an amazingly small numerical factor $N_h=1/2304$ again resulting
from a pile up of  `factors of 2'. In Eq.~(\ref{eq:diffxsecs}) we detailed the contribution of the non-vanishing polarization channels to 
the transverse scattering cross section (the dominant channels are $++\,\to\, ++$ and its complex conjugate).
The result of our analysis is synthesized in the plots of  Fig.~\ref{fig:WpWpTOWpWp}.
For the hard cross section (right plot, with $-3/4 < t/s < -1/4$) the signal wins over the SM background at $\sqrt{s}\sim 600$ GeV 
($a=0$), while for the inclusive cross section (left plot, with $-s + 4M_W^2 < t < -M_W^2$) one must even go above 1 TeV. 
This is consistent with the different $s$ dependence displayed in Eqs.~(\ref{soft}) and (\ref{hard}). These scales are both well above $M_W$ 
due to the big numerical factors $N_{h,s}$. Of course the interesting physical phenomenon, hard scattering of two longitudinal vector bosons, 
is better  isolated in the hard cross section, but at the price of an overall reduction of the rate.

\begin{figure}[t]
\begin{center}
\includegraphics[width=0.6\textwidth,clip,angle=0]{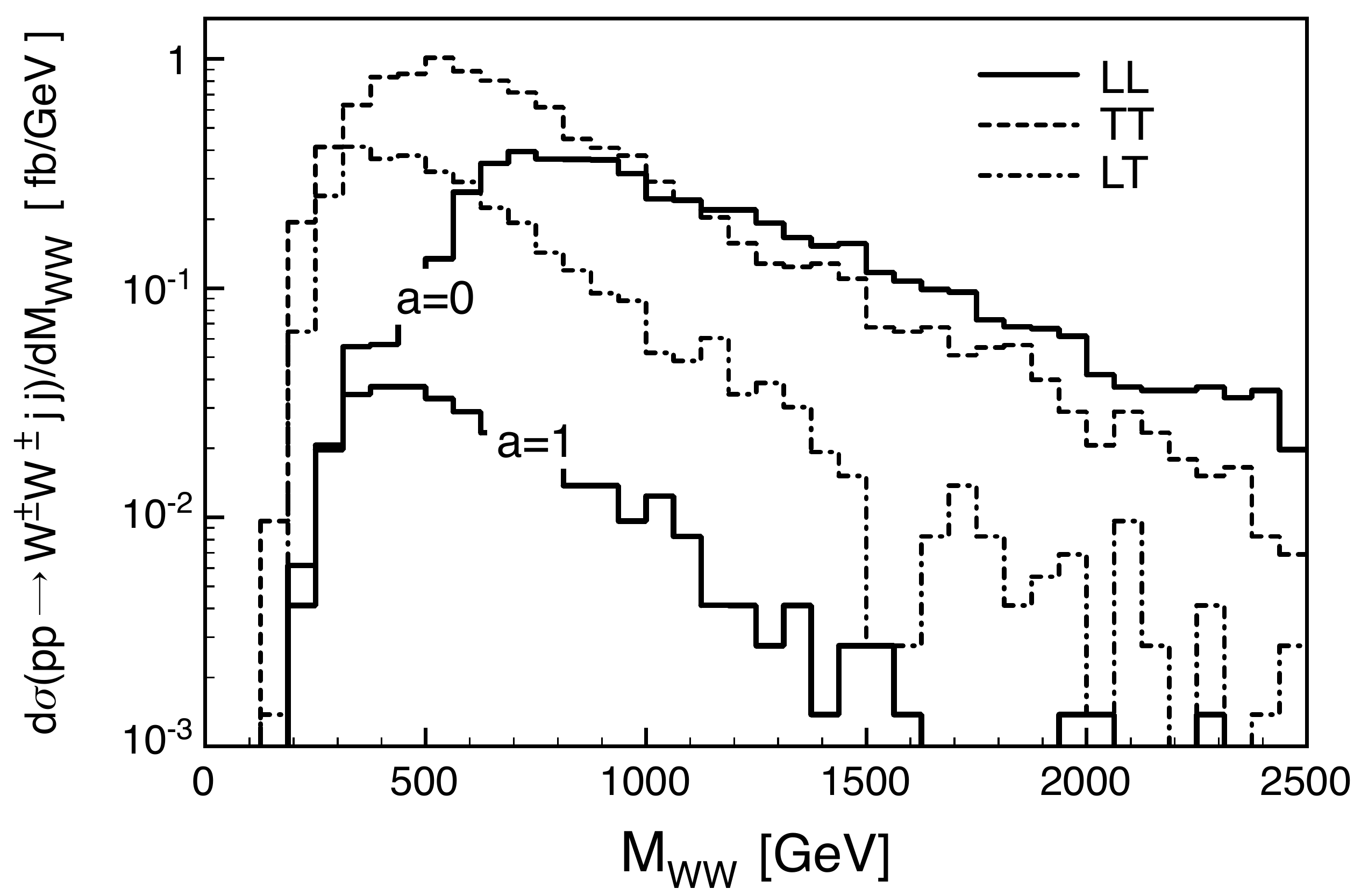}
\caption[]{
\label{fig:pptoWWjj}
\small
The differential cross section for $pp\to W^\pm W^\pm jj$ as a function of the invariant mass of the $WW$ pair, for different choices of the 
outgoing $W$ helicities.
All curves have been obtained by using Madgraph and imposing the following cuts: $M_{jj} > 500$ GeV, 
$p_{Tj} < 120$ GeV, $p_{TW} > 300$ GeV. The cut on $p_{Tj}$ exploits the forward jets always present in the signal. The cut on $p_{TW}$  
eliminates the forward region where the cross section is (trivially) dominated by the $Z$ and $\gamma$ $t$-channel exchange.
}
\end{center}
\end{figure}

It is this numerical accident that makes the study of strong vector boson scattering difficult at the LHC. The center of mass energy $m_{WW}$ of 
the vector boson system must be $\gsim 1$~TeV in order to have  a significant enhancement over the $TT\to TT$ background. But, taking 
into account the $\alpha_W$ price to radiate a $W$, $m_{WW}\sim 1$ TeV is precisely where the $W$ luminosity runs out of steam. This situation 
is depicted in Fig.~\ref{fig:pptoWWjj}. The case $a=0$ corresponds to the Higgsless case already studied 
in Ref.~\cite{bagger,ballestrero,Butterworth:2002tt,Han:2009em}. 
Our result, in spite of the different cuts, basically agrees with them: the cut in energy necessary to win over the $TT$ background reduces the 
cross section down to $\sigma(pp\to jj W_L^\pm W_L^\pm)\sim 2.5\,$fb~\footnote{This corresponds to $\sim O(10)$ events with
$100\,$fb$^{-1}$ in the fully leptonic final state $W^\pm W^\pm\to l^\pm \nu l^\pm \nu$.}.
Remarkably, a collider with a center of mass energy increased by about a factor of $2$
would do much better than the LHC. But this is an old story.
%

\subsection{$VV \to hh$ scattering}
\label{subsec:hhscatt}

\begin{figure}[tbp]
\begin{center}
\includegraphics[width=0.485\textwidth,clip,angle=0]{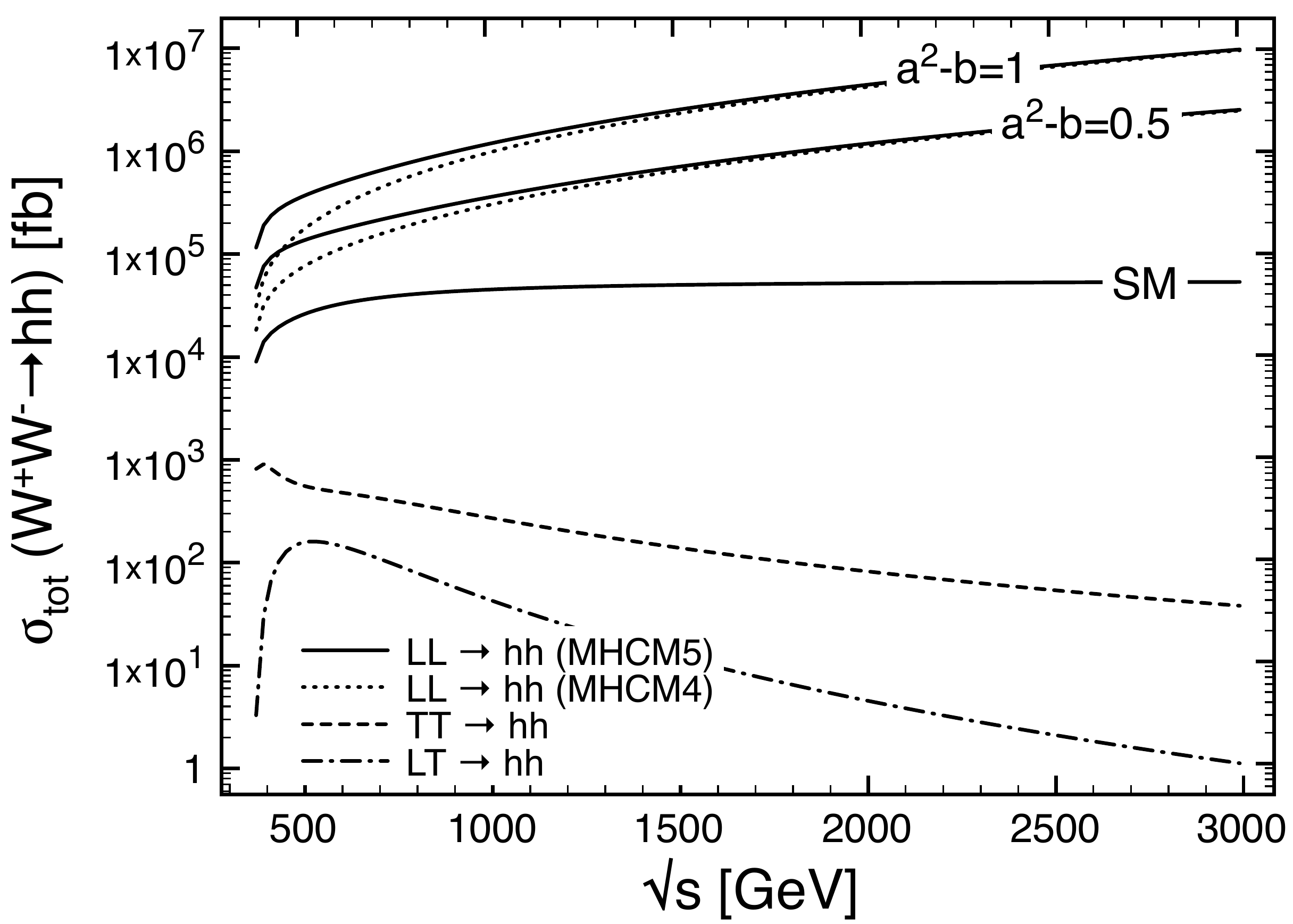}
\hspace{0.2cm}
\includegraphics[width=0.485\textwidth,clip,angle=0]{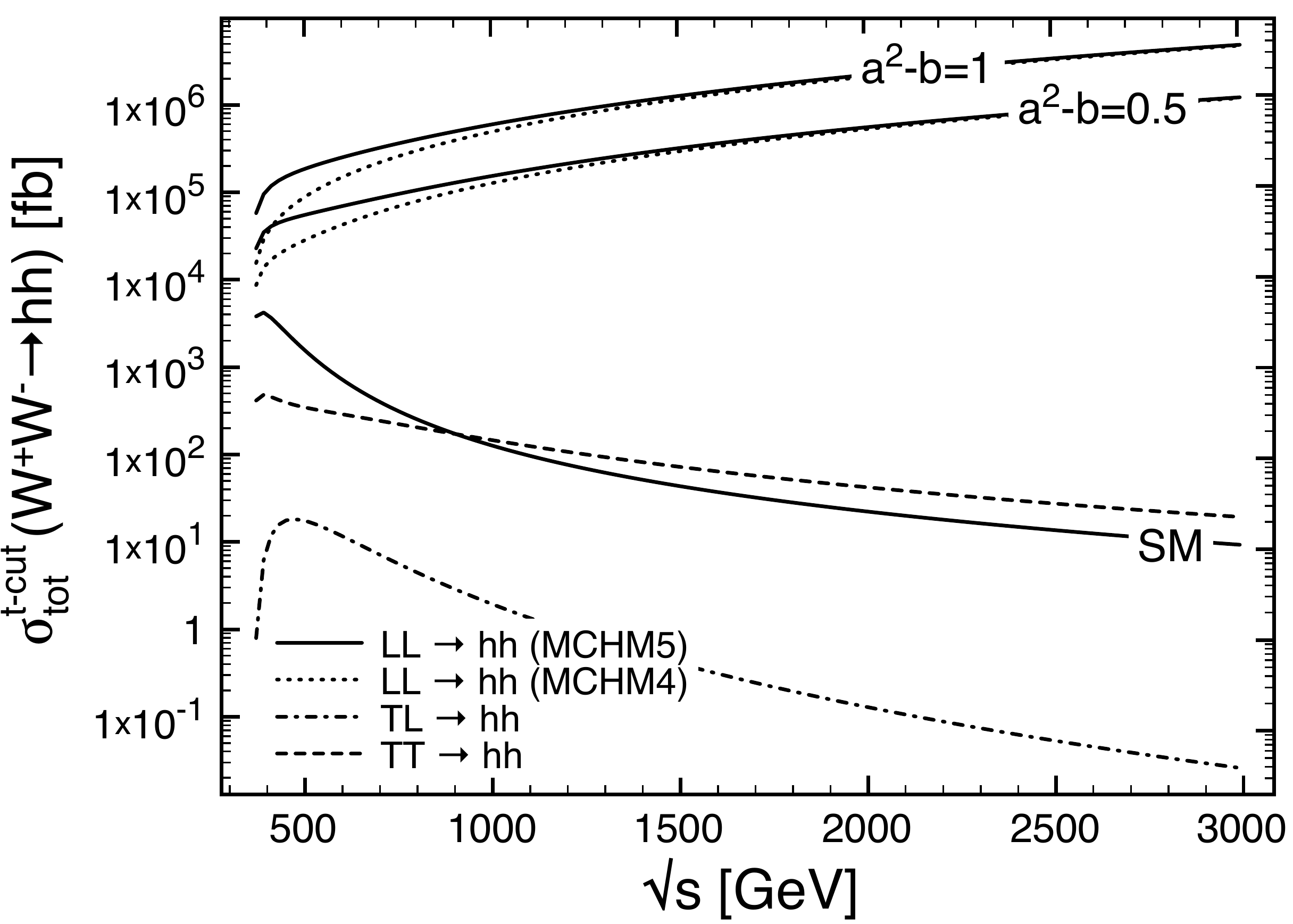}
\caption[]{
\label{fig:WpWmTOHH}
\small
Cross section for the hard scattering $W^+W^- \to hh$ with $m_h=180$~GeV. 
The left plot shows the inclusive cross section with  no cut on $t$. The right plot
shows the hard scattering cross section with a cut
$-s + 2 m_h^2 + 2 M_W^2 + Q_\text{min}^2 < t < - Q_\text{min}^2$, with
$Q_\text{min}^2 = s/2 - m_h^2 - M_W^2 - (s/4) \sqrt{(1-4m_h^2/s)(1-4M_W^2/s)}$.
This choice of $Q_\text{min}^2$ is compatible with the kinematical constraint close to threshold energies and
coincides with the cut applied in the right plot of Fig.~\ref{fig:WpWpTOWpWp} for $s\gg m_h^2$ (as~$Q_\text{min}^2\to s/4$).
Notice that differently from $WW\to WW$, the ratio of longitudinal over transverse scattering is not particularly enhanced by the cut. 
The behavior of the amplitudes near threshold is sensitive to the cubic self-coupling $d_3$ controlling the $s$-channel Higgs exchange.
The continuous and dotted $LL\to hh$  curves 
respectively correspond to the MHCM4 and MCHM5 models with $\xi = (a^2-b)$ and
$d_3$ as given in Eqs.~(\ref{eq:d3MCHM4}) and~(\ref{eq:d3MCHM5}).
}
\end{center}
\end{figure}

As illustrated by Fig.~\ref{fig:WpWmTOHH}, the situation is quite different for the $WW\to hh$ scattering. 
Here there is no equivalent of 
a fully transverse scattering channel, as the Higgs itself can be considered as a `longitudinal' mode,
being the fourth Goldstone from the strong dynamics.
The scatterings $W_T W_T \to hh$ and $W_L W_T \to hh$ 
never dominate over $W_L W_L \to hh$.
As previously, at large energy ($s \gg M_W^2$ with fixed $t$ and $u$)
the amplitude for the various polarization channels can be decomposed as: 
\begin{equation}
\mathcal{A} \simeq
\frac{A_W^t \, s}{t-M_W^2}
+ \frac{A_W^u\, s}{u-M_W^2}
+ A_\textrm{reg.} + A_s \, \frac{s}{v^2}\, ,
\label{eq:amplitudeWWHH}
\end{equation}
where the numerical constants $A$'s are given in Table~\ref{table:WWHH}.
\begin{table}[pt]
\begin{center}
$\begin{array}{|ccc|cccc|}
\hline
& \text{channel} & \textrm{weight} & A_W^t   & A_W^u  & A_\textrm{reg.} & A_s  \\
\hline
{\vrule height 18pt depth 10pt width 0pt}
LL \to hh  & &1/2 & a^2 g^2/2 &  a^2 g^2/2 
  & \frac{g^2 ( (4a^2-2b) M_W^2+(3 ad_3 -2 a^2) m_h^2)}{4M_W^2 }  & b-a^2 \\
\hline
\multirow{2}{*}{{\vrule height 20pt depth 8pt width 0pt} $TT\to hh$} & 
{\vrule height 15pt depth 8pt width 0pt}  \scriptstyle ++\;\to\; hh & 1& 0 & 0 & (b-a^2)g^2/2 & 0 \\
& {\vrule height 0pt depth 5pt width 0pt} 
\scriptstyle +-\;\to\; hh &1&  0 & 0 
& -a^2 g^2 /2 & 0 \\
\hline
\end{array}$
\end{center}
\caption[]{
\label{table:WWHH}
\small
$W^+W^-\to hh$ scattering: coefficients for the decomposition of the amplitude 
according to Eq.~(\ref{eq:amplitudeWWHH}). 
By crossing and complex conjugation there are only 4 independent polarization channels, one of which has vanishing
coefficients and is not shown.
When computing the cross section each channel has
to be weighted by the corresponding multiplicity factor reported in the third column.
}
\end{table}
%
The only scattering channel which can have in principle a Coulomb enhancement is also the one with the
energy-growing interaction, \textit{i.e.} the longitudinal to Higgs channel.
Furthermore, after deriving the differential cross sections for $s \gg v^2$ 
\begin{equation}
\label{eq:diffxsecWWhh}
\frac{d\sigma_{LL\to hh}}{dt} \simeq \frac{(b-a^2)^2}{32 \pi \, v^4}\, ,
\qquad \quad
\frac{d\sigma_{TT\to hh}}{dt} \simeq \frac{g^4(a^4+(b-a^2)^2)}{64 \pi \, s^2}\, ,
\end{equation}
one finds that in this case the naive estimate works well, and the onset of strong scattering is
at energies $\sqrt{s} \approx g v$.
Notice that the differential cross sections in Eq.~(\ref{eq:diffxsecWWhh}) are almost independent of $t$, 
except in the very forward/backward  regions where the longitudinal channels can be further enhanced by the $W$ exchange.

A final remark concerns the behavior of the $W_LW_L\to hh$ cross section close to threshold energies.
While at $s\gg v^2$ the cross section only depends on $(a^2-b)$, as expected
from the estimate performed in the previous section using the Goldstone boson approximation, at smaller energies
there is a significant dependence on the value of the trilinear coupling~$d_3$.
This is clearly shown in Fig.~\ref{fig:WpWmTOHH},  where the continuous and dotted curves 
respectively correspond to the MHCM4 and MCHM5 models with $\xi = (a^2-b)$ and
 $d_3$ as given in Eqs.~(\ref{eq:d3MCHM4}) and (\ref{eq:d3MCHM5}).
As we will see in the next sections, such model dependency is amplified by the effect of the parton distribution functions
and significantly affects the  total rate of signal events at the LHC, 
unless specific cuts are performed to select events with a large $M_{hh}$ invariant mass.

\section{The analysis}
\label{sec:analysis}

In this section we discuss the prospects to detect the production of a pair of Higgs bosons
associated with two jets at the LHC.
If the Higgs decays predominantly to $b \bar b $,  we have verified that 
the most important signal channel, 
$hhjj\to b\bar b b \bar bjj$,
is completely hidden by the huge QCD background.
We thus concentrate on the case in which the decay mode $h \to WW^{(*)}$
is large, and consider the final state $hhjj\to WW^{(*)}WW^{(*)}jj$.
As shown in Section~\ref{parametrization}, see Fig.~\ref{fig:BRs},  if the Higgs couplings to fermions are suppressed compared to the
SM prediction, the rate to $WW^{(*)}$ can dominate over $b\bar b$ even for light Higgses.
In our analysis we have set $m_h = 180\,$GeV  and considered as benchmark models the $SO(5)/SO(4)$  MCHM4 and MCHM5
discussed in the previous sections. All the values of the Higgs couplings are thus controlled by the ratio of the 
electroweak and strong scales $\xi = (v/f)^2$,
see Eqs.~(\ref{abgoldstone}), (\ref{c4goldstone}), (\ref{c5goldstone}), (\ref{eq:d3MCHM4}) and (\ref{eq:d3MCHM5}).
As anticipated, the two different models do not simply lead to different 
predictions for the Higgs decay fractions, but also to different $pp\to hhjj$ production rates as a consequence
of the distinct predictions for the Higgs cubic self-coupling $d_3$. For example, for $m_h = 180\,$GeV one has
\begin{center}
\begin{tabular}{l|cc}
$\sigma(pp\to hhjj)$ [fb] & MCHM4 & MCHM5 \\
\hline
$\xi =1$ &  9.3 & 14.0 \\
$\xi =0.8$ &  6.3 & 9.5 \\
$\xi =0.5$ &  2.9 & 4.2 \\
$\xi =0$ (SM) &  0.5 & 0.5 
\end{tabular}
\end{center}
where the acceptance cuts of Eq.~(\ref{eqn:acceptance}) have been imposed on the two jets.
Values of the signal cross section for the various final state channels
will be reported in the following subsections
for $\xi = 1,0.8,0.5$ in the MCHM4 and $\xi = 0.8,0.5$ in the MCHM5. We do not consider $\xi=1$ in the MCHM5 because 
the branching ratio $h\to WW^{(*)}$  vanishes in this limit.
Notice that the coupling $hWW$ formally vanishes for $\xi\to 1$  in both models, but in the MCHM4 all couplings are rescaled in the same
way, so that the  branching ratio $h\to WW^{(*)}$ stays constant to its SM value.
Cross sections for the SM backgrounds will be reported 
assuming SM values for the Higgs couplings and 
detailing possible (resonant) Higgs contributions as separate background processes whenever sizable. 
A final prediction for the total SM background in each model will  be presented at the end of the analysis 
in Section~\ref{sec:results} by properly rescaling the Higgs contributions to account for the modified Higgs couplings.

Throughout our analysis we have considered double Higgs production from
vector boson fusion only, neglecting the one-loop QCD contribution from gluon fusion in association with two jets. ÊThe latter is expected to have larger cross section than vector boson fusion~\cite{Glover:1987nx},Ê but it is insensitive to non-standard Higgs couplings to vector bosons. As discussed in the literature for single Higgs production with two jets~\cite{DelDuca:2001eu,DelDuca:2001fn},
event selections involving a cut on the dijet invariant mass and $\eta$ separation,
as the ones we are considering, strongly suppress the gluon fusion contribution.
We expect the same argument applies also to double Higgs production.

We concentrate on the three possible decay chains
that seem to be the most promising ones to isolate the signal from the background:
\begin{equation}
\begin{split}
{\cal S}_4 & =  p p \to hhjj \to l^+l^+l^-l^- 
 \etmiss + 2j \\[0.2cm]
{\cal S}_3 & =  p p \to hhjj \to l^{+}l^{-}l^{\pm}
\etmiss + 4j \\[0.2cm]
{\cal S}_2 & =  p p \to hhjj \to l^{+(-)}l^{+(-)} 
\etmiss + 5j\, (6j)\, ,
\end{split}
\label{eqn:channels}
\end{equation}
where $l^\pm=e^\pm/\mu^\pm$, $\etmiss$ denotes missing 
transverse energy due to the neutrinos and $j$ stands for
a final-state jet. A fully realistic analysis, including 
showering, hadronization  and detector simulation
is beyond the scope of the present paper. We will stick to the 
partonic level as far as possible, including showering 
effects only to provide a rough account of the jet-veto
benefit for this search. 
We perform a simple Gaussian smearing on the jets 
as a crude way to simulate detector effects.~\footnote{We have smeared both the
jet energy and momentum absolute value by $\Delta E/E = 100\%/\sqrt{E/\text{GeV}}$,
and the jet momentum direction using an angle resolution $\Delta\phi = 0.05$ radians
and $\Delta \eta = 0.04$.}
Signal events have been generated using \texttt{MADGRAPH}~\cite{MG-ME}, while both \texttt{ALPGEN}~\cite{alpgen}
and \texttt{MADGRAPH} have been used for the background.  A summary with information about the simulation
of each process, including the Montecarlo used, the choice of factorization scale and specific cuts applied at the 
generation level can be found in the Appendix~\ref{app:MC}.

Our event selection will be driven by simplicity as much as possible: 
we design a cut-based strategy by analyzing signal and background
distributions, cutting over the observable
which provides the best signal significance,
and reiterating the procedure until no further 
substantial improvement is achievable. 
As our starting point, we define the following 
set of {\em acceptance cuts}
\begin{equation}
\begin{aligned}
&  \ptj > 30\, \gev && \quad  | \etaj | < 5 \quad 
&& \Delta R_{jj'} > 0.7 \\
&  \ptl > 20\, \gev && \quad  | \etal | < 2.4  \quad 
&& \Delta R_{jl} > 0.4 \qquad \Delta R_{ll'} > 0.2\, ,
\end{aligned}
\label{eqn:acceptance}
\end{equation}
where $\ptj$ ($\ptl$) and $\etaj$ ($\etal$) are respectively 
the jet (lepton) transverse momentum and pseudorapidity, and 
$\Delta R_{jj'}$, $\Delta R_{jl}$, $\Delta R_{ll'}$ denote the jet-jet,
jet-lepton and lepton-lepton separations.

In the next sections we will present our analysis for each of the 
three channels of Eq.~(\ref{eqn:channels}) assuming
a value $m_h=180\, \gev$ for the Higgs mass. 
A qualitative discussion on the dependence of our results on the Higgs 
mass will be given in Section~\ref{sec:Higgsmass}.

\subsection{Channel ${\cal S}_3$: three leptons
plus one hadronically-decaying $W$}
\label{sec:sig3L}

\begin{figure}[tpb]
\begin{center}
\includegraphics[width=0.485\textwidth,clip,angle=0]{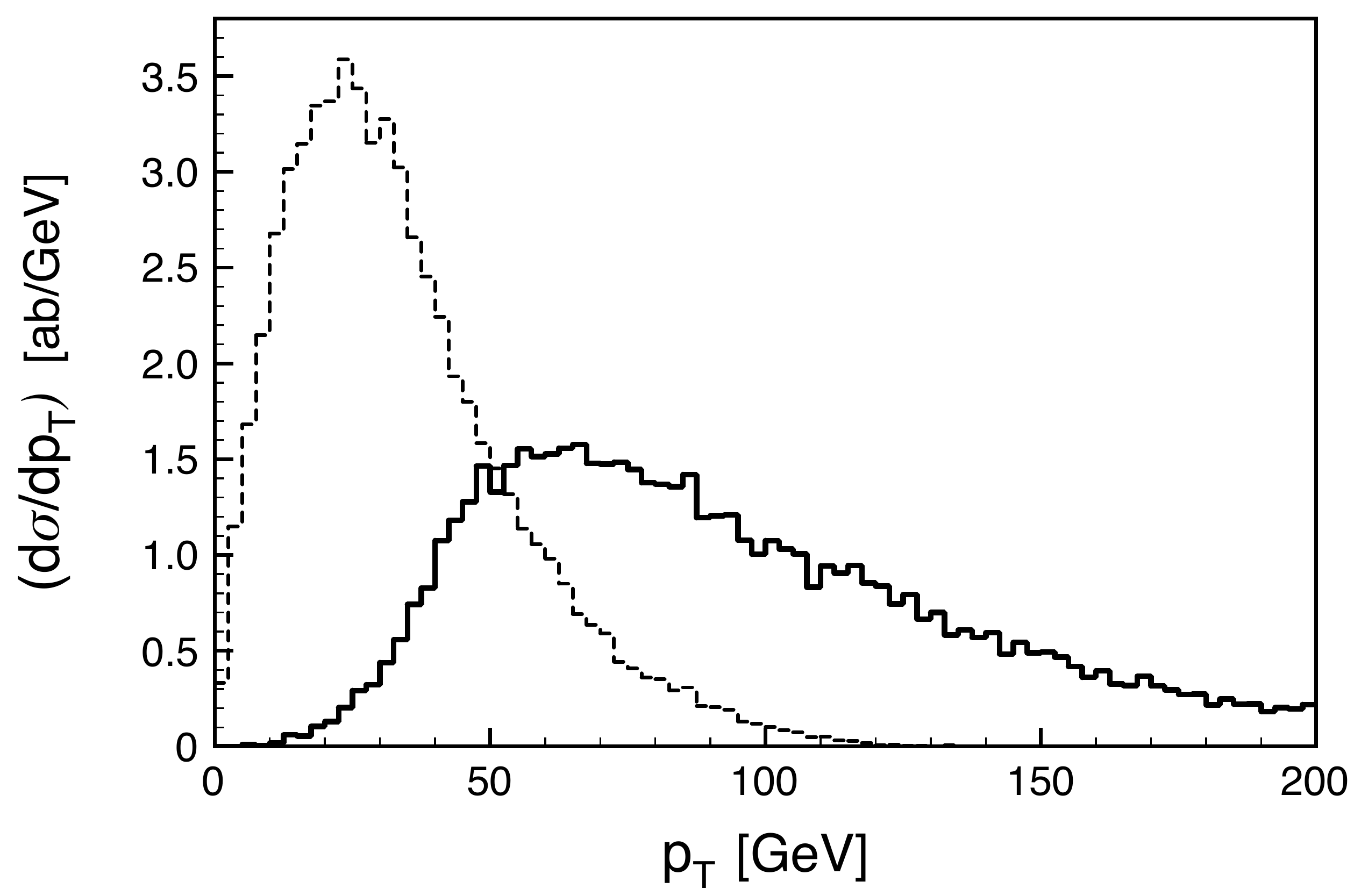}
\hspace{0.2cm}
\includegraphics[width=0.485\textwidth,clip,angle=0]{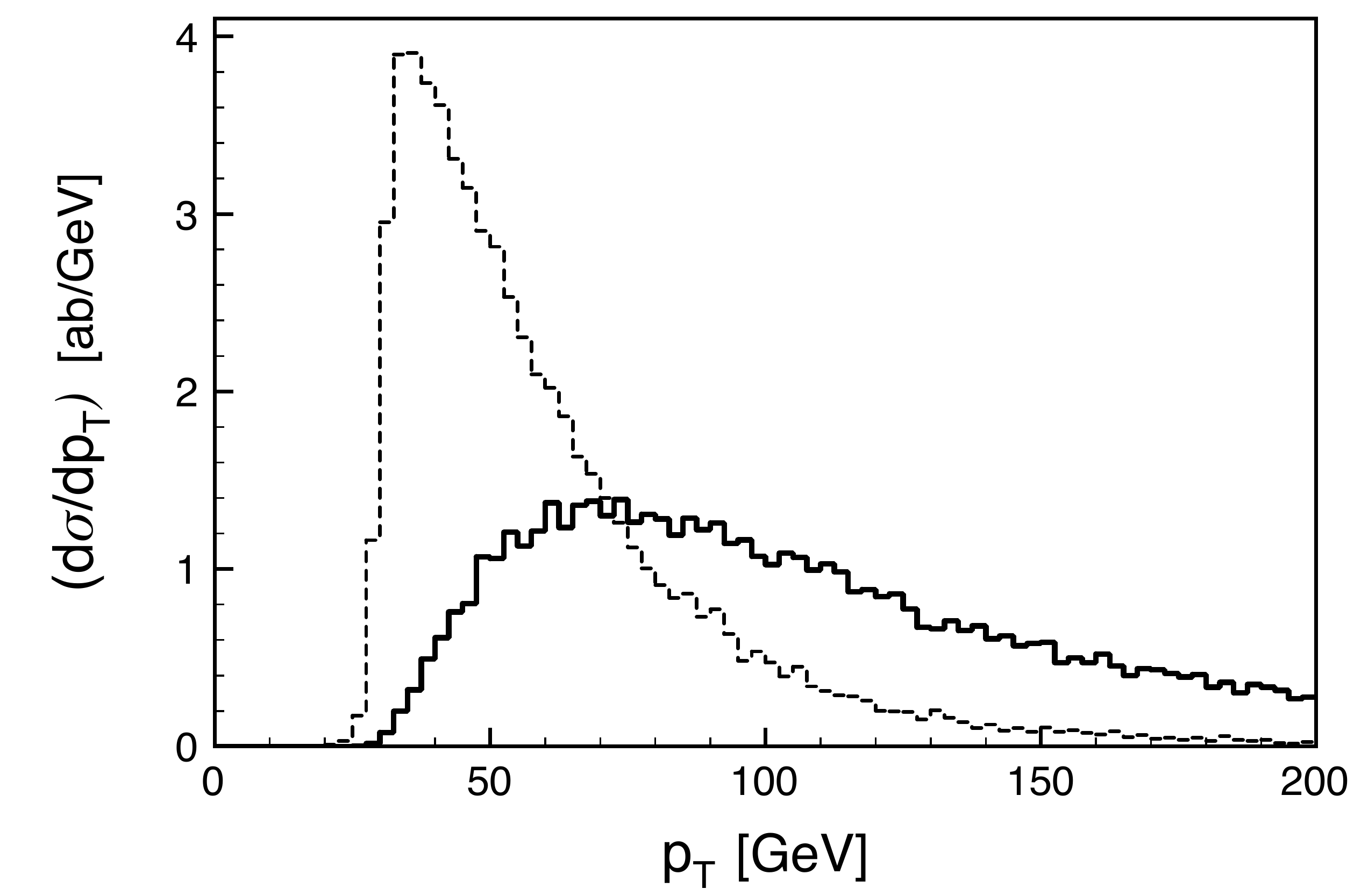}
 \caption[]{
 \label{fig:ptj3l}
\small
Differential cross section of the signal ${\cal S}_3$ in the MCHM4 with $\xi=1$ 
as a function of the transverse momentum of the jets.
On the left: jets from the $W$ decay; On the right: jets from the
primary interaction. Continuous line: hardest jet; Dashed line: second hardest jet.
Cuts as in Eq.~(\ref{eqn:acceptance}) have been applied, except no cut on the $\ptj$ of the 
jets from the $W$ decay has been applied on the left plot.  Jets from the primary interaction
on the right plot are required to satisfy $\ptj > 30\,$GeV.
}
\end{center}
\end{figure}

Perhaps the most promising final state channel is that with three leptons.
The signal is characterized by two widely separated jets (at least one in the forward region)
and up to two additional jets from the hadronically decaying $W$.
By using the definition of ``jet'' given in Eq.~(\ref{eqn:acceptance})
and working at the parton level, we find that the fractions of events with 4, 3 and 2~jets
are, respectively, $40\%$, $56\%$ and $4\%$.
Considering that the background  cross sections decrease by roughly a factor three for each additional jet,
{\em we will require at least 4 jets.}
In the case of the signal this choice allows the  reconstruction of  the  hadronic $W$, which gives an additional handle
to improve the signal to background ratio, as discussed in the following.
Signal events with less than 4 jets 
mostly arise when  some of the jets from the hadronic $W$ decay
are too soft to meet the $\ptj$ acceptance cut, while
only $\sim 30\%$ of the times two quarks merge into a single jet.
In  a more detailed and realistic analysis it is certainly 
worthwhile to explore the possibility of relaxing the
constraint on $\ptj$ at least for the softer jets.
Figure~\ref{fig:ptj3l} shows the signal cross
section as a function of the jets' transverse momentum.
%

In the second column of Table~\ref{tab:signbcks3L} we report the cross sections
after the acceptance cuts of Eq.~(\ref{eqn:acceptance}) for the signal and for
the main backgrounds that we have studied. 
%
\begin{table}[tb]
\begin{center}
\begin{tabular}{|l||l|l|l|l|l|}
\hline
{\tt Channel} & $\sigma_1$ & $\sigma_2$ & 
 $\sigma_3$  & $\sigma_4^{CMS}$  & $\sigma_4^{ATLAS}$   \\ 
\hline
\hline
${\cal S}_3$ $(\text{MCHM4} - \xi=1)$ & 30.4 & 27.7  & 16.8 & 16.7 & 16.4 \\
\hline
${\cal S}_3$ $(\text{MCHM4} - \xi=0.8)$ & 20.4 & 18.7 & 11.2 & 11.2 & 11.0 \\
\hline
${\cal S}_3$ $(\text{MCHM4} - \xi=0.5)$ & 9.45 & 8.64  & 5.26 & 5.24 & 5.14 \\
\hline
\hline
${\cal S}_3$ $(\text{MCHM5} - \xi=0.8)$ & 29.4 & 26.7 & 15.4 & 15.4 & 15.1 \\
\hline
${\cal S}_3$ $(\text{MCHM5} - \xi=0.5)$ & 14.8 & 13.6 & 7.88 & 7.85 & 7.71 \\
\hline
\hline
${\cal S}_3$ $(\text{SM} - \xi=0)$ & 1.73 & 1.34  & 0.75 & 0.75 & 0.73 \\
\hline
\hline
$ W l^+ l^- 4j$ & 12.0 $\times 10^3$  & 658 & 4.07 & 3.35 & 2.47 \\
\hline
$ W l^+ l^- 5j$ & 3.83 $\times 10^3$  & 16.6 & 0.13 & 0.08 & 0.00 \\
\hline
$  h l^+ l^- jj\to WW l^+ l^- jj$ & 102  & 29.7 & 0.50 & 0.50 & 0.49 \\
\hline
$ W W W  4j$ & 86.2 & 3.47 & 0.35 & 0.28 & 0.23 \\
\hline
$t \bar t Wjj $ & 408  & 11.3 & 0.66 & 0.55 & 0.37 \\
\hline
$t \bar t W jjj $ & 287 & 2.40 & 0.15 & 0.12 & 0.09 \\
\hline
$t \bar t W W  $ & 315 & 4.48 & 0.02 & 0.02 &  0.02\\
\hline
$t \bar t W W j $ & 817 & 28.1 & 1.40 & 1.16 & 0.89 \\
\hline
$t \bar t h jj  \to t \bar t W W j j $ & 610 & 8.89 & 0.65 & 0.52 & 0.38 \\
\hline
$t \bar t h jjj \to t \bar t W W j j j $ & 329 & 0.84  & 0.05 & 0.04 & 0.03 \\
\hline
$ W \tau^+ \tau^- 4j$ & 206 & 11.5 & 1.26 & 1.05 & 0.68 \\
\hline
\hline
Total background & 18.9 $\times 10^3$ & 775 & 9.23 & 7.66 & 5.65 \\
\hline
\end{tabular}
\caption[]{
\label{tab:signbcks3L} 
\small 
Cross sections, in  ab,  for the signal ${\cal S}_3$ (see
Eq.~(\ref{eqn:channels})) and for the main backgrounds after imposing
the cuts of Eq.~(\ref{eqn:acceptance}) ($\sigma_1$); 
of Eqs.~(\ref{eqn:acceptance}) and (\ref{eqn:0levcut}) ($\sigma_2$);
of Eqs.~(\ref{eqn:acceptance})--(\ref{eqn:1levcut}) ($\sigma_3$);
of Eqs.~(\ref{eqn:acceptance})--(\ref{eqn:mwcms}) ($\sigma_4^{CMS}$);
of Eqs.~(\ref{eqn:acceptance})--(\ref{eqn:1levcut}) and (\ref{eqn:mwatlas}) ($\sigma_4^{ATLAS}$).
For each channel, the proper branching fraction to a three-lepton final state (via $W\to l\nu, q\bar q$ and $\tau \to l\nu\nu_\tau$)
has been included.
}
\end{center}
\end{table}
A few comments are in order:
\bit 

\item The samples $t\bar t W (W)+ n\,  jets$ 
with $n$ larger than the minimal value are enhanced because of
two main reasons:
\begin{enumerate}

\item jets originating from the top decay can be too soft and fail
to satisfy the acceptance cut. Having extra available jets thus 
increases the efficiency of the acceptance cuts.

\item jets originating from the top decay are mostly central
in rapidity, which makes the occurrence of a pair with a large
dijet invariant mass and at least one jet forward
(one of the requirements that we will impose to improve the detectability 
of our signal) quite rare.
Additional jets from initial state radiation
are instead more likely to emerge with large rapidity.

\end{enumerate}

Notice that including all the samples $t\bar t W (W)+ n\, jets$ at the partonic level is 
redundant and in principle introduces a problem of double counting.
A correct procedure would be resumming soft and collinear emissions by means of a parton shower,
which effectively accounts for Sudakov form factors, and matching with the hard matrix element
calculation by means of some procedure to avoid double counting of jet emissions.
Here we retain all the $t\bar t W (W)+ n\, jets$ contributions,
as the cuts that we will impose on extra hadronic activity make the events with  additional jets almost completely negligible, 
solving in this way the problem of double counting.

\item Events with additional jets are much less important
for the $Wll$ backgrounds, where already at leading order the jets
can originate from a QCD interaction. This is clearly illustrated in 
Table~\ref{tab:signbcks3L} by the small cross section of $Wll5j$ after the cuts.

\item  For $m_h=180\,\gev$ the bulk of the contribution to $t\bar t W W+ n\, jets$
is via Higgs production and decay: $t\bar t h + n\, jets\to t\bar t W W^{(*)}+ n\, jets$.
Given the complexity of the final state, for $n=2,3$ we have computed this latter simpler signal 
as a good approximation of $t\bar t W W+ n\, jets$.

\item There is no overlap between $t\bar t W W$ and
$t\bar t W jj$, since the latter has been computed at order
$O(\alpha_{EW})$ and as such it does not
include contributions from intermediate $W^*\to jj$.

\item The process $WWW4j$ includes the resonant contributions $WWWWjj\to WWW4j$,
$hW4j\to WWW4j$ and $hWWjj\to WWWWjj\to WWW4j$.  
For simplicity, since $WWW4j$ represents only a small fraction of the total background at the end of the analysis,
the Higgs resonant contributions have not been separately reported in this case.

\item The process $Wl^+l^-4j$ includes the Higgs resonant contribution $hWjj\to ZZWjj$ with $ZZ\to l^+l^- jj$.
This accounts for less than $7\%$ of the total $Wl^+l^-4j$, and has not been reported separately for simplicity.

\item The process $W \tau^+ \tau^- 4j$ leads to a three-lepton
final state provided both $\tau$'s decay leptonically.
It is clearly subdominant compared to $W l^+ l^- 4j$, but
it is at the same time much less reduced by the cut on the 
dilepton invariant mass $\mSFOS$ which we impose in the following 
(see Eq.~(\ref{eqn:1levcut})). 
For this reason it must be included in the list of relevant backgrounds.

\eit

\begin{figure}[tbp]
\begin{center}
\includegraphics[width=0.485\textwidth,clip,angle=0]{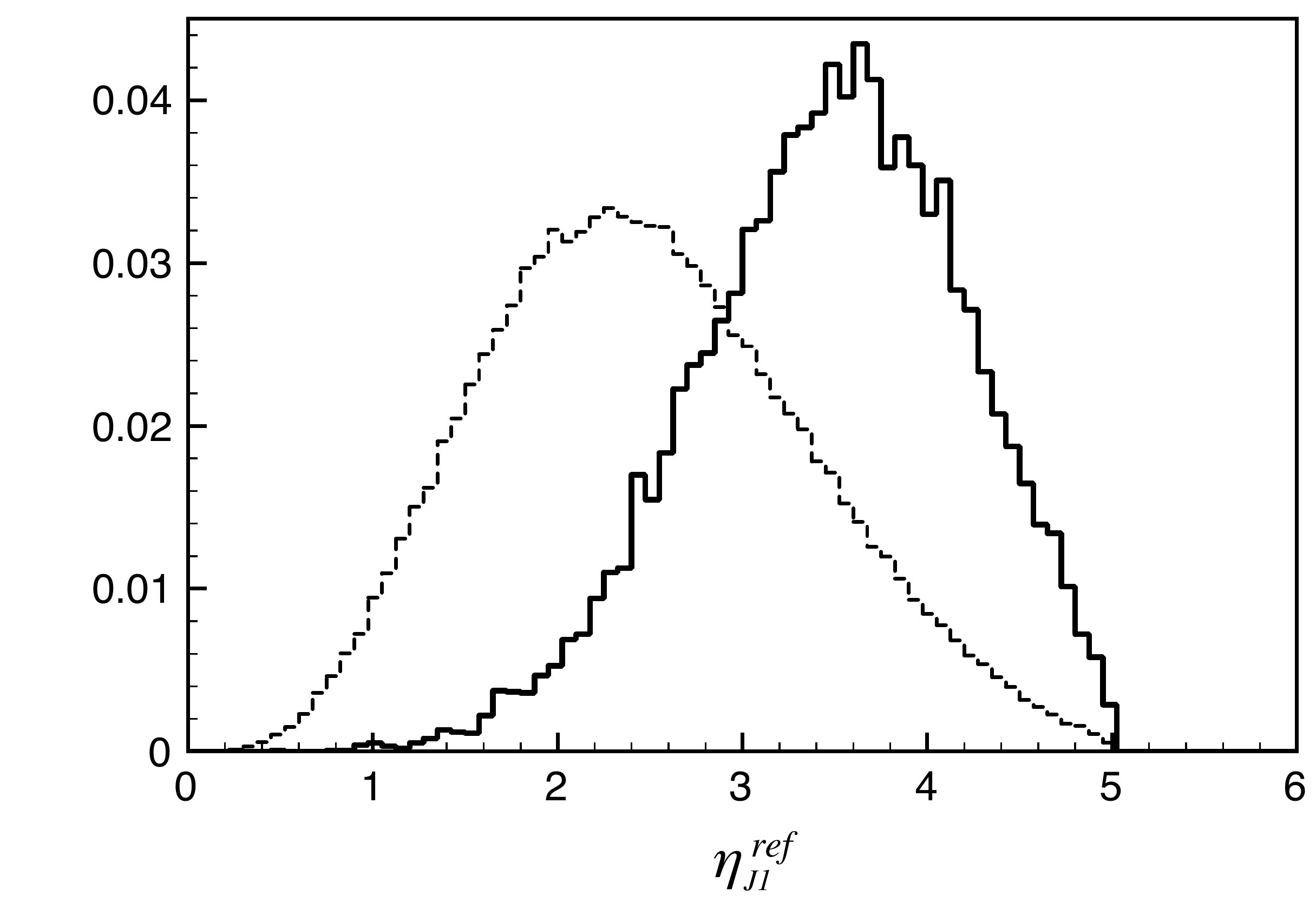}
\hspace{0.2cm}
\includegraphics[width=0.485\textwidth,clip,angle=0]{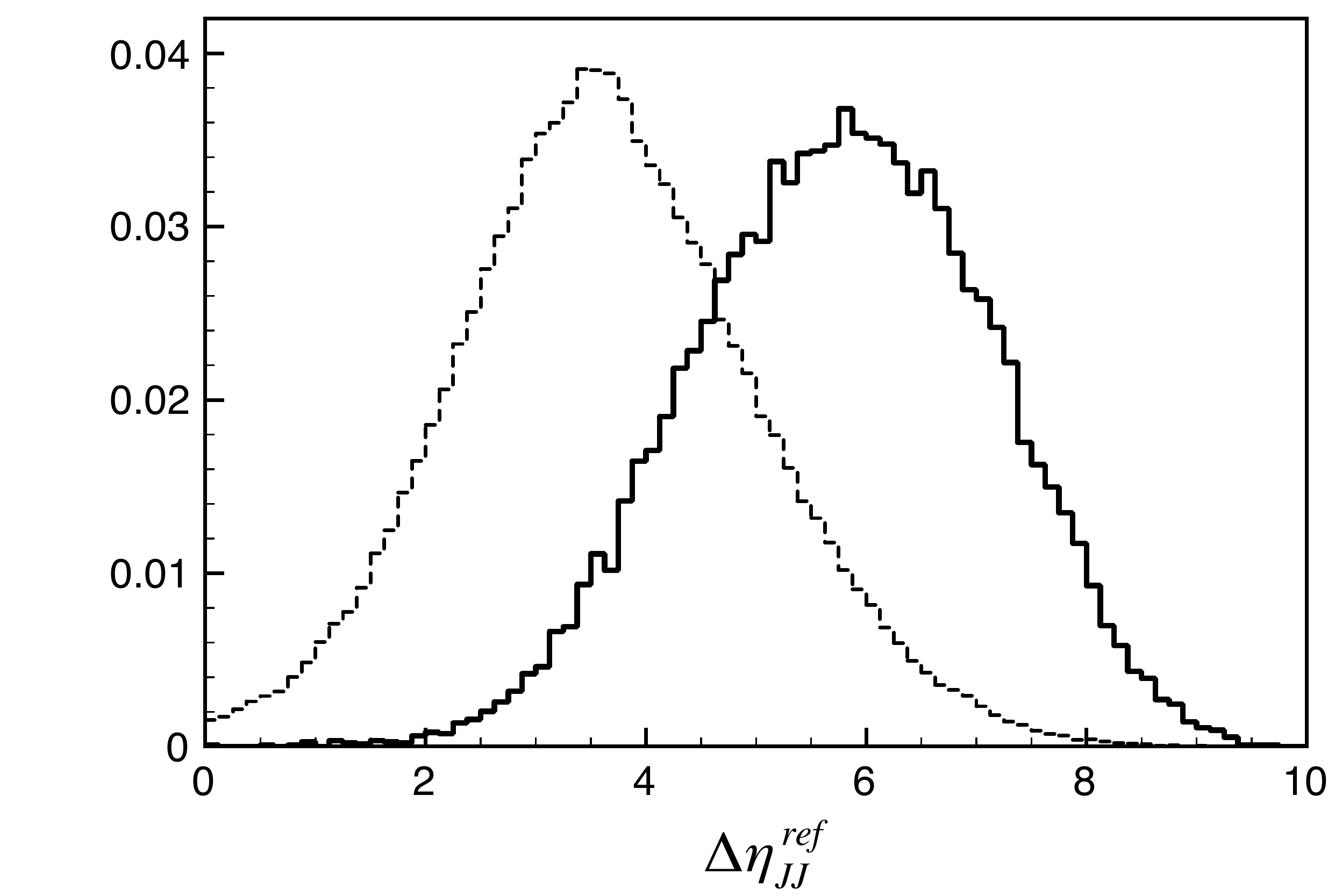}
\\[0.2cm]
\includegraphics[width=0.485\textwidth,clip,angle=0]{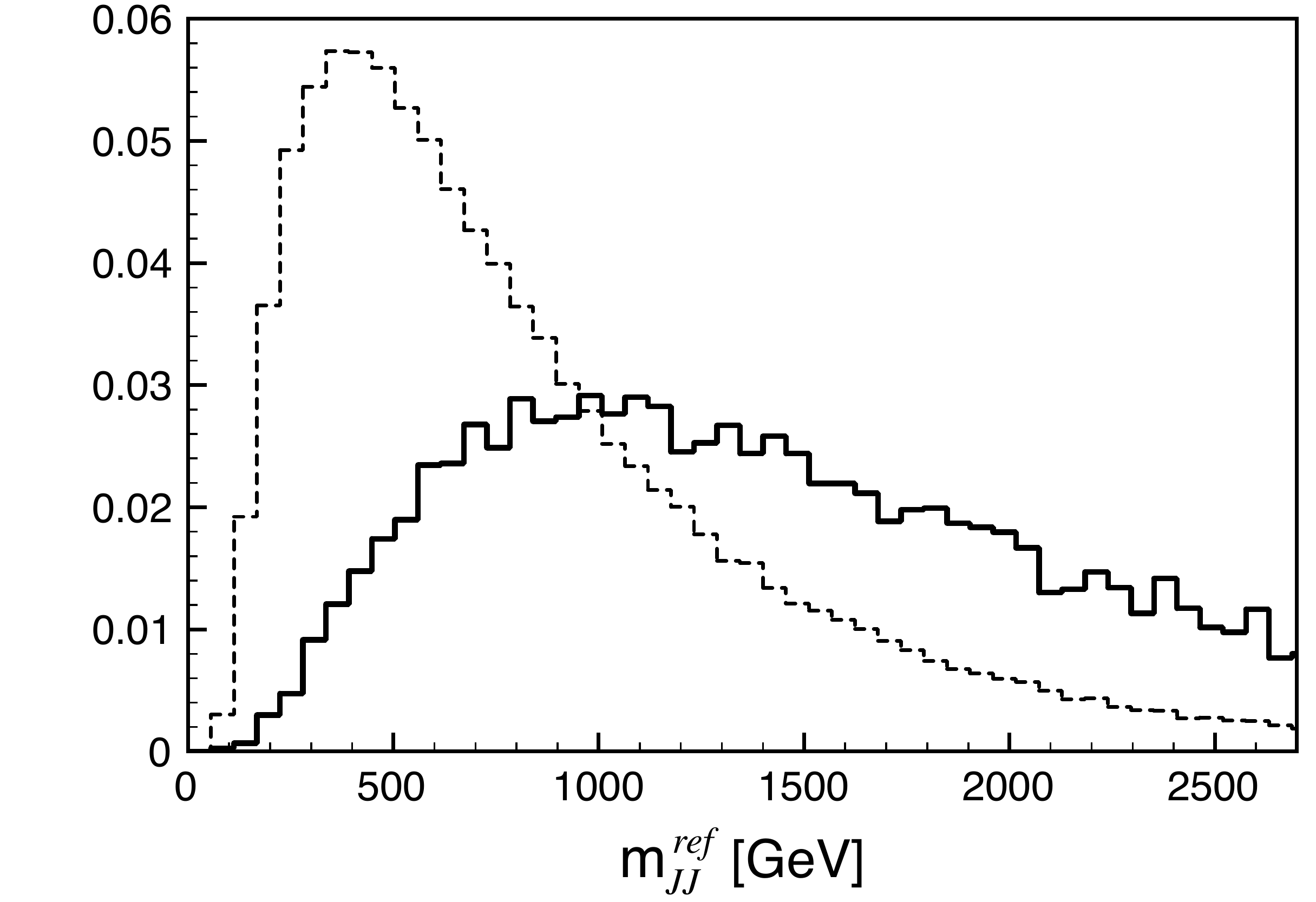}
 \caption[]{
 \label{fig:eta-deta-acccuts}
\small
Differential cross sections after the acceptance cuts of  Eq.~(\ref{eqn:acceptance}) for the 
signal ${\cal S}_3$ in the MCHM4 at $\xi=1$ (continuous line) and the background (dashed line).
Upper left plot: rapidity of the most forward jet (in absolute value);
Upper right plot: separation between the two reference jets;
Lower plot: invariant mass of the two reference jets.
All curves have been normalized to unit area.
}
\end{center}
\end{figure}

\begin{figure}[htbp]
\begin{center}
\includegraphics[width=0.485\textwidth,clip,angle=0]{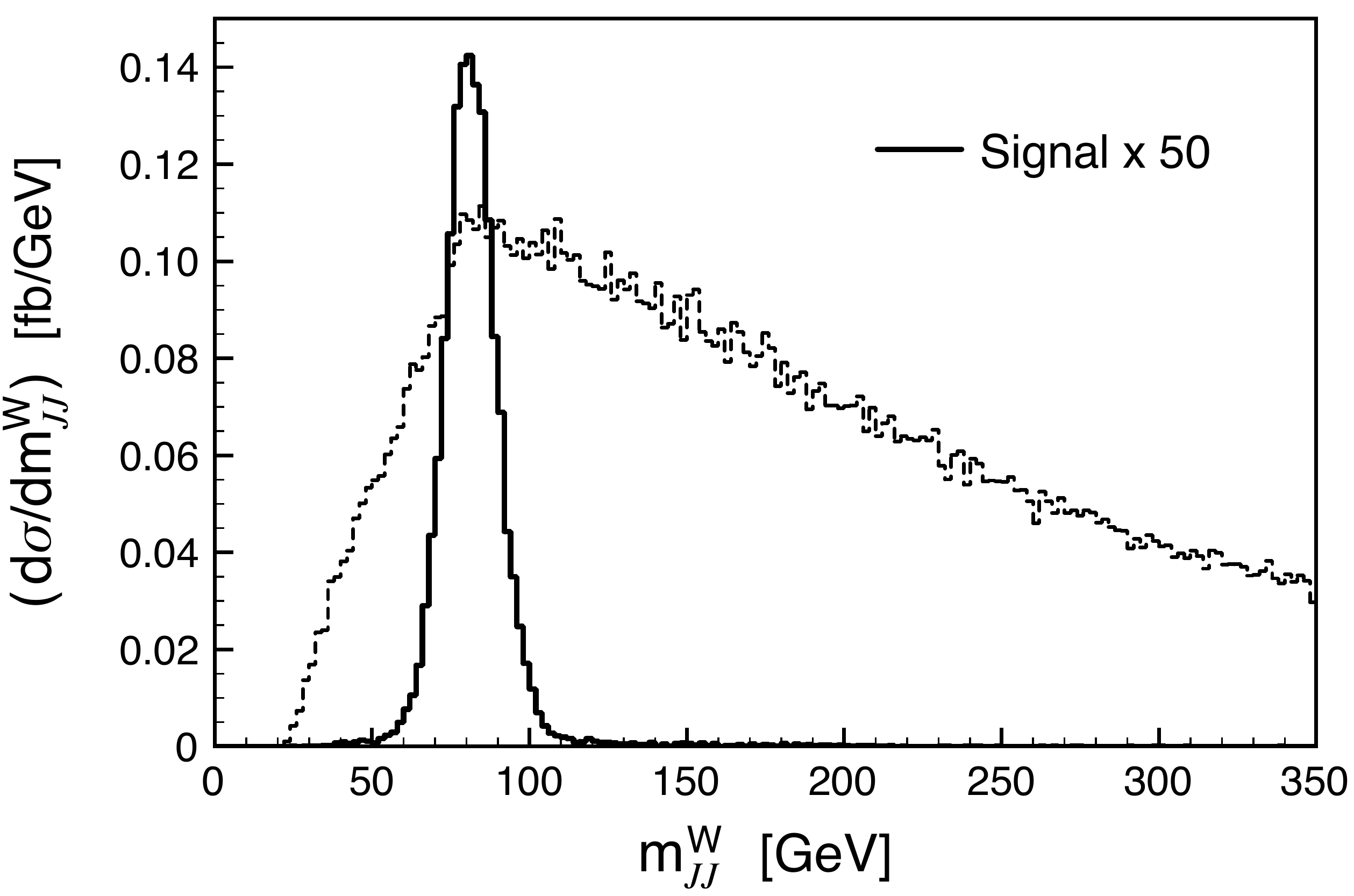}
\caption[]{
\label{fig:whadmass}
\small
Differential cross section  as a function of the invariant mass of all the non-reference jets
for the  signal ${\cal S}_3$ in the MCHM4 at $\xi=1$ (continuous line) and the background (dashed line)
after the acceptance cuts of  Eq.~(\ref{eqn:acceptance}).
}
\end{center}
\end{figure}

As clearly seen from Table~\ref{tab:signbcks3L}, after the acceptance cuts the  background is still by far dominating over the signal. 
We therefore try to exploit the peculiar kinematics of the signal, which is  
distinctive of vector boson fusion events: two widely separated jets with a least one at large rapidity.
We will refer to these two jets as ``reference'' jets in the following.  To identify them we first select the jet with the largest absolute rapidity, 
and we then compute the dijet invariant mass it forms with each one of the remaining jets: the two reference jets will be those
forming the largest dijet invariant mass.~\footnote{In the case of the signal this procedures selects, at the partonic level, the two jets which 
are not produced in the $W$ decay with an efficiency of $\sim 0.97$ ($\sim 0.90$) for $\xi \ge 0.5$ ($\xi=0$). A similar result is obtained using
$\Delta \eta_{JJ}$ to select the reference jets. At the partonic level $m_{JJ}$ looks slightly better,
although this has to be confirmed by a more detailed analysis.}
Figure~\ref{fig:eta-deta-acccuts} shows the
rapidity of the most forward jet (first reference jet), $\eta_{J1}^{ref}$,
the invariant mass of the two reference jets, $m_{JJ}^{ref}$,
and their separation, $\Delta \eta^{ref}_{JJ} = |\eta_{J1}^{ref}-\eta_{J2}^{ref}|$.
%
%
In the case of the signal, the remaining jets will reconstruct  a $W$ boson. In Fig.~\ref{fig:whadmass} we plot the invariant mass of all 
the jets other than the reference ones, $m^W_{JJ}$, for both the signal~\footnote{Obviously, the distribution 
for the signal has a Breit-Wigner peak with a small continuous tail due to events where jets from the decay of the $W$ have been
chosen as reference jets.  The  experimental resolution on the dijet mass is much larger than the $W$ width, and this has to be
properly taken into account if we wish to use this observable to improve the significance of the signal.
At the rough level of our analysis, this will be  taken into account by selecting an appropriate mass window around the $W$ mass.}
and the background.
%
%

\begin{figure}[tbp]
\begin{center}
\includegraphics[width=0.485\textwidth,clip,angle=0]{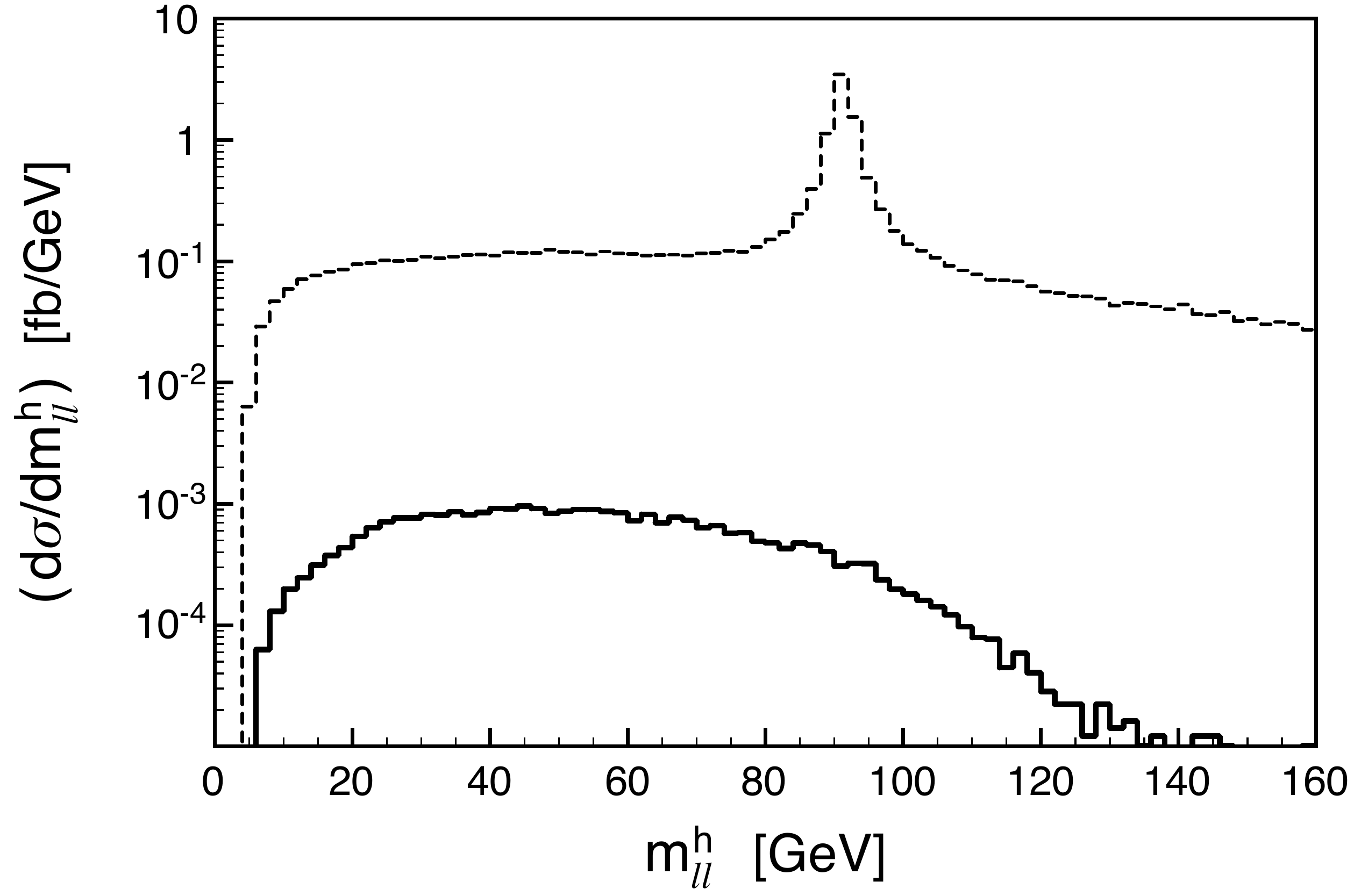}
\hspace{0.2cm}
\includegraphics[width=0.485\textwidth,clip,angle=0]{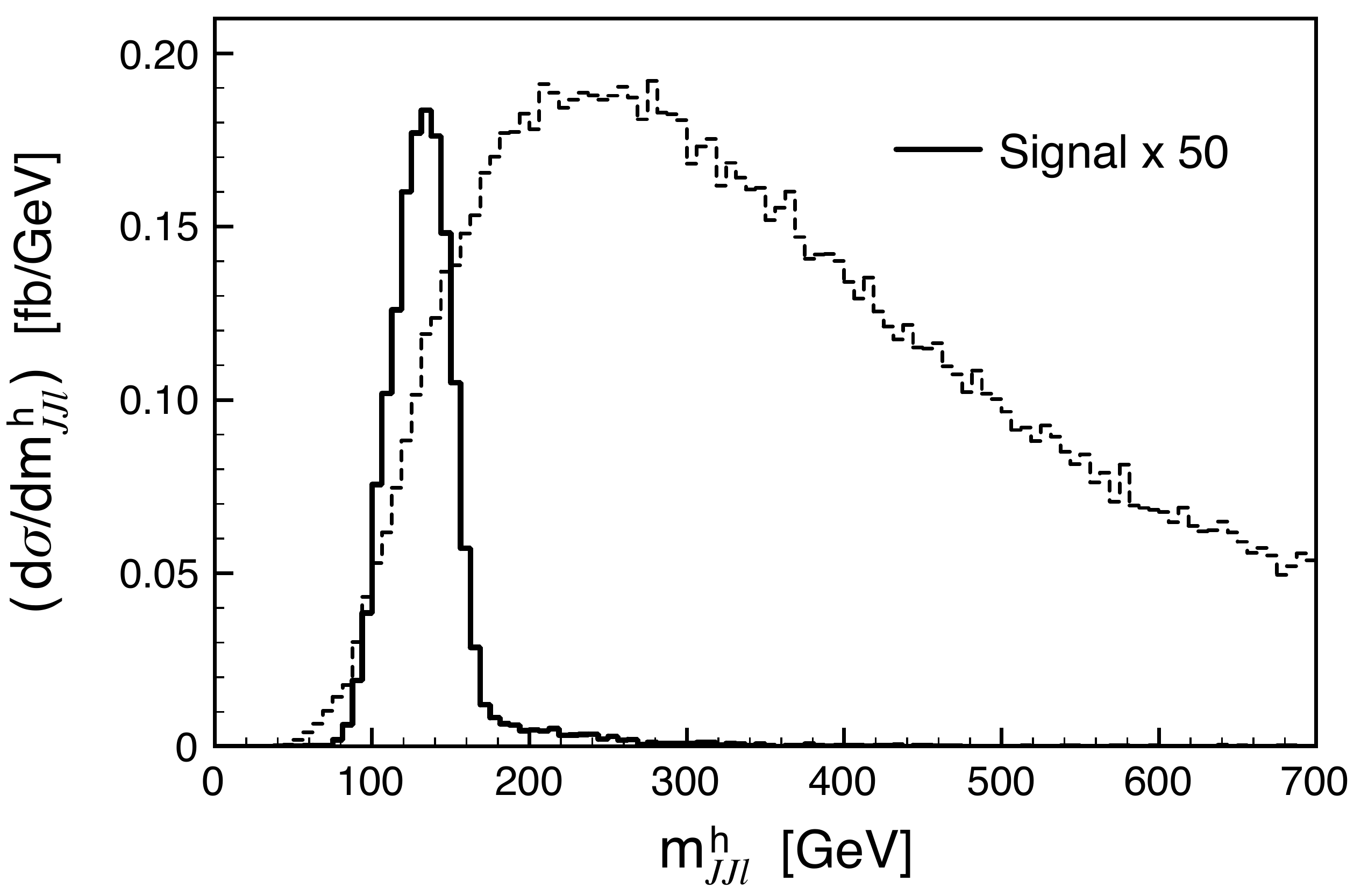}
\caption[]{
\label{fig:mh1mh2}
\small
Differential cross sections after the acceptance cuts of  Eq.~(\ref{eqn:acceptance}) for the 
signal ${\cal S}_3$ in the MCHM4 at $\xi=1$ (continuous line) and the background (dashed line).
Left plot: invariant mass of the two leptons forming the first Higgs candidate;
Right plot: invariant mass of the lepton plus jets forming the second Higgs candidate.
}
\end{center}
\end{figure}

A second crucial feature of the signal is that there are two Higgs bosons in the final state: one decaying fully
leptonically, the other semileptonically. The two leptons from the leptonically-decaying Higgs can be 
identified as those forming the opposite-charge pair with the smallest relative angle.
Both lepton spin correlations and the boost of the Higgs in the laboratory frame
favour this configuration. For example, for a final state $e^+ \mu^+ e^- X$, we compute $\cos \theta_{e^+e^-}$
 and $\cos \theta_{\mu^+e^-}$ and we pick up the pair with the largest cosine.~\footnote{Here $\theta_{i j}$ is defined as the angle between the 
directions of  particle $i$ and particle $j$.} 
Figure~\ref{fig:mh1mh2} shows the mass of this lepton pair, $m_{ll}^h$,  for both the signal and
the background. The other Higgs boson candidate is reconstructed as the sum of the
remaining lepton plus all the jets different from the reference ones; its mass,
$m_{JJl}^h$, is also shown in Fig.~\ref{fig:mh1mh2}.
%
%

As a first set of cuts, we use the observables discussed above and require that each individual
cut reduces the signal by no more than $\sim 2\%$. We demand:
\begin{equation}
\begin{aligned}
& |\eta_{J1}^{ref}| \ge  1.8 \quad && m^{ref}_{JJ} \ge 320 \ \gev \quad 
 && \Delta \eta^{ref}_{JJ}  \ge  2.9 \\[0.2cm]
& |m^W_{JJ}-m_W| \le  40 \ \gev   \quad && m_{ll}^h  \le  110 \ \gev  \quad 
 && m_{JJl}^h \le  210 \ \gev
\end{aligned}
\label{eqn:0levcut}
\end{equation}
Signal and background cross sections after this set of cuts are
reported as $\sigma_2$ in  Table~\ref{tab:signbcks3L}. 
We first notice that all the backgrounds with a number of jets 
larger than four have been strongly reduced: this is mostly due to the cuts on $m_{JJ}^W$ and
on $m_{JJl}^h$, that heavily penalize events with a large available jet energy.
This is the reason why we can neglect the problem of double counting
introduced by including samples with arbitrary number of jets: after the
cuts of Eq.~(\ref{eqn:0levcut}) are imposed, the events with a too large number 
of jets are essentially rejected.

\begin{figure}[tbp]
\begin{center}
\includegraphics[width=0.485\textwidth,clip,angle=0]{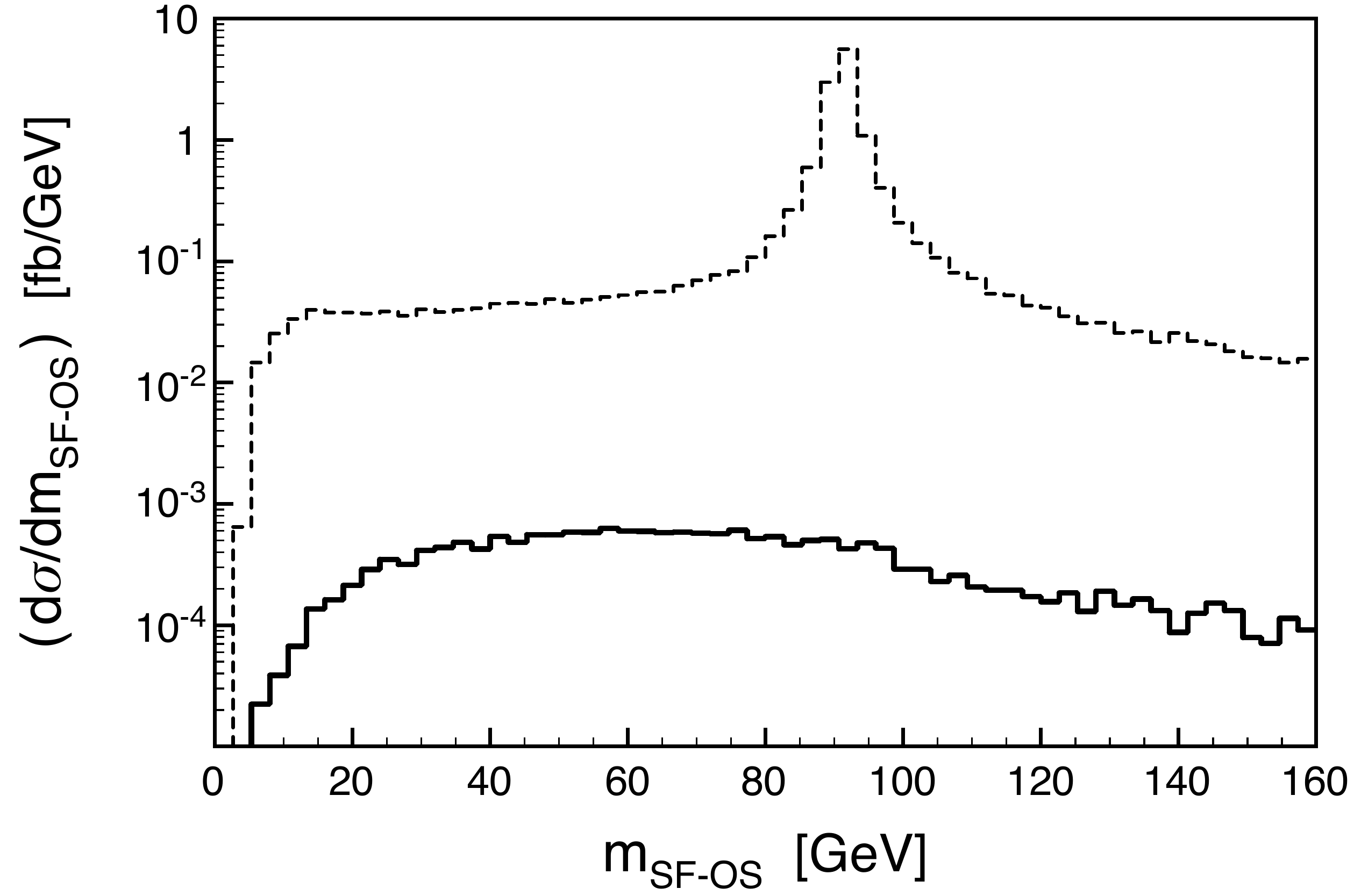}
\caption[]{
\label{fig:zmass}
\small
Differential cross section  after the acceptance cuts of  Eq.~(\ref{eqn:acceptance})
as a function of the invariant mass, $\mSFOS$, of the $e^+e^-$ or $\mu^+\mu^-$ pair.
Whenever two such pairings are possible the mass closer to $M_Z$ is selected. 
Continuous line:  signal ${\cal S}_3$ in the MCHM4 at $\xi=1$; Dashed line: background.
}
\end{center}
\end{figure}

We now proceed to identify the cuts which are most effective for improving
the significance of our signal. We first notice that the largest background,
$W l^+ l^- 4j$, has a dominant contribution from the $Z$ resonance. 
In Fig.~\ref{fig:zmass} we plot the invariant mass, $\mSFOS$, of the $e^+e^-$ or $\mu^+\mu^-$ pair
found in the event. If two such pairings are possible 
(this is the case when the three leptons in the final state all have the same flavor),
the invariant mass closer to $M_Z$ is selected.
%
%
It is clear that the significance of the signal can be largely improved by
excluding values of $\mSFOS$ that are in a window around the $Z$ pole
or close to the photon pole.

We searched for the optimal set of cuts on $\mSFOS$ and other possible distributions
(including all those mentioned above and shown in Figs.~\ref{fig:eta-deta-acccuts}--\ref{fig:zmass} 
by following an iterative procedure: at each
step we cut over the observable which provides the largest enhancement of the signal significance,
until no further improvement is possible. The significance has been computed performing 
a goodness-of-fit test of the background-only hypothesis with Poisson statistics.~\footnote{
\label{fot:signif}
Given the number of signal and background events a p-value is computed using the Poisson distribution.
The significance is  defined as the number of standard deviations that a Gaussian variable would fluctuate
in one direction to give the same p-value.
For example,  a p-value = $2.85 \times 10^{-7}$ corresponds to a $5\sigma$ significance.}
We assumed 300 fb$^{-1}$ (3000 fb$^{-1}$) of integrated luminosity at the LHC 
(at the LHC luminosity upgrade).
We end up with the following set of additional cuts:
\begin{equation}
\begin{gathered}
\mSFOS \ge 20 \, \gev \qquad  |\mSFOS-M_Z| \geq 7\, \Gamma_Z \\[0.2cm]
\Delta \eta_{JJ}^{ref} \ge 4.5  \qquad
 m_{JJ}^{ref} \ge 700\, \gev \qquad m_{JJl}^h \leq  160 \, \gev  \, ,
\end{gathered}
\label{eqn:1levcut}
\end{equation}
 $M_Z$ and $\Gamma_Z$ being respectively the $Z$ boson mass and width.
The cross sections for signal and backgrounds after these cuts 
are reported as $\sigma_3$ in Table~\ref{tab:signbcks3L}.

As a final set of cuts, we consider a further restriction on 
$m^W_{JJ}$ around the $W$ pole:
\begin{align}
|m^W_{JJ}- M_W| & <  30 \, \gev \label{eqn:mwcms} \\[0.2cm]
|m^W_{JJ}- M_W| & <  20 \, \gev \label{eqn:mwatlas} 
\end{align}
The cuts in Eqs.~(\ref{eqn:mwcms})-(\ref{eqn:mwatlas}) correspond to twice the expected invariant 
dijet mass resolution respectively for the {\tt CMS} and {\tt ATLAS} detector resolution. 
The corresponding final cross sections are denoted as $\sigma_4^{CMS}$ and $\sigma_4^{ATLAS}$ 
in Table~\ref{tab:signbcks3L}. 
An additional veto on $b$-jets has a relatively small impact, since it would reduce the $t\bar t W (W) + jets$ backgrounds 
which are however already subdominant.
Assuming for example a $b$-jet tagging efficiency of $\epsilon_b = 0.55$ for $\eta_b < 2.5$, the signal significances
increase by approximately $10\%$.

\subsubsection{Estimate of showering effects}

There is still one feature of the signal which has not been exploited yet. A unique signature
of vector boson fusion events is a very small hadronic  activity  in the central region (rapidities
between the first and second reference jet)~\cite{rapiditygap}. 
This is not the case for the backgrounds, especially after
imposing the cuts on $\Delta \eta^{ref}_{JJ}$ and $m_{JJ}^{ref}$ in Eq.~(\ref{eqn:1levcut}),
which imply a large total invariant mass $\sqrt{\hat s}$ for the event and therefore a stronger
radiation probability (the radiation probability is  proportional to $\log^2(\hat s/ \lambda^2)$, where
$\lambda$ is the infrared/collinear cut-off).
By vetoing this activity in the central region, one can then obtain an additional
suppression of the background without affecting much the signal.
For our event selection, the effect of the showering on the
background is twofold: a large number of jets appears in
the final state and, as a consequence,
both $m^W_{JJ}$ and $m^h_{JJl}$ are shifted towards larger
values.~\footnote{Let us denote as $X$ the system of final state jets 
other than the reference jets. If the additional radiation is from the $X$ system, $M_X$ will be unaffected, if instead
it is from initial state or from the reference jets the momentum of the radiation will add to that of the 
$X$ system increasing its mass.}
In order to assess the relative impact of  these effects, we have processed both the signal and the
most relevant background, $Wl^+l^-4j$, through the  parton shower {\tt PYTHIA}~\cite{pythia}, and we have reconstructed the final-state jets 
using a cone algorithm \`a la UA1, as implemented in the {\tt GETJET} \cite{getjet} routine.
To avoid mixing different and
unrelated effects, we have studied only {\em the relative efficiencies}
of the various cuts compared to the partonic level analysis. 

\begin{figure}[tbp]
\begin{center}
\includegraphics[width=0.485\textwidth,clip,angle=0]{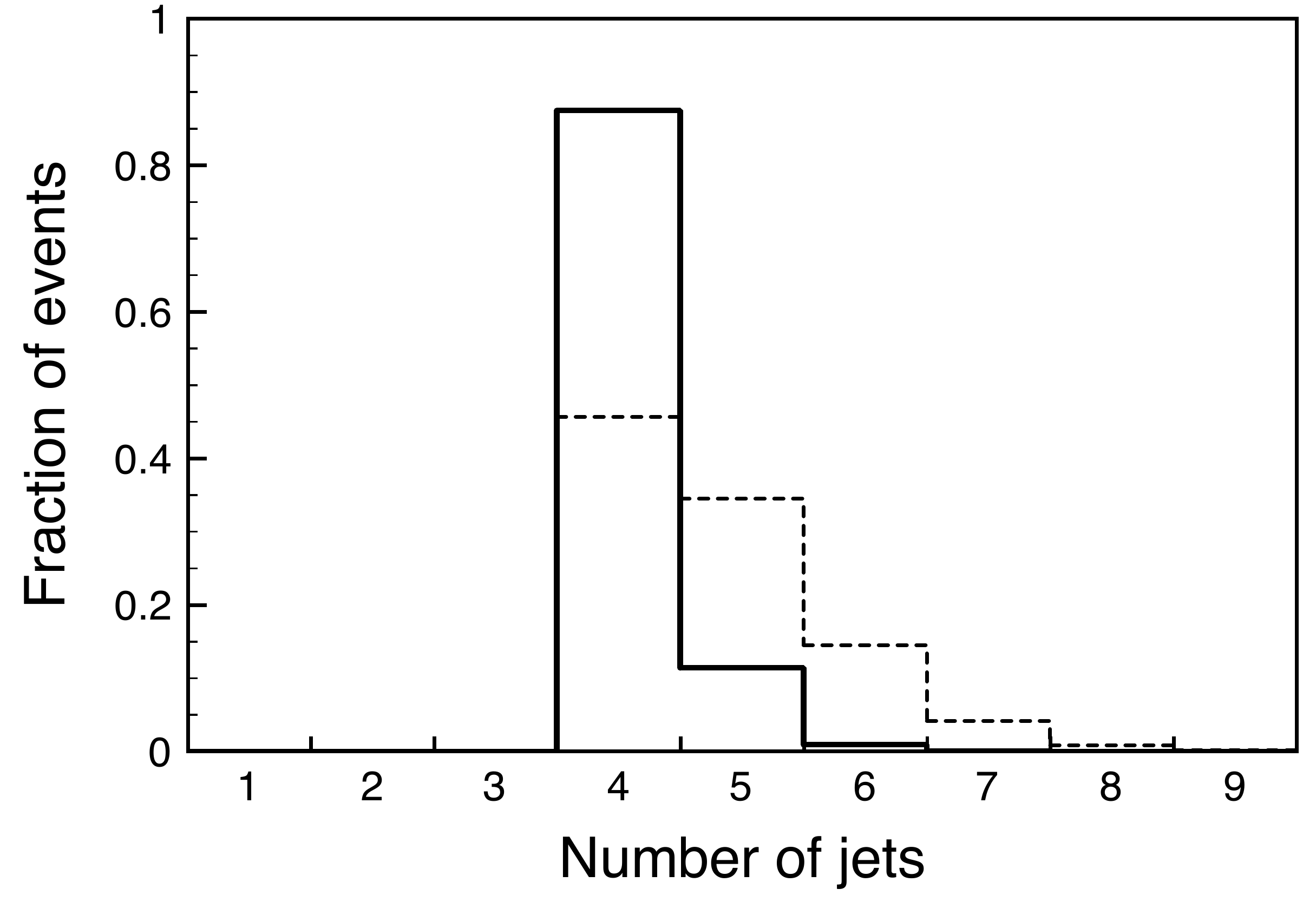}
\caption[]{
\label{fig:njets}
\small
Number of jets after showering with {\tt PYTHIA }
and imposing the acceptance cuts of Eq.~(\ref{eqn:acceptance}). 
Continuous line:  signal ${\cal S}_3$ in the MCHM4 with $\xi=1$; Dashed line: background.
Jets are reconstructed using the cone algorithm implemented in the {\tt GETJET} routine.
}
\end{center}
\end{figure}

\begin{figure}[tbp]
\begin{center}
\includegraphics[width=0.485\textwidth,clip,angle=0]{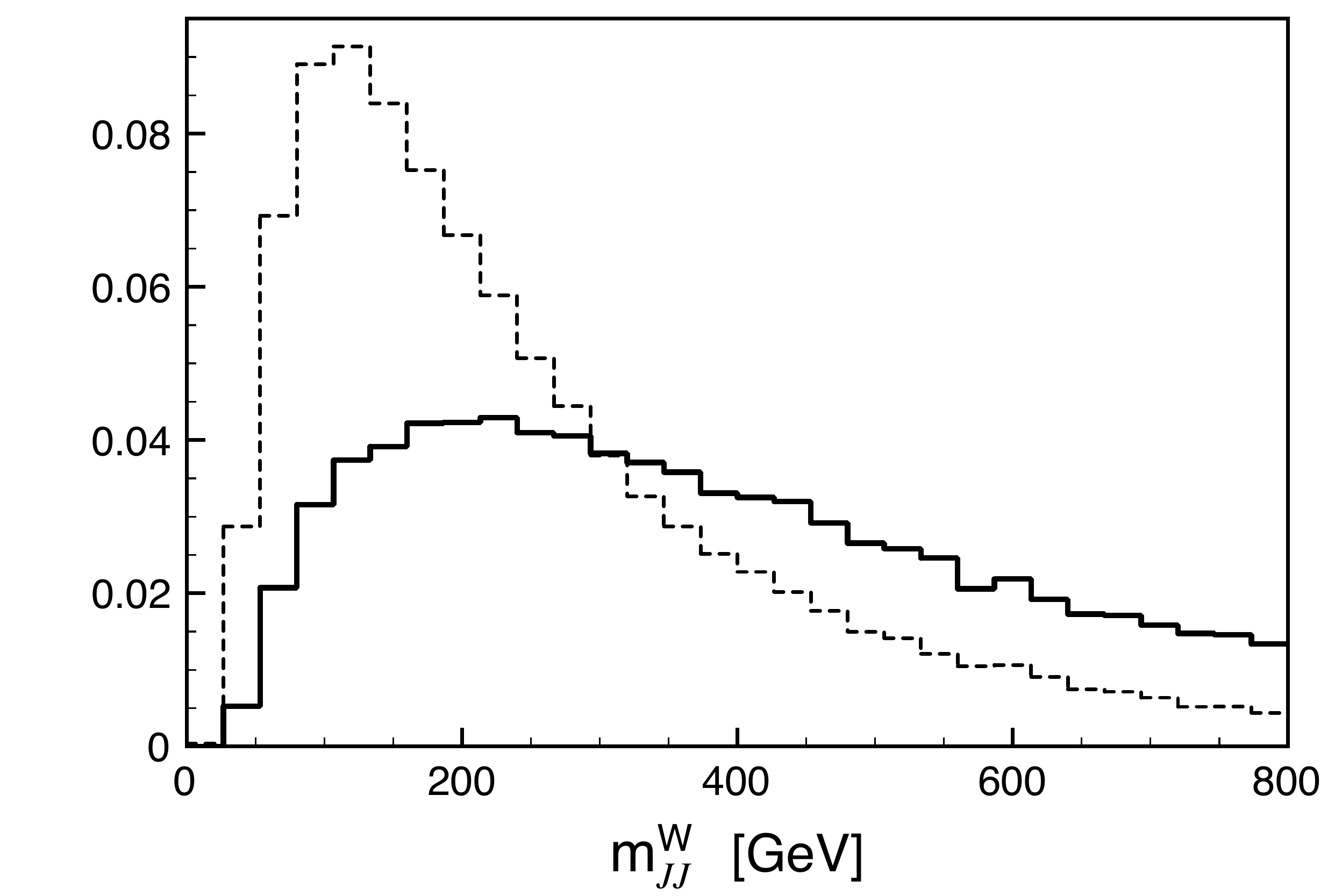}
\hspace{0.2cm}
\includegraphics[width=0.485\textwidth,clip,angle=0]{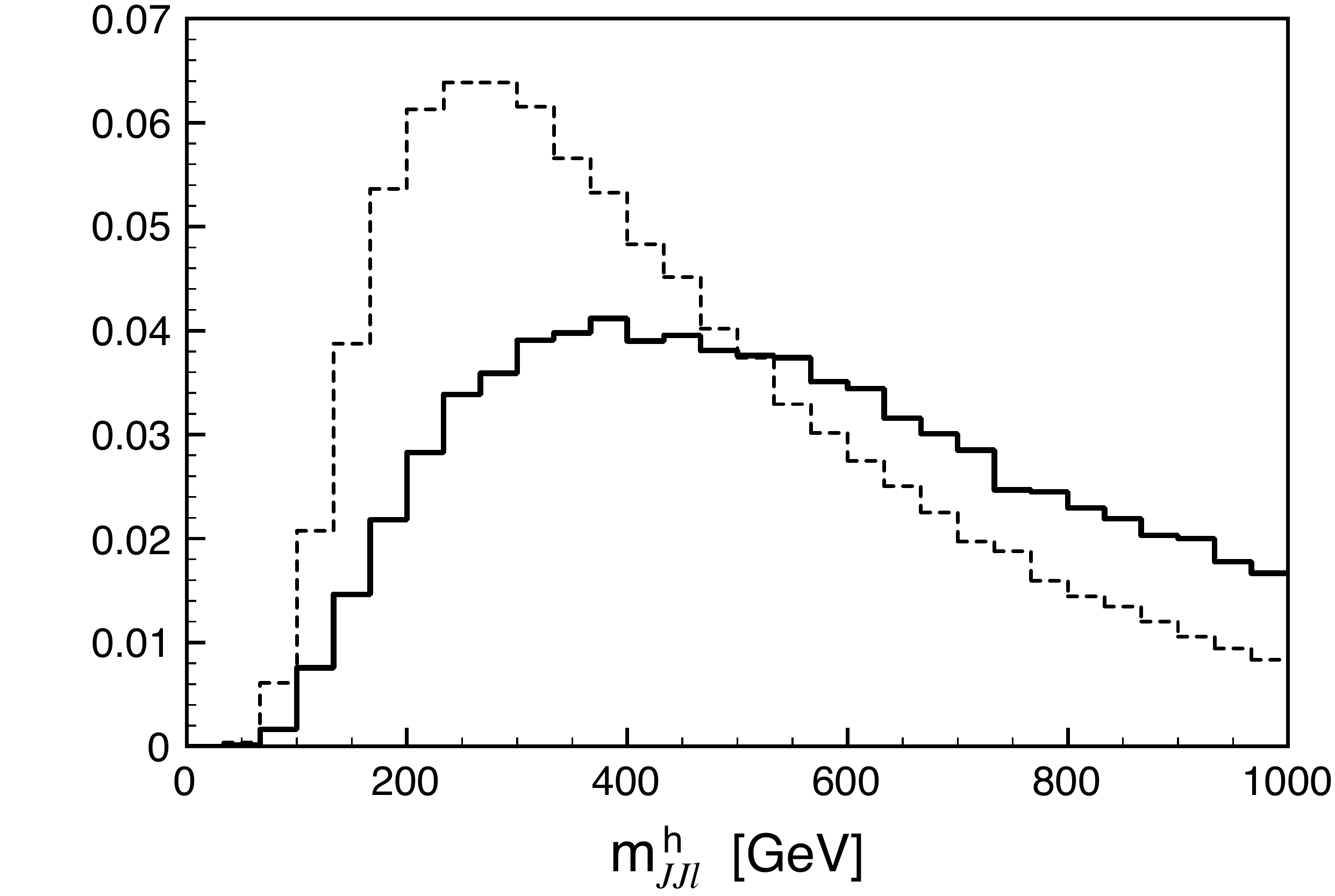}
\caption[]{
\label{fig:mwjjshower}
\small
Differential cross section for the background  $Wl^+l^- 4j$ after
the showering (continuous line) and at the parton level (dashed line)
as as a function of $m^W_{JJ}$ (left plot) and $m^h_{JJl}$ (right plot).
Only events which pass the acceptance cuts of Eq.~(\ref{eqn:acceptance}) and those 
on $\Delta \eta^{ref}_{JJ}$ and $m_{JJ}^{ref}$ of Eq.~(\ref{eqn:1levcut}) have been included.
}
\end{center}
\end{figure}

Figure~\ref{fig:njets} shows the
distribution of the number of jets for both the signal and
the $Wl^+l^-4j$ background after showering and imposing the acceptance
cuts of Eq.~(\ref{eqn:acceptance}). 
%
%
%
%
A comparison between the $m^W_{JJ}$ and $m^h_{JJl}$ background distributions,
as reconstructed at the parton and shower level
after imposing the cuts on $\Delta \eta_{JJ}^{ref}$ and $m_{JJ}^{ref}$ 
of Eq.~(\ref{eqn:1levcut}), is shown in Fig.~\ref{fig:mwjjshower}. 
%
%
Notice that these observables,  as well as the jet multiplicity, are strongly correlated, so that applying a cut on any one of them
strongly diminishes the efficiency on the others. 

A rough estimate of the  effect of the showering can be obtained by monitoring the collective efficiency of the cuts 
on $m_{JJ}^W$ (Eq.~(\ref{eqn:0levcut})) and on  $\Delta \eta_{JJ}^{ref}$, $m_{JJ}^{ref}$, $m_{JJl}^h$
(Eq.~(\ref{eqn:1levcut})). After showering, we find the following additional reduction on the 
signal and background rates compared to the partonic level:
\begin{center}
\begin{tabular}{c|cc}
   & ${\cal S}_3$ $(\xi=1,0.8,0.5)$ & \quad $ W l^+ l^- 4j$ \\[0.1cm]
\hline
$ \epsilon_{\mathrm shower}/\epsilon_{\mathrm parton}$ 
 & 0.8 & \quad 0.6
\end{tabular}
\end{center}
A further veto on events with more than 5 jets has a negligible impact, both for the signal
and the background, as the cuts on $m_{JJ}^W$ and $m_{JJl}^h$ effectively act like a veto
on extra hadronic activity. 
Although a full inclusion of showering effects can only be obtained by using matched
samples,  yet we expect that our rough estimate captures the bulk of the effect.

\subsubsection{Additional backgrounds from fake leptons}
\label{sec:fakelep3L}

Since the number of signal events at the end of our analysis is very small, it is important to check if there are additional
potential sources of reducible backgrounds. Here we consider the possibility that a jet is occasionally 
identified as a lepton, in which case we speak of a ``fake'' lepton from a jet.
We find that the effect of such jet mistagging is likely to  be negligible in the three lepton case as follows.

As shown in Table~\ref{tab:signbcks3L}, the dominant background in this case is $W l l 4j$. After the acceptance cuts
we have $\sigma_{pp \to W l^+ l^- 4j} = 12\,$fb. A first possibility is that a fake lepton (most likely an electron)
originates from the misidentification of a ``light jet'' (originated either from gluons or from a light quark).
In this case the most serious potential source of background is $ l l + 5j$. Since the relative cross section
after the acceptance cuts is $\sigma_{pp \to l^+ l^- + 5j} \simeq 2.8\,$pb, even a modest mistagging probability $\lesssim 10^{-3}$ 
(according to both CMS and ATLAS collaborations~\cite{CMS:TDR1,ATLAS:CERNOPEN}, 
rejection factors as small as $10^{-5}$ can be achieved by making the jet reconstruction 
algorithm tight enough) is sufficient to suppress this source of background.

A second possibility is that a heavy quark ($b$ or $c$) decays semileptonically and the resulting lepton is isolated.
Backgrounds of this type are $l^+ l^- b \bar b + 3j$ and $l^+ l^- c \bar c + 3j$, which have similar cross sections.
To estimate the first process we have computed the cross section for $pp \to l^+ l^- b \bar b + 3j$ where 
one of the two $b$'s is randomly chosen and assumed to be mistagged as a lepton.
After applying  the cuts of  Eqs.~(\ref{eqn:acceptance})--(\ref{eqn:mwcms}) we obtain a cross section of $1.2\,$fb. 
A $b$ mistagging probability $\sim 10^{-3}$ is therefore sufficient to keep this background 
below the irreducible background. This level of rejection seems feasible
at the LHC: in Ref.~\cite{CHMISID} a mistagging probability of $7 \times 10^{-3}$ is estimated
for a lepton with $p_T>10$ GeV, rapidly decreasing (by a factor 10 to 30 for $p_T>20$ GeV) with increasing~$p_T$.
A potentially more problematic contribution is $t \bar t + 3j$, whose cross
section after acceptance cuts is $\sigma_{pp \to t \bar t+ 3j}= 770\,$fb. 
A $b$ mistagging probability $\lesssim 10^{-3}$
makes this background at most as important as the other $t\bar t$ channels
in Table~\ref{tab:signbcks3L}, which however turn out to be subdominant at the end of the analysis.

We thus conclude that the effect of fake leptons is expected to be negligible in the three lepton case.

\subsection{Channel ${\cal S}_2$:  two  same-sign leptons
plus two hadronically-decaying $W$'s}
\label{sec:sig2L}

In the case of a two-lepton final state, in order to keep the 
background at a manageable level, and avoid the otherwise overwhelming $t\bar t$ background,
we are forced to select only events with two leptons with the same charge.

Along with the two leptons, the signal is characterized by two widely 
separated jets and up to four additional jets from the two hadronically-decaying $W$'s.
Using the definition of ``jet'' given in Eq.~(\ref{eqn:acceptance})
and working at the parton level, we find that 
in the majority of the events at least one quark from a $W$ decay is either
too soft to form a jet or it merges with another quark to form one single jet.
The fractions of signal events with 6, 5, 4 and 3 jets are respectively 
$0.16$, $0.43$, $0.37$ and $0.04$.
{\em We choose to retain events with at least 5 jets}.
Including events with a lower jet multiplicity is not convenient, as the
background increases by a factor $\sim 3$ for each jet less, and the identification of the
Higgs daughters in the signal becomes less effective.

In order to suppress the otherwise overwhelming $W l^{+} l^{-} \!+ jets$
background, we forbid the presence of extra hard isolated leptons: we require to have
\bit

\item[] {\em exactly two leptons} (with the same charge)
satisfying the acceptance cuts of Eq.~(\ref{eqn:acceptance}). 

\eit
In this way the resonant contribution $W Z + jets \to W l^{+} l^{-} \!+ jets$ is strongly suppressed. 
Other backgrounds that can have 3 leptons in their final 
state at the partonic level are also reduced.~\footnote{These backgrounds are:
$t \bar t W jjj$, $t \bar t W W j$, $t \bar t h jj  \to t \bar t W W jj$,
$t \bar t h jjj \to t \bar t W W jjj$ and $ W \tau^+ \tau^- 5j$.}

In the second column of Table~\ref{tab:signbcks2L}, we report
the cross sections after the acceptance cuts of Eq.~(\ref{eqn:acceptance})
for the signal ${\cal S}_2$ and for the main backgrounds we have studied. 
%
\begin{table}[!tp]
\begin{center}
\begin{tabular}{|l||l|l|l|l|l|}
\hline
{\tt Channel} & $\sigma_1$ & $\sigma_2$ & $\sigma_3$ & $\sigma_4^{CMS}$  & $\sigma_4^{ATLAS}$\\ 
\hline
\hline
${\cal S}_2$ $(\text{MCHM4} -\xi=1)$ & 69.4 & 62.8 & 51.8 & 51.3 & 49.9 \\
\hline
${\cal S}_2$ $(\text{MCHM4} - \xi=0.8)$ & 47.0 & 42.6 & 34.9 & 34.6 & 33.7 \\
\hline
${\cal S}_2$ $(\text{MCHM4} - \xi=0.5)$ & 22.2 & 20.1 & 16.9 & 16.7 & 16.2 \\
\hline
\hline
${\cal S}_2$ $(\text{MCHM5} - \xi=0.8)$ & 68.5 & 61.8 & 50.0 & 49.4 & 47.8 \\
\hline
${\cal S}_2$ $(\text{MCHM5} - \xi=0.5)$ & 35.5 & 32.2 & 26.4 & 26.1 & 25.3 \\
\hline
\hline
${\cal S}_2$ $(\text{SM} - \xi=0)$ & 4.51 & 3.52 & 2.87 & 2.84 & 2.76 \\
\hline
\hline
$ W l^+ l^- 5j$ & 2.23 $\times 10^{3}$  & 200 & 61.8 & 55.1 & 42.1 \\
\hline
$ W^{+(-)} W^{+(-)}   5j$ & 700 & 53.3 & 13.8 & 11.7 & 8.91 \\
\hline
$ W W W jjj$ & 194 & 29.5 & 8.65 & 8.49 & 8.18 \\
\hline
$ h W jjj$ & 97.0 & 29.2 & 12.5 & 12.4 & 12.1 \\
\hline
$ W W W W j$ & 5.94 & 0.63 & 0.11 & 0.11 & 0.11 \\
\hline
$ W W W W jj$ & 10.9 & 1.40 & 0.52 & 0.52 & 0.49 \\
\hline
$t \bar t W j$ & 929  & 89.0 & 13.4 & 12.9 & 12.0 \\
\hline
$t \bar t W jj$ & 1.64 $\times 10^3$ & 134 & 25.7 & 23.2 & 19.6 \\
\hline
$t \bar t W jjj$ & 1.18 $\times 10^3$ & 44.6 & 8.52 & 7.48 & 6.04 \\
\hline
$t \bar t W W$ & 886 & 24.1 & 1.27 & 1.24 & 1.15 \\
\hline
$t \bar t W W j$ & 1.65 $\times 10^3$ & 173 & 28.4 & 26.6 & 23.2 \\
\hline
$t \bar t h jj  \to t \bar t W W jj$  & 1.27 $\times 10^3$ & 98.6 & 18.7 & 17.2 & 14.5 \\
\hline
$t \bar t h jjj \to t \bar t W W jjj$  & 732 & 21.3 & 3.99 & 3.64 & 3.07 \\
\hline
$ W \tau^+ \tau^- 4j$ & 655 & 78.2 & 22.8 & 20.0 & 15.9 \\
\hline
$ W \tau^+ \tau^- 5j$ & 463 & 31.7 & 8.77 & 7.88 & 6.19 \\
\hline 
\hline
Total Background & 12.7 $\times 10^3$ & 1.01 $\times 10^3$ & 229 & 209 & 174 \\
\hline
\end{tabular}
\caption[]{
\label{tab:signbcks2L} 
\small
Cross sections, in ab, for the signal ${\cal S}_2$ (see Eq.~(\ref{eqn:channels}))
and for the main backgrounds after imposing the cuts of
Eq.~(\ref{eqn:acceptance}) ($\sigma_1$); of 
Eqs.~(\ref{eqn:acceptance}) and (\ref{eqn:0levcut2L}) ($\sigma_2$);
of Eqs.~(\ref{eqn:acceptance}) and (\ref{eqn:0levcut2L})-(\ref{eqn:1levcut2L}) ($\sigma_3$);
of Eqs.~(\ref{eqn:acceptance}) and (\ref{eqn:0levcut2L})--(\ref{eqn:mwcms-SS2L}) ($\sigma_4^{CMS}$);
of Eqs.~(\ref{eqn:acceptance}), (\ref{eqn:0levcut2L}), (\ref{eqn:1levcut2L}) and (\ref{eqn:mwatlas-SS2L}) ($\sigma_4^{ATLAS}$).
For each channel the proper branching fraction to a same-sign dilepton
final state (via $W\to l\nu, q\bar q$ and $\tau \to l\nu\nu_\tau, q\bar q \nu_\tau$) has been included.
For the decay modes of the taus, see text.
In the case of the background $Wl^+l^-5j$, the lepton with different
sign is required to fail the acceptance cuts of Eq.~(\ref{eqn:acceptance}), see text.}
\end{center}
\end{table}
A few comments are in order (comments made for Table~\ref{tab:signbcks3L} also apply and will not be repeated here):
\bit 

\item While the cross section for $WW$ production is obviously much larger than the cross section for
$WWW$ production, those for $WWW$ and $W^{+(-)}W^{+(-)}$ (equal sign) are comparable,
so that both these latter backgrounds must be included.

\item The background 
$WWWWj$ includes the resonant contribution $WWhj \to WWWWj$. For simplicity, since $WWWWj$
represents only a small fraction of the total background at the end of the analysis, the Higgs resonant
contribution has not been reported separately.
There is no overlap between $WWWWj$ and $W W W jjj$, since the latter has been generated at order $O(\alpha_{EW}^3)$
and as such it does not include contributions from intermediate $W^* \to jj$. 

\item The process $W\tau^+\tau^- 4j$ leads to a dilepton final state if one $\tau$ decays leptonically and the other
is mistagged as a QCD jet.~\footnote{We thank James Wells for pointing out to us the importance of the processes
$W\tau^+\tau^- 4j$ and $W\tau^+\tau^- 5j$ as potential backgrounds.}
We have conservatively assumed that the momentum of the mistagged jet is equal to that of the parent $\tau$,
and we have included a mistagging probability at the end of our analysis.

\item The process $W\tau^+\tau^- 5j$ leads to a dilepton final state
if one $\tau$ decays leptonically and the other is either not detected 
(independently of its decay mode), or it decays hadronically and it is mistagged as a QCD jet.
We include the mistagging probability at the end of our analysis, when we
impose a veto on hadronic taus in the event.
The momentum of the mistagged jet has been assumed to be equal to that of the parent~$\tau$.

\item Included in the cross sections of the processes $t\bar t WWj$, $t\bar thjj$, $t\bar thjjj$ and $t\bar tWjjj$ is the contribution 
of the three leptons final state where both tops decay leptonically and the wrong-sign lepton fails the acceptance cut.
The analog contribution from $t\bar tW4j$ has been computed and found to be very small, and for simplicity is not reported here.

\item If required for trigger issues, the cut on the hardest lepton can be increased to $p_T > 30$~GeV at basically no cost for the signal (the efficiency relative to the acceptance cuts of Eq.~(\ref{eqn:acceptance}) is 97\%). ÊThis should be sufficient to pass the high-level trigger at CMS and ATLAS even during the high-luminosity phase of the LHC.
Furthermore, the presence of a huge amount of hadronic energy in the signal might help to reduce the trigger requirements on the 
$p_T$ of the leptons.

\eit

\begin{figure}[tbp]
\begin{center}
\includegraphics[width=0.485\textwidth,clip,angle=0]{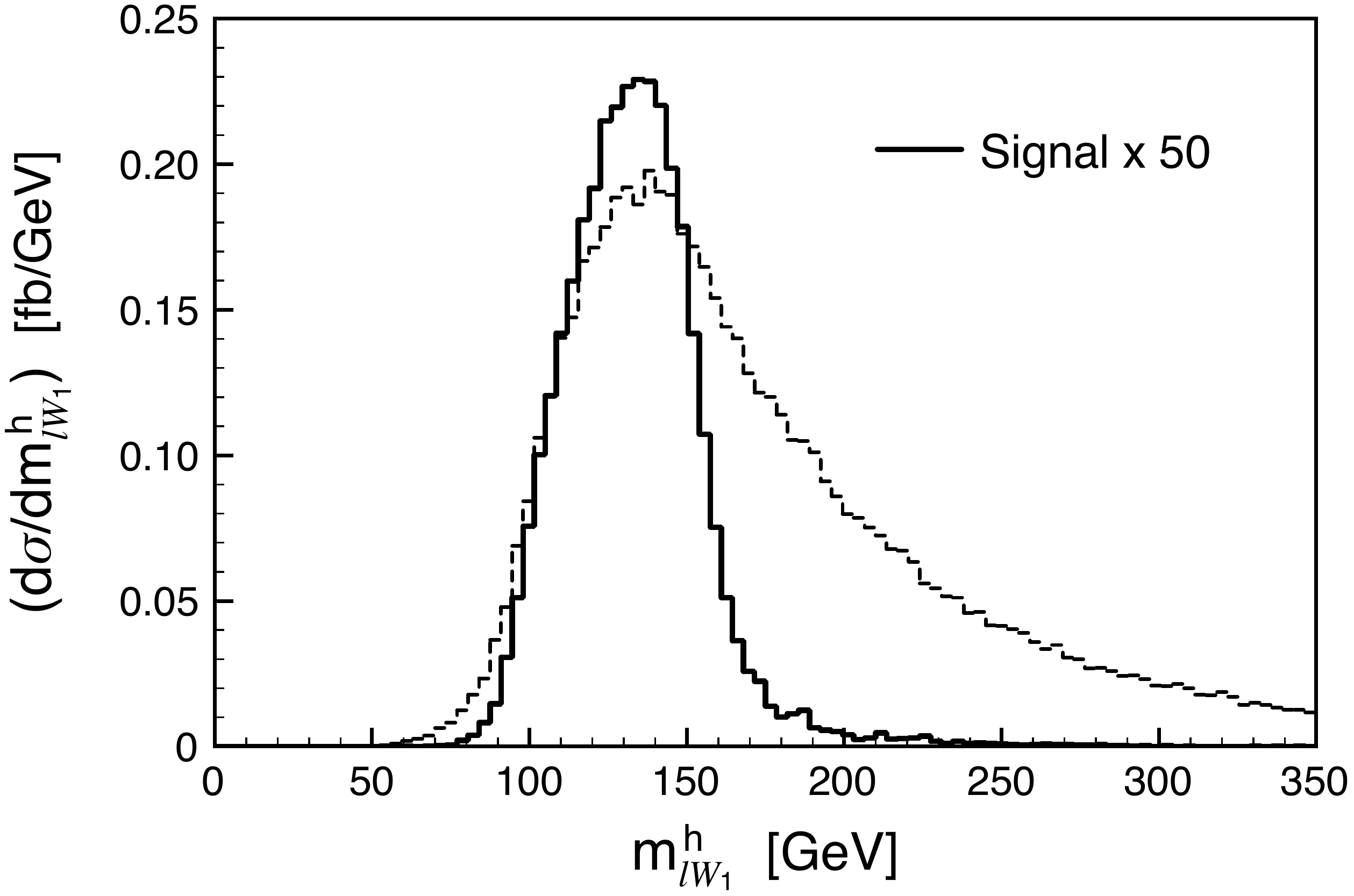}
\hspace{0.2cm}
\includegraphics[width=0.485\textwidth,clip,angle=0]{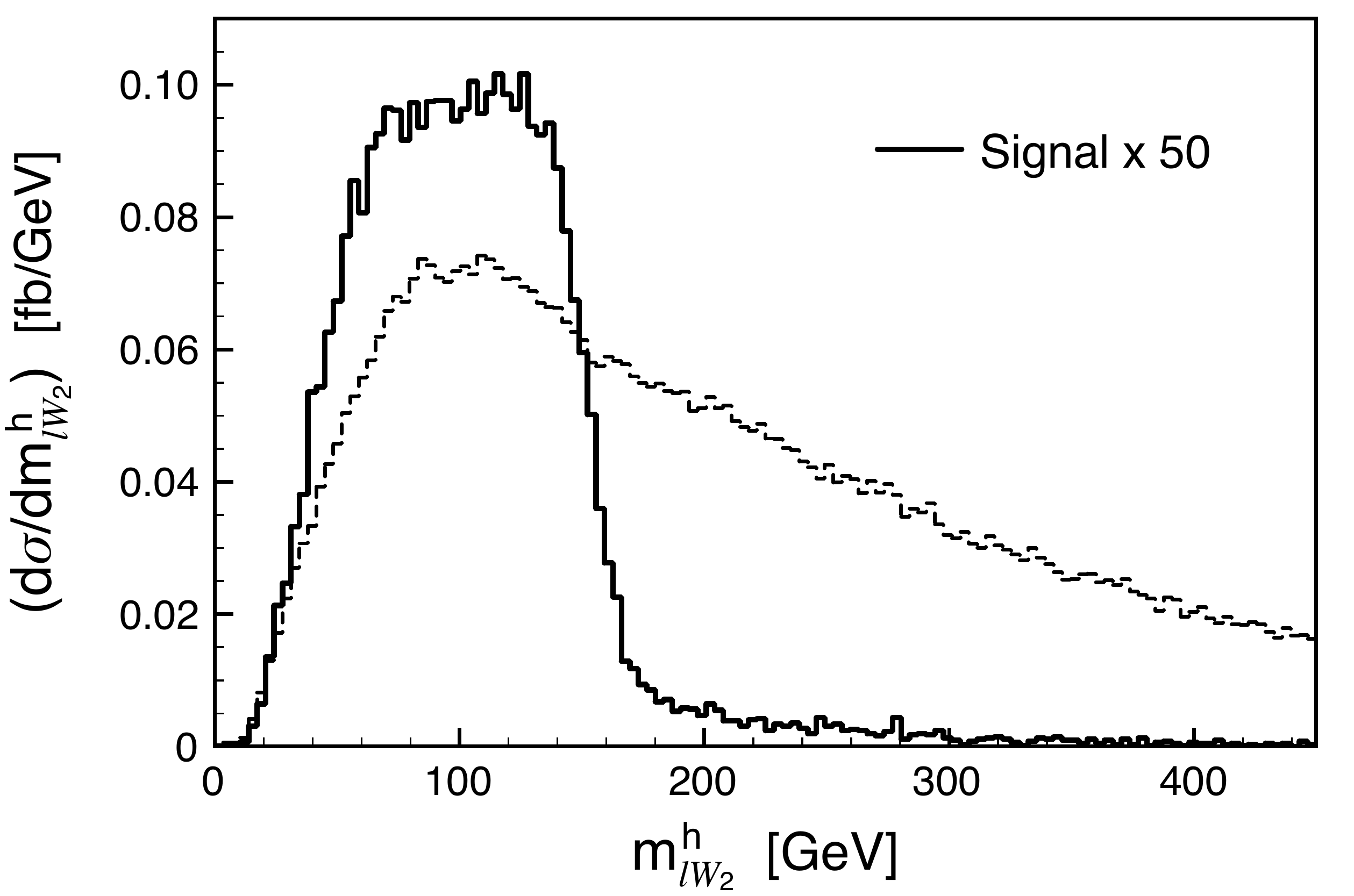}
\caption[]{
\label{fig:mh1mh2-SS2L}
\small
Differential cross section  after the acceptance cuts of  Eq.~(\ref{eqn:acceptance}) 
as a function of the invariant mass of the two leptons forming the first Higgs candidate (left plot)
and of the  invariant mass of the lepton plus jets forming the second Higgs candidate (right plot).
Continuous curve: signal ${\cal S}_2$ in the MCHM4 at $\xi=1$; Dashed curve: background.
}
\end{center}
\end{figure}

As one can see from Table~\ref{tab:signbcks2L}, after the acceptance cuts 
the background dominates by far over the signal.
In order to select our first set of additional cuts we proceed 
in close analogy to the three-lepton case.
We first identify the two ``reference'' jets as described in Section~\ref{sec:sig3L}.
The distributions of the rapidity of the first reference jet, $\eta_{J1}^{ref}$,
the invariant mass of the two reference jets, $m_{JJ}^{ref}$, and their separation, $\Delta \eta_{JJ}^{ref}$, 
are quite similar to those of Fig.~\ref{fig:eta-deta-acccuts} and are thus not reported here.
Next, we reconstruct one hadronic $W$ as follows:
using all the non-reference jets,
we select the pair with invariant mass $m^W_{JJ}$ closer to the $W$
mass. If $|m^W_{JJ}-M_W|<40\,\gev$ we label these two jets as $j_1^{W_1}$ and
$j_2^{W_1}$, otherwise the event is rejected.
All the remaining jets will be labelled as belonging to the other
hadronic $W$, $j_k^{W_2}$.
We then proceed to identify the decay products of the two Higgs bosons.
As a criterion to select the lepton and the $W$ from the same Higgs,
we use the separation $\Delta R$ between them, as they will tend to emerge collimated
due to the Higgs boost. More explicitly, by defining 
\begin{equation*}
p_{W_i}= \sum_n p_{j_n^{W_i}} \, ,
\end{equation*}
we compute $\Delta R_{l_1 W_1}$ and $\Delta R_{l_2 W_1}$. 
If $\Delta R_{l_1 W_1} < \Delta R_{l_2 W_1}$, we  assign $l_1$ and $j_k^{W_1}$ to the first Higgs and the
remaining jets and lepton to the second one; otherwise we form the first Higgs boson candidate with $l_2$ and $j_k^{W_1}$
and the other one with the remaining jets and lepton. We denote by $m^h_{lW_1}$ and $m^h_{lW_2}$ the invariant mass of the 
Higgs system containing respectively the jet $j_k^{W_1}$ and $j_k^{W_2}$.
They are plotted in Fig.~\ref{fig:mh1mh2-SS2L}  for both the signal and the background. 
%

As a first set of cuts we use the observables discussed above and
require that each individual cut reduces the signal by no more than $\sim 2\%$.
We demand:
\begin{equation} \label{eqn:0levcut2L}
\begin{aligned}
& |\eta_{J1}^{ref} | \ge 1.9 && \quad m_{JJ}^{ref} \ge 320\,\gev 
&& \quad \Delta \eta_{JJ}^{ref} \ge 2.8 \\[0.2cm]
& |m^W_{J_1^{W_1}J^{W_1}_2}- M_W| \le 40 \, \gev && \quad m_{lW_1}^h \le 185 \, \gev 
&& \quad m_{lW_2}^h \le 210 \,\gev  \, .
\end{aligned}
\end{equation}
Signal and background cross sections after these cuts are reported as $\sigma_2$ in 
the third column of Table~\ref{tab:signbcks2L}.  Notice that similarly to the three-lepton case,
all the backgrounds with a large number of jets have been strongly reduced.

As done for the three-lepton channel we search for the optimal set of cuts 
by following an iterative procedure: at each step we cut
on the observable which leads to the largest increase in the signal significance,
until no further improvement is possible.
We end up with the  following set of additional cuts:
\begin{equation}
\Delta \eta_{JJ}^{ref} \ge 4.5 \qquad 
m^h_{lW_1} \leq 180 \, \gev \qquad m^h_{lW_2} \leq 180 \, \gev \, .
\label{eqn:1levcut2L}
\end{equation}
Signal and background rates after these cuts are reported as $\sigma_3$ in Table~\ref{tab:signbcks2L}.

As a final cut, we require $m^W_{J_1^{W_1}J^{W_1}_2}$ to deviate
from $M_W$ by no more than twice the CMS or ATLAS dijet mass resolution:
\begin{align}
|m^W_{J_1^{W_1}J^{W_1}_2}- M_W| & <  30 \, \gev \label{eqn:mwcms-SS2L} \\[0.2cm]
|m^W_{J_1^{W_1}J^{W_1}_2}- M_W| & <  20 \, \gev \, .\label{eqn:mwatlas-SS2L} 
\end{align}
The resulting cross sections are denoted respectively as $\sigma_4^{CMS}$ and $\sigma_4^{ATLAS}$
in Table~\ref{tab:signbcks2L}.
We do not impose an analog cut on the invariant mass of the second hadronic $W$ candidate, formed by all
the remaining jets, since 
the previous cuts already strongly suppress the backgrounds with large jet multiplicities,
so that in the majority of the events, the second $W$ system is formed by a single jet and hence has a small
invariant mass.

A further reduction of the $Wl^+l^-5j$ background can be achieved by vetoing
events which contain soft leptons ($ 1\,\gev \le p_{Tl} \le 20\,\gev$) 
that are isolated from any jet ($\Delta R_{jl} > 0.4$) and form a same-flavor opposite-sign 
pair with at least one of the two hard leptons.
In order to estimate the efficiency of such veto on the signal, we showered and hadronized
the events with \texttt{PYTHIA}. We find that the majority of the additional leptons
originates from the decay of the final-state hadrons, especially from the leptonic
decay of charmed mesons. The fraction of signal events rejected is quite small, less than $4\%$, 
and we will neglect it. For simplicity, the effect of the veto on all the backgrounds 
with exactly two leptons at the parton level~\footnote{These backgrounds are: $W^{+(-)} W^{+(-)} 5j$,
$WWW jjj$, $hW jjj$, $WWWW j$, $WWWW jj$, $t \bar t W j$, $t \bar t W jj$, $t \bar t W W$ and
$ W \tau^+ \tau^- 4j$.}
will also be neglected.
The cross sections after this veto are reported in Table~\ref{tab:signbcks2L-2} as $\sigma_5^{CMS}$ and 
$\sigma_5^{ATLAS}$, respectively after the cut of Eq.~(\ref{eqn:mwcms-SS2L}) and 
Eq.~(\ref{eqn:mwatlas-SS2L}).
%
\begin{table}[t]
\begin{center}
\begin{tabular}{|l||l|l||l|l|l|}
\hline
{\tt Channel} & $\sigma_5^{CMS}$ & $\sigma_6^{CMS}$ & $\sigma_5^{ATLAS}$ & $\sigma_6^{ATLAS}$ \\ 
\hline
\hline
${\cal S}_2$ $(\text{MCHM4} - \xi=1)$ & 51.3 & 51.3 & 49.9 & 49.9  \\
\hline
${\cal S}_2$ $(\text{MCHM4} - \xi=0.8)$ & 34.6 & 34.6 & 33.7 & 33.7  \\
\hline
${\cal S}_2$ $(\text{MCHM4} - \xi=0.5)$ & 16.7 & 16.7 & 16.2 & 16.2  \\
\hline
\hline
${\cal S}_2$ $(\text{MCHM5} - \xi=0.8)$ & 49.4 & 49.4 & 47.8  & 47.8 \\
\hline
${\cal S}_2$ $(\text{MCHM5} - \xi=0.5)$ & 26.2 & 26.2 & 25.3 & 25.3  \\
\hline
\hline
${\cal S}_2$ $(\text{SM} - \xi=0)$ & 2.84 & 2.84 & 2.76 & 2.76 \\
\hline
\hline
$ W l^+ l^- 5j$ & 21.8 & 21.8 & 16.3 & 16.3  \\
\hline
$ W^{+(-)} W^{+(-)}   5j$ & 11.7 & 11.7 & 8.91 & 8.91  \\
\hline
$ W W W jjj$ & 8.49 & 8.49 & 8.18 & 8.18  \\
\hline
$ h W jjj$ & 12.4 & 12.4 & 12.1 & 12.1  \\
\hline
$ W W W W j$ & 0.11 & 0.11 & 0.11 & 0.11  \\
\hline
$ W W W W jj$ & 0.52 & 0.52 & 0.49 & 0.49  \\
\hline
$t \bar t W j$ & 12.9  & 3.65 & 12.0 & 3.43  \\
\hline
$t \bar t W jj$ & 23.2 & 8.47 & 19.6 & 7.34  \\
\hline
$t \bar t W jjj$ & 7.10 & 3.23 & 5.77 & 2.62  \\
\hline
$t \bar t W W$ & 1.24 & 0.51 & 1.15 & 0.49 \\
\hline
$t \bar t W W j$ & 25.6 & 10.2 & 22.2 & 8.91  \\
\hline
$t \bar t h jj  \to t \bar t W W jj$  & 15.4 & 7.39 & 13.0 & 6.22  \\
\hline
$t \bar t h jjj \to t \bar t W W jjj$  & 3.11 & 1.70 & 2.65 & 1.44 \\
\hline
$ W \tau^+ \tau^- 4j$ & 20.0 & 5.24 & 15.9 & 4.19  \\
\hline
$ W \tau^+ \tau^- 5j$ & 6.28 & 4.86 & 5.02 & 3.79  \\
\hline
\hline
Total Background & 170 & 100  & 144 & 84.6  \\
\hline
\end{tabular}
\caption[]{
\label{tab:signbcks2L-2} 
\small
Cross sections, in ab, for the signal ${\cal S}_2$ and for the main backgrounds after the cuts of 
Eqs.~(\ref{eqn:acceptance}) and (\ref{eqn:0levcut2L})--(\ref{eqn:mwcms-SS2L})
plus a veto on soft leptons ($\sigma_5^{CMS}$) or a veto on soft leptons, taus and $b$-jets ($\sigma_6^{CMS}$);
of Eqs.~(\ref{eqn:acceptance}), (\ref{eqn:0levcut2L}), (\ref{eqn:1levcut2L}) and (\ref{eqn:mwatlas-SS2L})
plus a veto on soft leptons ($\sigma_5^{ATLAS}$) or a veto on soft leptons, taus and $b$-jets 
($\sigma_6^{ATLAS}$).
}
\end{center}
\end{table}

A final reduction of the background can be obtained by resorting to $b$-jet and tau tagging
and vetoing these particles in the final state. 
We assume a $b$-tagging efficiency $\epsilon_b = 0.55$ within $\eta_b < 2.5$ 
($\epsilon_b = 0$ otherwise), and a $\tau$ veto efficiency of $80\%$ within  $\eta_\tau < 2.5$
(zero otherwise). The resulting cross sections after vetoing both taus and $b$-jets are reported as $\sigma_6^{CMS}$ 
and $\sigma_6^{ATLAS}$ in Table~\ref{tab:signbcks2L-2}, respectively after the cut of 
Eq.~(\ref{eqn:mwcms-SS2L}) 
and Eq.~(\ref{eqn:mwatlas-SS2L}).

\subsubsection{Estimate of showering effects}

\begin{figure}[tbp]
\begin{center}
\includegraphics[width=0.485\textwidth,clip,angle=0]{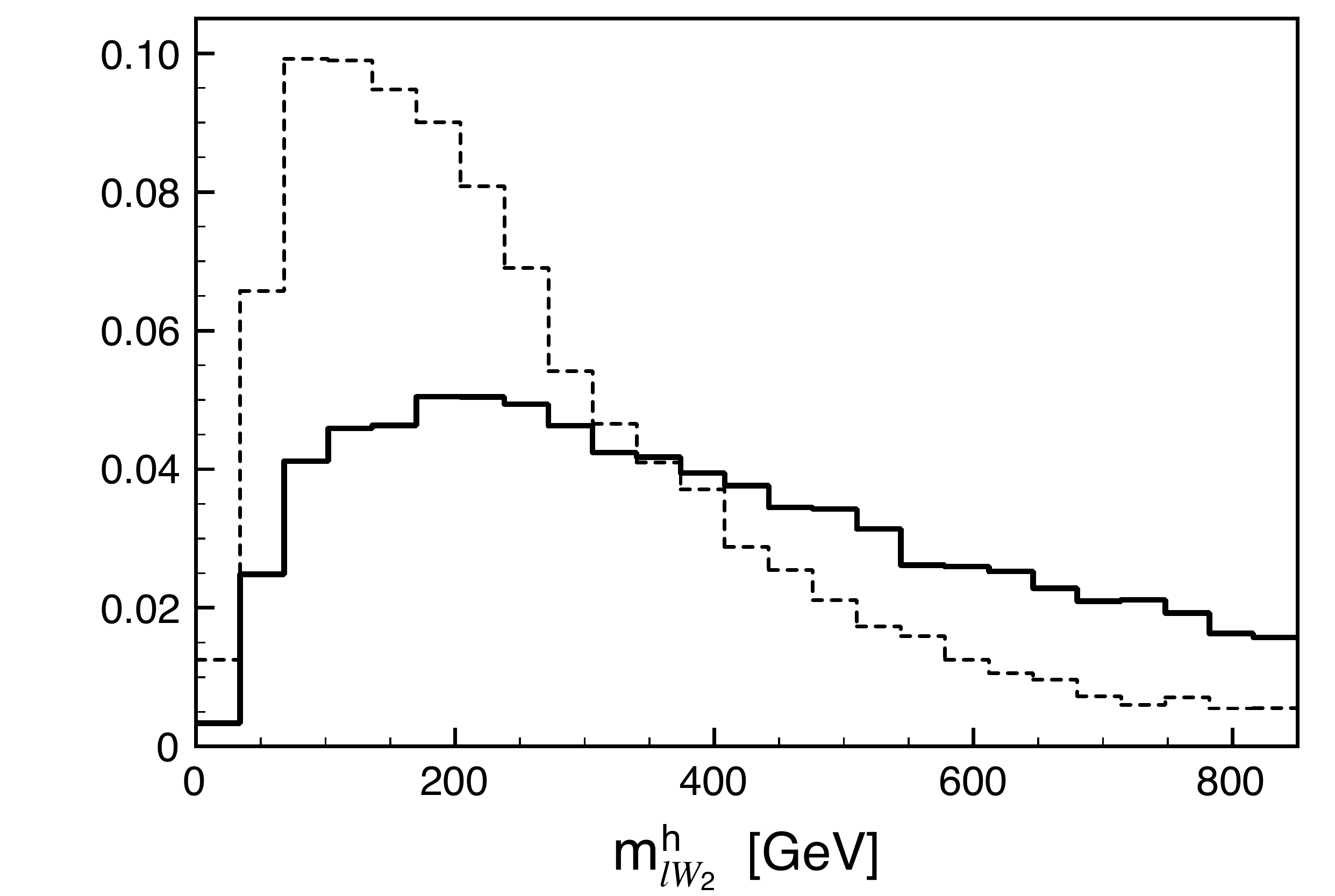}
\caption[]{
\label{fig:mhljjshower}
\small
Differential cross section as a function of $m^h_{lW_2}$  after the showering (continuous line) and at the parton
level (dashed line) of the sum of the backgrounds
$Wl^+l^-5j$, $WWW jjj$, $hW jjj$, $t \bar t W jj$ and $t \bar t W W j$.
Only events which pass the acceptance cuts and those on $m_{JJ}^{ref}$ (Eq.~(\ref{eqn:0levcut2L})) and on $\Delta \eta_{JJ}^{ref}$
(Eq.~(\ref{eqn:1levcut2L})) have been included.
}
\end{center}
\end{figure}

As for the three leptons channel, background events have generically a
larger hadronic activity in the central region, compared to the signal,
once the showering is turned on.
In this case, the main effect is that of shifting the $m^h_{lW_2}$ distribution
towards larger values. This is clearly illustrated by Fig.~\ref{fig:mhljjshower}, which reports
the sum of the cross sections of the main backgrounds, 
$Wl^+l^-5j$, $WWW jjj$, $hW jjj$, $t \bar t W jj$ and $t \bar t W W j$, as a function of
$m^h_{lW_2}$ after imposing the acceptance cuts, $\Delta \eta_{JJ}^{ref}> 4.5$ and $m_{JJ}^{ref}>320\,\gev$.
%
%

We derive a rough estimate of the effect of the showering by monitoring the collective efficiency
of the cuts on $m_{JJ}^{ref}$ (Eq.~(\ref{eqn:0levcut2L})) and on $\Delta \eta_{JJ}^{ref}$, $m_{lW_2}^h$
(Eq.~(\ref{eqn:1levcut2L})). After the showering, we find the following additional reduction on the 
rates of the signal and of the main backgrounds:
\begin{center}
\begin{tabular}{c|cccccc}
   & ${\cal S}_2$ $(\xi=1,0.8,0.5)$ & \quad $ W l^+ l^- 5j$ 
                                    & $WWW jjj$ & $hW jjj$  & $t \bar t W jj$ & $t \bar t W W j$ \\[0.1cm]
\hline
$ \epsilon_{\mathrm shower}/\epsilon_{\mathrm parton}$ 
 & 0.86 & \quad 0.34 & 0.32 & 0.80 & 0.55 & 0.95
\end{tabular}
\end{center}
A further veto on events with 7 or more jets has a negligible impact, both for the signal
and the background, as the cut on $m_{lW_2}^h$ effectively acts like a veto
on extra hadronic activity. 
In the case of events with 6 jets one could think of keeping only those where
the two jets associated with the second Higgs candidate reconstruct an hadronic $W$:
$|m^W_{J^{W_2}_1 J^{W_2}_2} - M_W | < 40\, \gev$.
We find, however, that even this additional constraint has little impact on the background,
as $m^W_{J^{W_2}_1 J^{W_2}_2}$ is already forced to be small after the cut on $m_{lW_2}^h$ is met.
As for the three-lepton case, it is worth stressing that these results should be confirmed
by a full treatment of showering effects using matched samples.

\subsubsection{Fake leptons and  lepton charge misidentification}
\label{sec:fakelep2L}

Differently from the three lepton case, we expect
the effect of fake leptons from jet misidentification
to be much more relevant for same-sign dilepton events.
The reason is that the cross section for the production of two same-sign $W$'s is about
two orders of magnitude smaller than that for $W^+W^-$. 
It might then turn out to be more convenient to 
produce one $W$
and pay the misidentification probability factor for a fake lepton from an extra jet
than having a second leptonically-decaying $W$ with the same sign.
Moreover, an additional source of background comes in this case
from events where the charge of a primary lepton is misidentified.
A precise estimate of all these effects
is beyond the scope of the present paper, since it would require a full
detector simulation as well as a dedicated strategy
designed to minimize the effect while keeping the lepton
reconstruction efficiency as high as possible.
We will  limit ourselves to performing a crude estimate
and quoting the rejection factors required to make such backgrounds
negligible.

The most serious potential source of background with fake leptons
from light jets is $W+ 6j$.
Table~\ref{tab:fakeblep} reports the relative cross section
after all the cuts imposed in our analysis (without including any 
mistagging probability factor). The quoted number is
obtained by computing the cross section for
$pp \to W+ 6j$, picking up randomly one jet and
assuming it is mistagged as a lepton, and multiplying
by a factor 6 to account for the six different possibilities
to mistag a jet.
A rejection factor of $\sim 10^{-5}$,  quoted as achievable by both collaborations~\cite{CMS:TDR1,ATLAS:CERNOPEN}, 
is sufficient to reduce
this background down to a manageable level.
A dedicated experimental study is however required to 
establish whether this can be obtained without
reducing too much the lepton identification efficiency.
The largest background with fake leptons from heavy quarks is 
$t \bar t  jj$, with one $b$ from a top decay
tagged as a lepton. Table~\ref{tab:fakeblep} reports the cross sections for
$t \bar t  jj$ and $t \bar t 3j$ after all the cuts
plus a $b$-jet veto. For simplicity, we have approximated the ``fake'' lepton momentum to be equal to that
of its parent $b$ quark. This is a conservative, reasonable assumption as 
the requirement of having a hard, isolated lepton forces the remaining hadronic
activity from the $b$ decay to be quite soft to escape detection~\cite{CHMISID}.
As Table~\ref{tab:fakeblep} shows, our rough estimate seems to indicate
that rejection factors as small as $10^{-4}$ are required to make this background
comparable to those studied in the previous sections.

\begin{table}
\begin{center}
\begin{tabular}{|l||l||l|}
\hline
{\tt Channel} & $\sigma_6^{CMS}$  & $\sigma_6^{ATLAS}$  \\ 
\hline
\hline
$ W +6 j$ &  3.8$\,\times\, 10^{4}$ & 3.0$\,\times\, 10^4$  \\
\hline
\end{tabular} \hspace{0.5cm}
\begin{tabular}{|l||l||l|}
\hline
{\tt Channel} & $\sigma_6^{CMS}$  & $\sigma_6^{ATLAS}$  \\ 
\hline
\hline
$ t \bar t jj$ &  46.0$\,\times\, 10^{4}$ & 44.3$\,\times\, 10^4$  \\
\hline
$ t \bar t 3j$ & 17.9$\,\times\, 10^4$ & 15.8$\,\times\, 10^4$ \\
\hline
\end{tabular}
\caption[]{
\label{tab:fakeblep} 
\small
Cross sections, in ab, for the most important backgrounds with fake leptons from
light jets (table on the left) and from heavy jets (table on the right).
In both tables, $\sigma_6^{CMS}$ and $\sigma_6^{ATLAS}$  indicate the cross section after
respectively the cuts of Eqs.~(\ref{eqn:acceptance}) and (\ref{eqn:0levcut2L})--(\ref{eqn:mwcms-SS2L}) plus a veto on $b$-jets,
and Eqs.~(\ref{eqn:acceptance}), (\ref{eqn:0levcut2L}), (\ref{eqn:1levcut2L}) and (\ref{eqn:mwatlas-SS2L})
plus a veto on $b$-jets.
}
\end{center}
\end{table}

\begin{figure}[tbp]
\begin{center}
\includegraphics[width=0.485\textwidth,clip,angle=0]{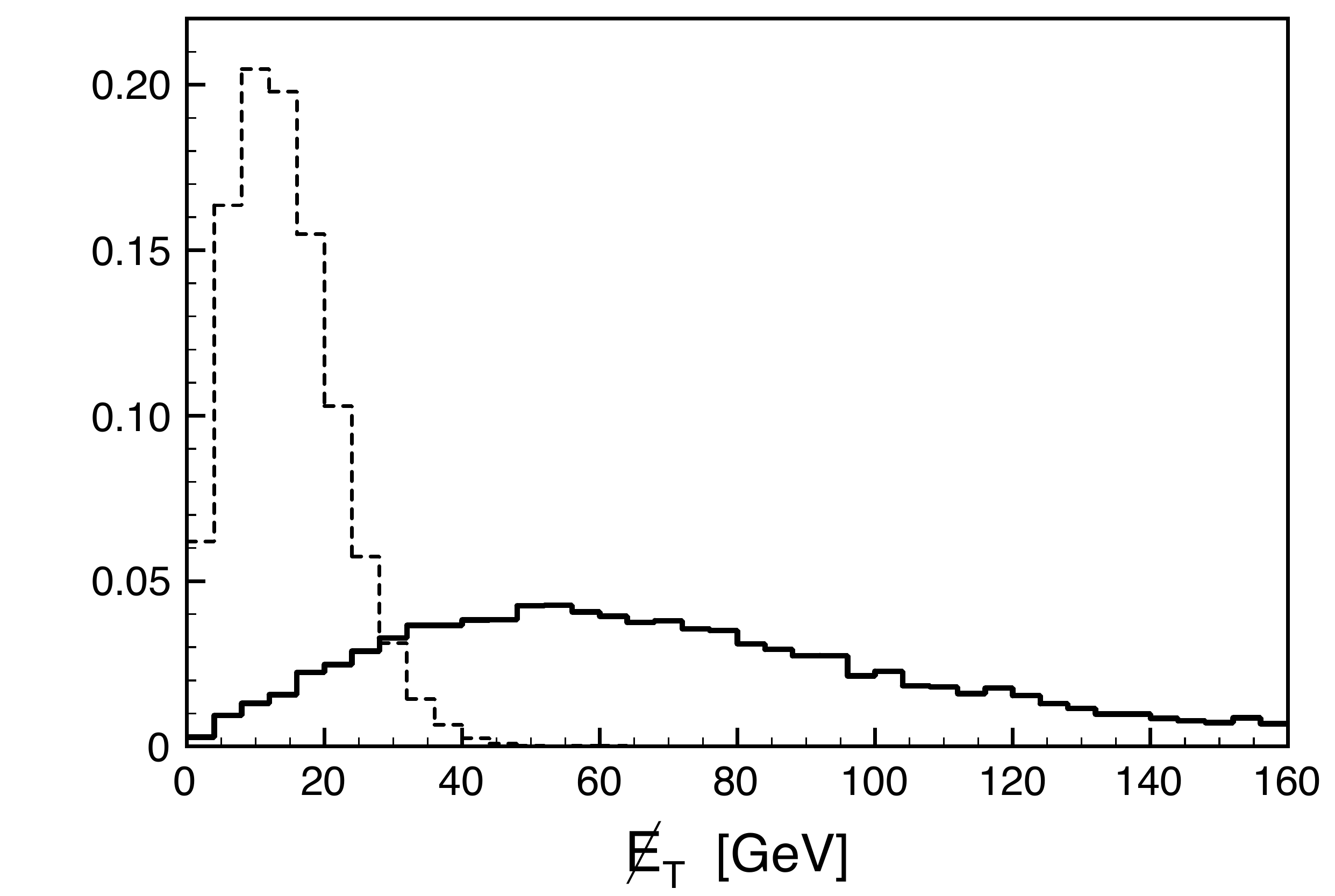}
\caption[]{
\label{fig:etmiss}
\small
$\etmiss$  distribution for the signal ${\cal S}_2$ in the MCHM4 at $\xi=1$
(continuous curve) and the $l^+l^-5j$ background (dashed curve) 
after the cuts of  Eqs.~(\ref{eqn:acceptance}) and (\ref{eqn:0levcut2L})--(\ref{eqn:mwatlas-SS2L}).
The missing transverse energy is computed including a Gaussian resolution 
$\sigma (\etmiss) = 0.55\cdot \sqrt{ \sum E_T/\text{GeV}}$, see text.
The curves have been normalized to unit area.
}
\end{center}
\end{figure}

Finally, we consider the most dangerous backgrounds where
the charge of a primary lepton is not correctly measured. 
The size of this effect  strongly  depends on the 
algorithm used to reconstruct the leptons, and it is in general
larger for electrons than for muons. 
Table~\ref{tab:chargemisid} reports the cross
sections for $t \bar t 3j$, $t \bar t 4j$ and $l^+l^-5j$ after all the cuts
imposed in our analysis plus a $b$-jet veto,
assuming that the charge of one lepton has not been correctly measured. 
%
\begin{table}
\begin{center}
\begin{tabular}{|l||c||c||c|}
\hline
{\tt Channel} & $\sigma_6^{CMS}$  & $\sigma_6^{ATLAS}$ 
  & $\sigma_6^{ATLAS}\times \eps_{CH}\times \eps_{\;\;\etmiss > 25\,\text{GeV}}$ \\ 
\hline
\hline
$ \xi =1 $  &  51.3 & 49.9 & 45.1 \\
\hline
\hline
$ t \bar t 3j$ &  11.9$\,\times\, 10^{3}$ & 9.2$\,\times\, 10^{3}$ & 11.1 \\
\hline
$ t \bar t 4j$ &  4.0$\,\times\, 10^{3}$ & 3.2$\,\times\, 10^{3}$ & 3.96 \\
\hline
$ l^+l^- 5j$ &  112.4$\,\times\, 10^{3}$ & 88.4$\,\times\, 10^{3}$ & 11.5  \\
\hline
\end{tabular}
\caption[]{
\label{tab:chargemisid} 
\small
Cross sections, in  ab, for the most important backgrounds where the charge of a primary lepton is misidentified.
The first two columns show the cross sections  after  the 
cuts of respectively  Eqs.~(\ref{eqn:acceptance}) and (\ref{eqn:0levcut2L})--(\ref{eqn:mwcms-SS2L})  plus a veto on $b$-jets
($\sigma_6^{CMS}$) and  Eqs.~(\ref{eqn:acceptance}), (\ref{eqn:0levcut2L}), (\ref{eqn:1levcut2L}) and (\ref{eqn:mwatlas-SS2L}) 
plus a veto on $b$-jets ($\sigma_6^{ATLAS}$), without including any charge misidentification efficiency.
The last column reports the cross section $\sigma_6^{ATLAS}$ multiplied by the efficiency of
a cut $\etmiss > 25\,$GeV and a  charge misidentification probability equal to $10^{-3}$
for electrons and $3 \times 10^{-4}$ for muons (collectively indicated as $\eps_{CH}$).
For convenience, values of the cross section for the  signal ${\cal S}_2$ in the MCHM4 at $\xi=1$ are also shown.
}
\end{center}
\end{table}
Even after applying a charge misidentification probability 
$10^{-3}$ for electrons and a few$\,\times 10^{-4}$ for muons
as quoted in the ATLAS TDR~\cite{ATLAS:CERNOPEN} for leptons with $p_T \sim 100\,$GeV,
the $l^+l^- 5 j$  background is still sizable, while the $t\bar t+jets$ channels are smaller.
Since however $l^+l^- 5 j$ has no neutrinos, while the signal has two of them,
$\not \!\! E_T$ provides an important handle to reduce this background. 
In Fig.~\ref{fig:etmiss} we plot $\etmiss$ for both the  signal ${\cal S}_2$ in the MCHM4 at $\xi=1$ and $l^+l^-5 j$.
We have computed $\etmiss$ by including a  Gaussian resolution $\sigma (\etmiss)= a\cdot \sqrt{ \sum E_T/\text{GeV}}$,
where $\sum E_T$ is the total transverse energy deposited in the calorimeters (from electrons
and jets). We chose $a=0.55$, which is expected to be a good fit for the ATLAS detector~\cite{ATLAS:CERNOPEN}.
Assuming a charge misidentification probability equal to $10^{-3}$ for electrons and $3\times 10^{-4}$ for muons, 
we find that a cut $\etmiss > 25\,$GeV provides the best sensitivity, the efficiency on the signal being $\simeq 0.9$.
The corresponding  cross sections after this cut
(including the charge misidentification probabilities)
are reported in the last column of Table~\ref{tab:chargemisid},
which shows that the background has been reduced to a manageable level although
it remains non-negligible.

To summarize, our estimates show that both backgrounds with fake leptons from jets
and those with misidentification of the charge of a primary lepton are expected
to have an important impact on the same-sign dilepton channel.
A detailed experimental study is therefore needed 
to  determine the precise relevance of these backgrounds and 
fully assess the signal significance in this case.

\subsection{Channel ${\cal S}_4$: four leptons}
\label{sec:sig4L}

The last channel  we have considered is the one with four leptons.
In this case, the signal is characterized by two widely separated jets with no further
hadronic activity in between, four hard leptons from the  decay of the two Higgses and  missing energy.
The second column of Table~\ref{tab:signbcks4L} reports the cross sections ($\sigma_1$)
for the signal ${\cal S}_4$ and for the main backgrounds we have studied 
after the acceptance cuts of Eq.~(\ref{eqn:acceptance}).
\begin{table}[t]
\begin{center}
\begin{tabular}{|l||l|l|l|l|l|}
\hline
{\tt Channel} & $\sigma_1$ & $\sigma_2$ & $\sigma_3$  & $\sigma_4$  & $\sigma_5$   \\ 
\hline
\hline
${\cal S}_4$ $(\text{MCHM4} - \xi=1)$ & 6.64 & 6.16  & 5.10 & 4.33 & 4.33 \\
\hline
${\cal S}_4$ $(\text{MCHM4} - \xi=0.8)$ & 4.40 & 4.10  & 3.38 & 2.86 & 2.86 \\
\hline
${\cal S}_4$ $(\text{MCHM4} - \xi=0.5)$ & 1.99 & 1.86  & 1.52 & 1.30 & 1.30 \\
\hline
\hline
${\cal S}_4$ $(\text{MCHM5} - \xi=0.8)$ & 6.06 & 5.59  & 4.52 & 3.76 & 3.76 \\
\hline
${\cal S}_4$ $(\text{MCHM5} - \xi=0.5)$ & 3.00 & 2.79  & 2.26 & 1.90 & 1.90 \\
\hline
\hline
${\cal S}_4$ $(\text{SM} - \xi=0)$ & 0.32 & 0.24  & 0.19 & 0.15 & 0.15 \\
\hline
\hline
$l^+ l^- l^+ l^- jj$ & 1.73 $\times 10^3$ & 171 & 0.04 & 0.00 & 0.00 \\
\hline
$ l^+ l^- \tau^+ \tau^- jj$ & 44.6 & 4.28  & 0.55 & 0.11 & 0.11 \\
\hline
$ hjj \to  l^+ l^- \tau^+ \tau^- jj$ & 1.03 & 0.57  & 0.12 & 0.06 & 0.06 \\
\hline
$WW l^+ l^- jj $ & 105 & 0.78  & 0.10 & 0.03 & 0.03 \\
\hline
$h l^+ l^- jj \to WW l^+ l^- jj $ & 41.4 & 11.2  & 1.30 & 0.75 & 0.75 \\
\hline
$hWWjj \to WWWWjj$  & 0.79 & 0.07  & 0.06 & 0.04 & 0.04 \\
\hline
$t\bar t l^+ l^-$  & 558 & 6.15 & 0.90 & 0.02 & 0.01 \\
\hline
$t\bar t l^+ l^- j$  & 624 & 57.3 & 1.26 & 0.24 & 0.13 \\
\hline
$t\bar t WW$ & 67.5 & 0.48 & 0.34 & 0.02 & 0.01 \\
\hline
$t\bar t WW j$ & 83.3 & 6.58 & 0.84 & 0.14 & 0.08 \\
\hline
$t\bar t h jj \to t\bar t WW jj$ & 46.0 & 8.19 & 0.08 & 0.02 & 0.02 \\
\hline
$t\bar t h jjj \to t\bar t WW jjj$ & 22.9 & 6.15 & 0.00 & 0.00 & 0.00 \\
\hline
\hline
Total background & 3.32 $\times 10^3$ & 272 & 5.59 & 1.44 & 1.25 \\
\hline
\end{tabular}
\caption[]{
\label{tab:signbcks4L} 
\small
Cross sections, in  ab, for the signal ${\cal S}_4$ (see Eq.~(\ref{eqn:channels})) and for the main backgrounds after imposing
the cuts of Eq.~(\ref{eqn:acceptance}) ($\sigma_1$); 
of Eqs.~(\ref{eqn:acceptance}) and (\ref{eq:MC4L}) ($\sigma_2$);
of Eqs.~(\ref{eqn:acceptance}), (\ref{eq:MC4L}), (\ref{cutfor4ljj}) plus a veto on extra jets, ($\sigma_3$);
of Eqs.~(\ref{eqn:acceptance}), (\ref{eq:MC4L}), (\ref{cutfor4ljj}), (\ref{Detaopt}) plus a veto on extra jets ($\sigma_4$);
of Eqs.~(\ref{eqn:acceptance}), (\ref{eq:MC4L}), (\ref{cutfor4ljj}), (\ref{Detaopt}) plus a veto on extra jets and on $b$-jets ($\sigma_5$).
For each channel the proper branching fraction to a four-lepton
final state (via $W\to l\nu$ and $\tau \to l\nu\nu_\tau$) has been included.
}
\end{center}
\end{table}
We notice that (comments made for Tables~\ref{tab:signbcks3L} and~\ref{tab:signbcks2L} also apply and will not be
repeated here):
\begin{itemize}
\item The Higgs resonant contributions $hl^+l^- jj\to WWl^+l^-jj$  and $hjj \to \tau^+\tau^-l^+l^-jj$ 
are separately reported and are thus not included in the backgrounds $\tau^+\tau^-l^+l^-jj$ and $WWl^+l^- jj$.
\item The  background $l^+l^-l^+l^-jj$ includes the Higgs resonant contribution $hjj\to l^+l^-l^+l^-jj$.
The latter has not been separately reported in this case  since the entire background is negligible at the end of the analysis.
\item The background $WWWWjj$ is largely dominated by its resonant  subprocess $hWWjj \to WWWWjj$.
The non-resonant contribution  is negligible and it has not been reported in the table.
\end{itemize}

As for the three- and two-lepton case, the two reference jets have been identified as the pair with
the largest invariant mass containing the most forward jet.
We identify  the pair of leptons coming from the first Higgs, $(l_1^+ l_1^-)$, 
and that from the second Higgs,  $(l_2^+ l_2^-)$, by using the angular separation 
as a criterion: there are two ways of combining the initial four leptons in two opposite-sign pairs, and
we choose the combination which maximizes $\cos\theta_{l_1^+ l_1^-} + \cos\theta_{l_2^+ l_2^-}$,
as leptons from the same Higgs tend to emerge collimated due to both the Higgs boost and
the spin correlations.
We will refer to $(l_1^+ l_1^-)$ and $(l_2^+ l_2^-)$ defined in this way as our two Higgs candidates.

As a first set of cuts, we use the invariant mass and rapidities of the two reference jets
as well as the invariant masses of the two Higgs candidates. 
The corresponding  distributions for the signal have the same shape as those in 
Figs.~\ref{fig:eta-deta-acccuts} and~\ref{fig:mh1mh2} (left plot).
Similarly to the previous two analyses, we require that each individual cut reduces the signal 
by no more than $\sim 2\%$. We demand:
\begin{equation}
\begin{gathered}
|\eta_{J1}^{ref}| \ge  1.8 \qquad m^{ref}_{JJ} \ge 320 \ \gev \qquad 
  |\Delta \eta^{ref}_{JJ}|  \ge  2.9 \\[0.25cm]
 m_{l_1l_1}^h  \le  110 \ \gev \quad \qquad m_{l_2l_2}^h  \le  110 \ \gev \, .
\end{gathered}
\label{eq:MC4L}
\end{equation}
Signal and background cross sections after this set of cuts are reported in Table~\ref{tab:signbcks4L} as $\sigma_2$.  

\begin{figure}[tbp]
\begin{center}
\includegraphics[width=0.485\textwidth,clip,angle=0]{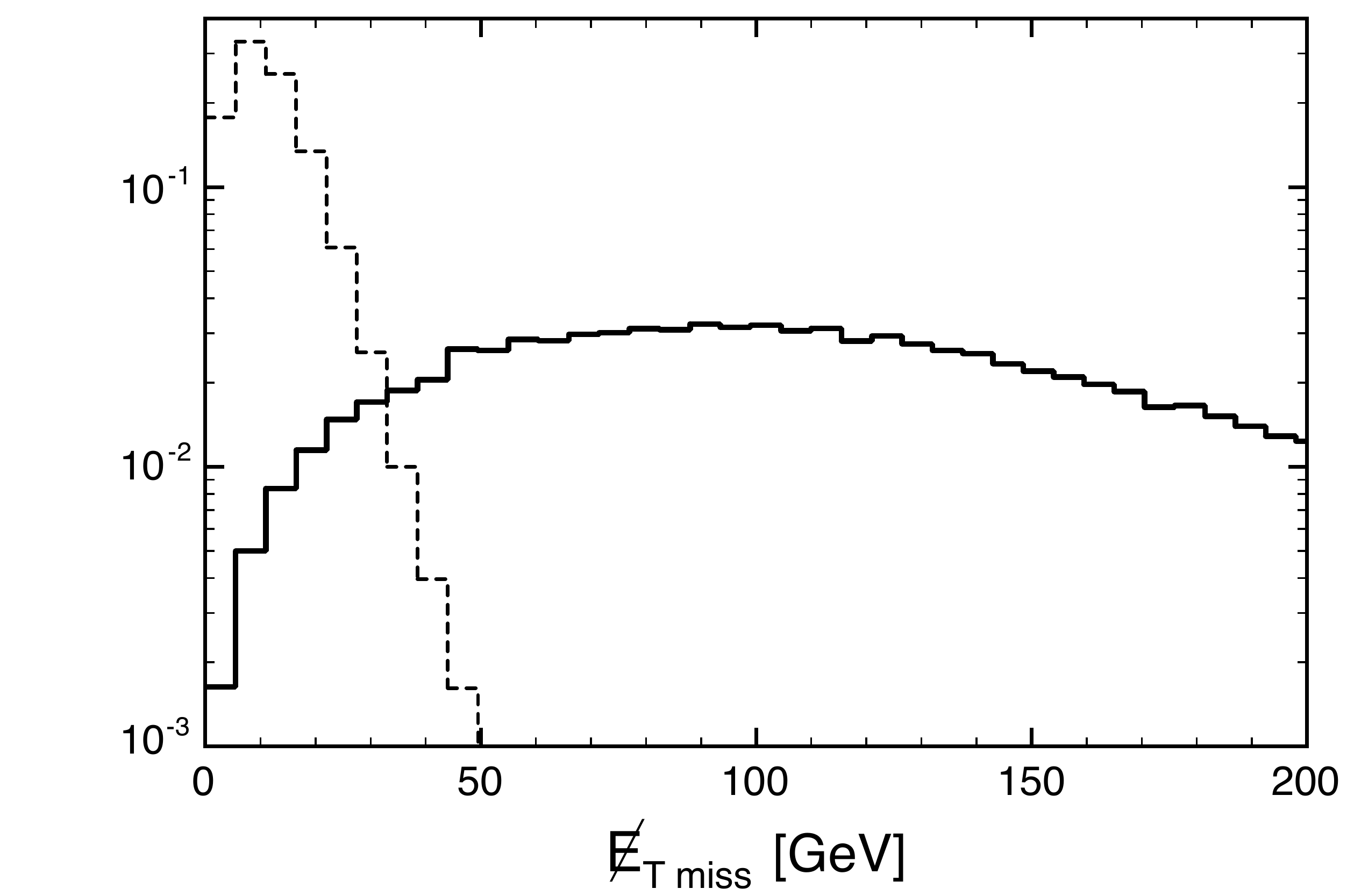}
\caption[]{
\label{fig:etmissFL}
\small
$\etmiss$  distribution for the signal ${\cal S}_4$ in the MCHM4 at $\xi=1$
(continuous curve) and the $l^+l^-l^+l^-jj$ background (dashed curve) 
after the cuts of  Eqs.~(\ref{eqn:acceptance}) and (\ref{eq:MC4L}).
The missing transverse energy is computed including a Gaussian resolution 
$\sigma (\etmiss) = 0.55\cdot \sqrt{ \sum E_T/\text{GeV}}$, see text.
The curves have been normalized to unit~area.
}
\end{center}
\end{figure}

At this level, the $l^+ l^- l^+ l^- jj$ background is much larger than the signal. 
It can be drastically reduced, however, by exploiting the fact that the signal has four neutrinos,
hence a substantial amount of missing energy, while $l^+ l^- l^+ l^- jj$ has none, see Fig.~\ref{fig:etmissFL}.
Here as before, the missing energy of each event has been computed by including a Gaussian resolution
$\sigma (\etmiss) = 0.55\cdot \sqrt{ \sum E_T/\text{GeV}}$ to account for calorimeter effects, where
$\sum E_T$ is the total transverse energy of jets and electrons.
A further reduction is obtained by cutting on the invariant mass  of same-flavor opposite-sign
lepton pairs, $\mSFOS$, excluding values around $M_Z$. This clearly suppresses $l^+ l^- l^+ l^- jj$,
as well as all the processes with  resonant $Z$ contributions.  
We find that optimized values for these cuts, which almost completely eliminate the $l^+ l^- l^+ l^- jj$ background,
are as follows:
\begin{equation}
\label{cutfor4ljj}
\etmiss \geq 40\,\text{GeV}\qquad\quad  |\mSFOS-M_Z| \geq 2\, \Gamma_Z \, .
\end{equation}
The individual efficiencies on $l^+ l^- l^+ l^- jj$  are of $\sim 5\times 10^{-3}$ for the cut on $\etmiss$ 
and $\sim 0.05$ for that on $\mSFOS$.
The other feature of the signal that can be exploited to further reduce the background is its small hadronic activity 
in the central region. We have thus imposed a  
\bit

\item[] {\em veto on any extra hard and isolated  jet} \ (in addition to the two reference jets) 

\eit
satisfying the acceptance cuts of Eq.~(\ref{eqn:acceptance}).  
Signal and background cross sections after this veto and the cuts in Eq.~(\ref{cutfor4ljj}) 
are reported in Table~\ref{tab:signbcks4L} as $\sigma_3$. 

Next, as for the other channels, we have monitored the observables of Eqs.~(\ref{eq:MC4L}) and (\ref{cutfor4ljj})
in search for the optimal set of cuts. We find that the best improvement in the 
signal efficiency is obtained by strengthening the cut on the separation between the reference jets as follows:
\begin{equation}
\label{Detaopt}
|\Delta \eta^{ref}_{JJ}| > 4.5 \, .
\end{equation}
Signal and background cross sections after this cut are reported in Table~\ref{tab:signbcks4L} as $\sigma_4$. 

A final reduction of the background is obtained by  imposing a  veto on $b$-jets in the central region $\eta_b < 2.5$.
Assuming a $b$-tagging efficiency $\eps_b =0.55$ we find the signal and background cross sections 
reported in Table~\ref{tab:signbcks4L} as $\sigma_5$.

\subsection{Results}
\label{sec:results}

%
\begin{table}[!h]
\begin{center}
\begin{tabular}{llcccccccc}
 & & \multicolumn{2}{c}{3 leptons} & & \multicolumn{2}{c}{2 leptons} & & \multicolumn{2}{c}{4 leptons} \\[0.2cm]
\cline{3-4} \cline{6-7} \cline{9-10} \\[-0.4cm]
\multicolumn{2}{l}{\# Events with $300\,\text{fb}^{-1}$ \hspace*{1cm}} & signal &  bckg. &
  & signal &  bckg. & & signal &  bckg. \\[0.2cm]
\hline \\[-0.2cm]
\multirow{3}{*}{MCHM4}\hspace{0.5cm} 
                                       & $\xi =1$    & 4.9 & 1.1 & & 15.0 & 16.6 & & 1.3 & 0.08 \\[0.1cm]
                                       & $\xi =0.8$ & 3.3 & 1.2 & & 10.1 & 18.3 & & 0.9 & 0.14 \\[0.1cm]
                                       & $\xi =0.5$ & 1.5 & 1.4 & & 4.9   & 21.0 & & 0.4 & 0.23 \\[0.5cm]
\multirow{2}{*}{MCHM5} & $\xi =0.8$ & 4.5 & 1.8 & & 14.3 & 26.0 & & 1.1 & 0.19 \\[0.1cm]
                                       & $\xi =0.5$ & 2.3 & 1.2 & & 7.6   & 18.4 & & 0.6 & 0.21 \\[0.5cm]
                        SM           & $\xi =0$    & 0.2 & 1.7 & & 0.8   & 25.4 & & 0.05 & 0.37\\
\end{tabular}
\caption[]{
\label{tab:finalnumbers}
\small
Number of events with $300\,\text{fb}^{-1}$ of integrated luminosity 
based on the cross sections predicted in each channel at the end of the analysis ($\sigma_4^{ATLAS}$, $\sigma_6^{ATLAS}$ and
$\sigma_5$  for the channels with respectively three, two and four leptons).
Values for the background have been obtained by properly rescaling the Higgs contributions to account for its
modified couplings in each model.
}
\end{center}
\end{table}
\begin{table}[!h]
\begin{center}
\begin{tabular}{llccccccc}
 & & \multicolumn{3}{c}{SM hypothesis}  & & \multicolumn{3}{c}{CHM hypothesis}  \\[0.2cm]
\cline{3-5} \cline{7-9}  \\[-0.4cm]
\multicolumn{2}{l}{Significance \hspace*{1.7cm}} & ${\cal S}_3$ & ${\cal S}_2$
  & ${\cal S}_4$ &  & ${\cal S}_3$ & ${\cal S}_2$ &  ${\cal S}_4$ \\[0.2cm]
\hline \\[-0.2cm]
\multirow{3}{*}{MCHM4}\hspace{0.2cm} 
                                       & {\small $\xi =1$}    &  {\small 2.7 (9.0)} &  {\small 2.7 (8.6)} & {\small 1.3 (4.8)}     &  & {\small 3.1 (10.3)}  & {\small 3.2 (10.3)} & {\small 2.0 (7.1)}  \\[0.1cm]
                                       & {\small $\xi =0.8$} &  {\small 1.9 (6.4)} &  {\small 1.8 (6.0)} & {\small 0.8 (3.5)}     &   & {\small 2.1 (7.2)}    & {\small 2.1 (6.9)}   & {\small 1.2 (4.7)} \\[0.1cm]
                                       & {\small $\xi =0.5$} &  {\small 0.8 (3.2)} &  {\small 0.9 (3.0)} & {\small 0.0    (1.7)}  &   & {\small 0.9  (3.4)}   & {\small 1.0 (3.2)}   & {\small 0.0    (2.0)} \\[0.5cm]
\multirow{2}{*}{MCHM5} & {\small $\xi =0.8$} &  {\small 2.5 (8.3)} &  {\small 2.6 (8.3)} & {\small 1.1 (4.2)}      &   & {\small 2.5 (8.2)}    & {\small 2.5 (8.2)}  &  {\small 1.3 (5.1)} \\[0.1cm]
                                       & {\small $\xi =0.5$} &  {\small 1.3 (4.7)} &  {\small 1.4 (4.5)} & {\small 0.0    (2.5)}  &   &  {\small 1.5 (5.3)}   & {\small 1.6 (5.2)}   & {\small  0.0    (3.0)}
\end{tabular}
\caption[]{
\label{tab:sigsig}
\small
Signal significance with 300 fb$^{-1}$ in the channels with three (${\cal S}_3$), two (${\cal S}_2$) and four (${\cal S}_4$) leptons
assuming two statistical  hypotheses: Higgs with SM couplings (SM hypothesis)
and Higgs with modified couplings (CHM hypothesis), see text. Numbers in parenthesis correspond to the 
significance with 3 ab$^{-1}$.
}
\end{center}
\end{table}
%
We collect here our final results for the three channels and the statistical significance of the signal in each case.
Table~\ref{tab:finalnumbers} reports the  final number of events with $300\,\text{fb}^{-1}$ of integrated luminosity 
based on the cross sections predicted in each channel at the end of the analysis ($\sigma_4^{ATLAS}$, $\sigma_6^{ATLAS}$ and
$\sigma_5$  for the channels with respectively three, two and four leptons).
Values for the background have been obtained by properly rescaling the Higgs contributions to account for its
modified couplings in each model (see Eqs.~(\ref{abgoldstone})--(\ref{c5goldstone})).  The Higgs decay branching
fractions that have been used in the case of the MCHM5 are those shown in the right plot of Fig.~\ref{fig:BRs}.
Backgrounds from fake leptons and charge misidentification, for which we provided an estimate in 
Sections~\ref{sec:fakelep3L} 
and \ref{sec:fakelep2L}, have not been included.  In the case of the same-sign dilepton channel their inclusion is likely to decrease the
signal significance.

The signal significance is shown in Table~\ref{tab:sigsig} for two statistical 
hypotheses:~\footnote{The way in which the significance has been computed from the number of events
is explained in footnote~\ref{fot:signif}.}
in the first hypothesis (dubbed as SM in the Table), we assume the Standard Model and compute the significance of  
the observed excess of events 
compared to its expectation.  This means in particular that the number of background
events assumed in this case is that for the SM, ie, $\xi=0$.
In the second hypothesis (dubbed as CHM in the Table), we assume that the Higgs couplings have been already measured 
by means of single production processes, and that the underlying model has been identified.
In this case the assumed number of background events  is that predicted by taking into account the modified Higgs couplings.
%

\section {Features of strong double Higgs production}

\begin{figure}[tp]
\begin{center}
\includegraphics[width=0.485\textwidth,clip,angle=0]{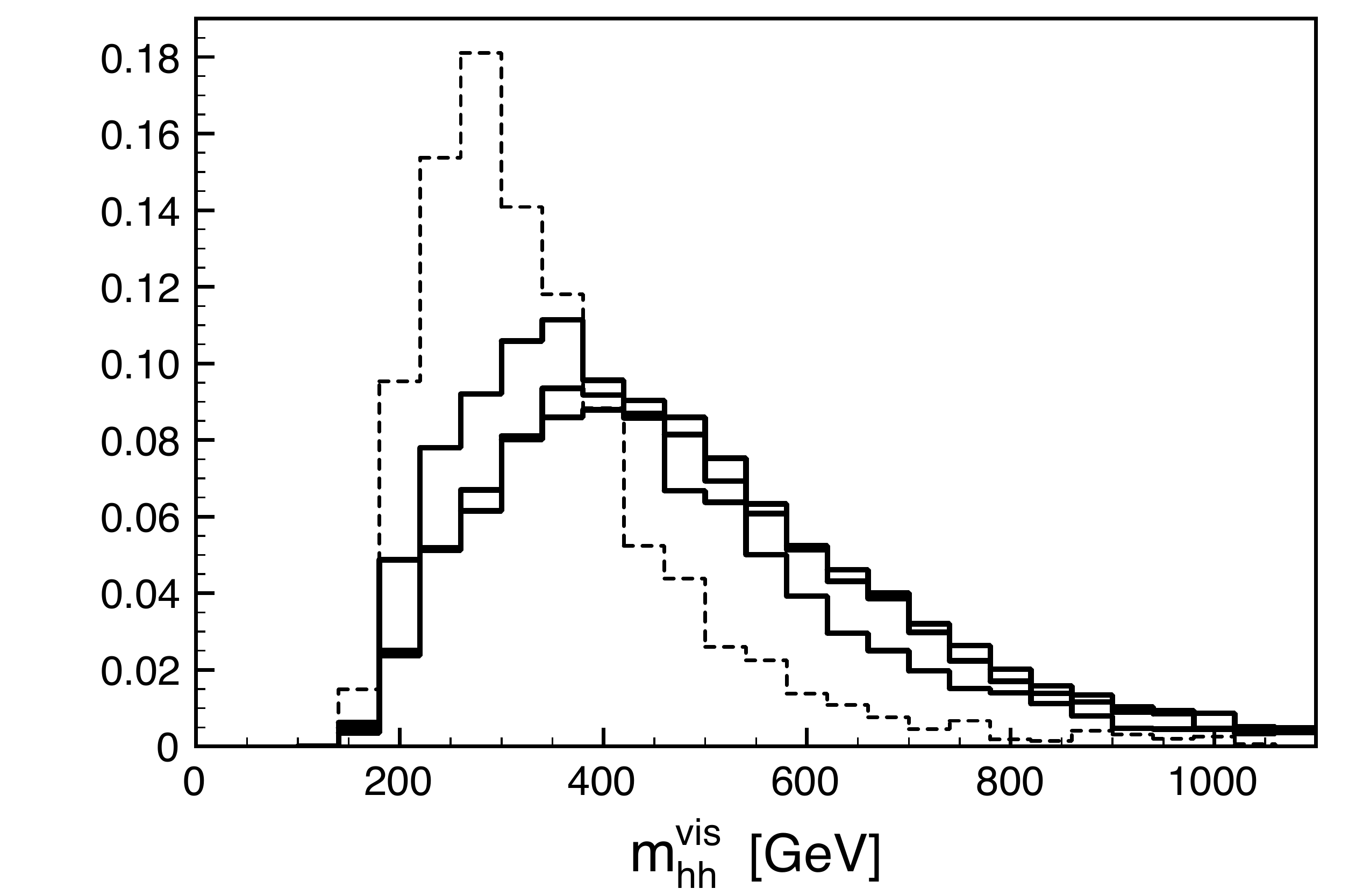}
\hspace{0.2cm}
\includegraphics[width=0.485\textwidth,clip,angle=0]{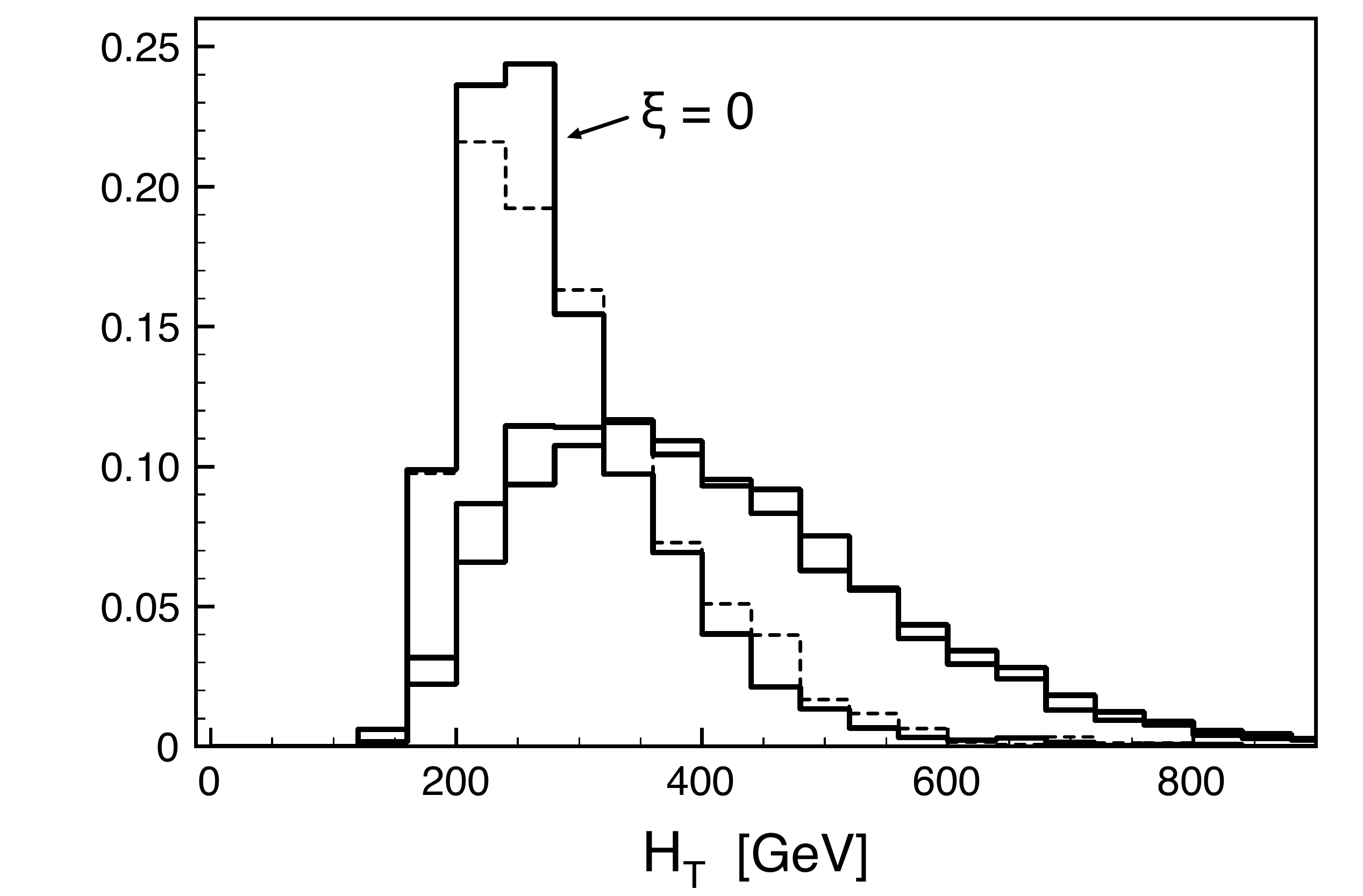}
\\[0.1cm]
\includegraphics[width=0.485\textwidth,clip,angle=0]{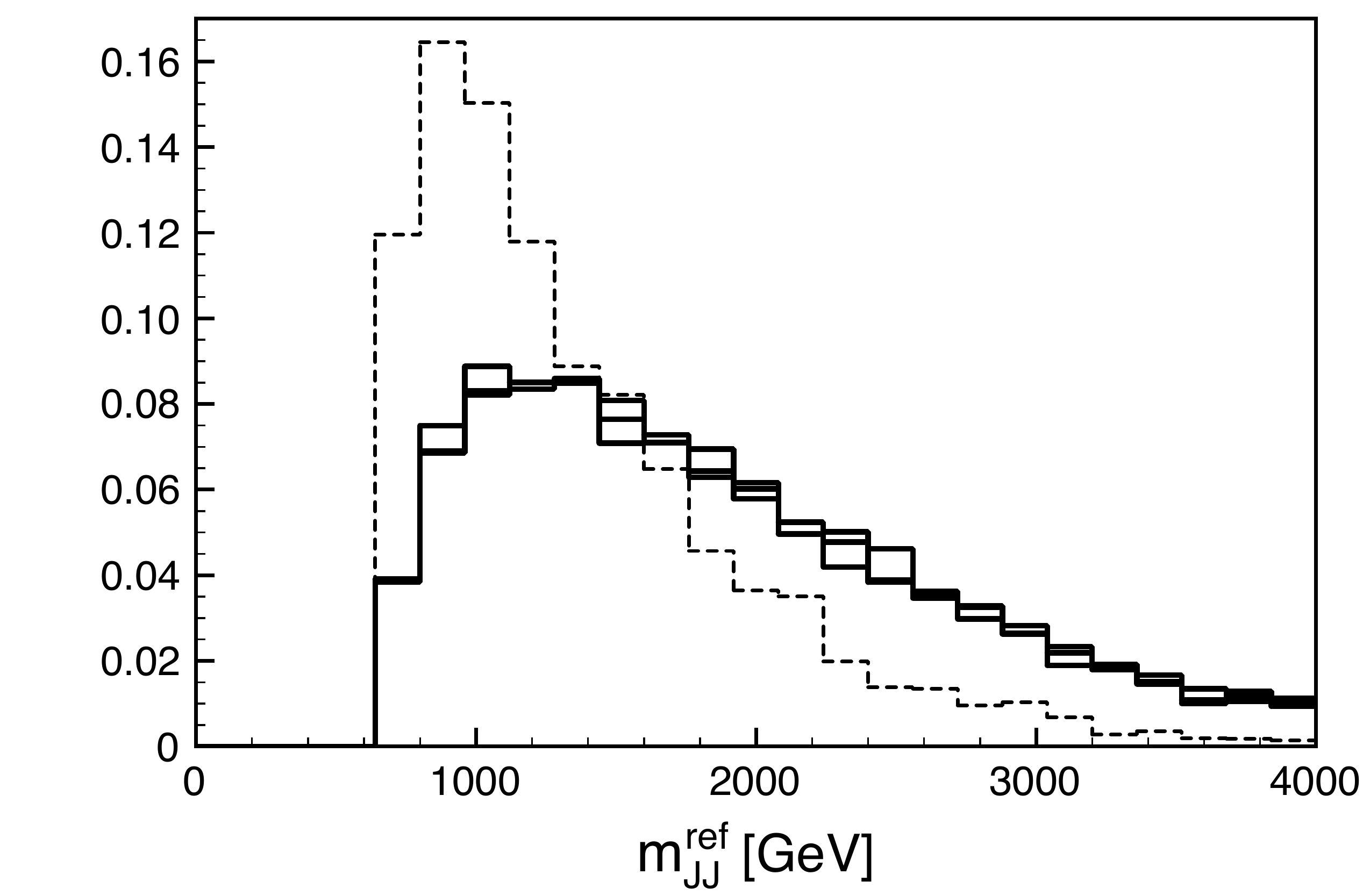}
\hspace{0.2cm}
\includegraphics[width=0.485\textwidth,clip,angle=0]{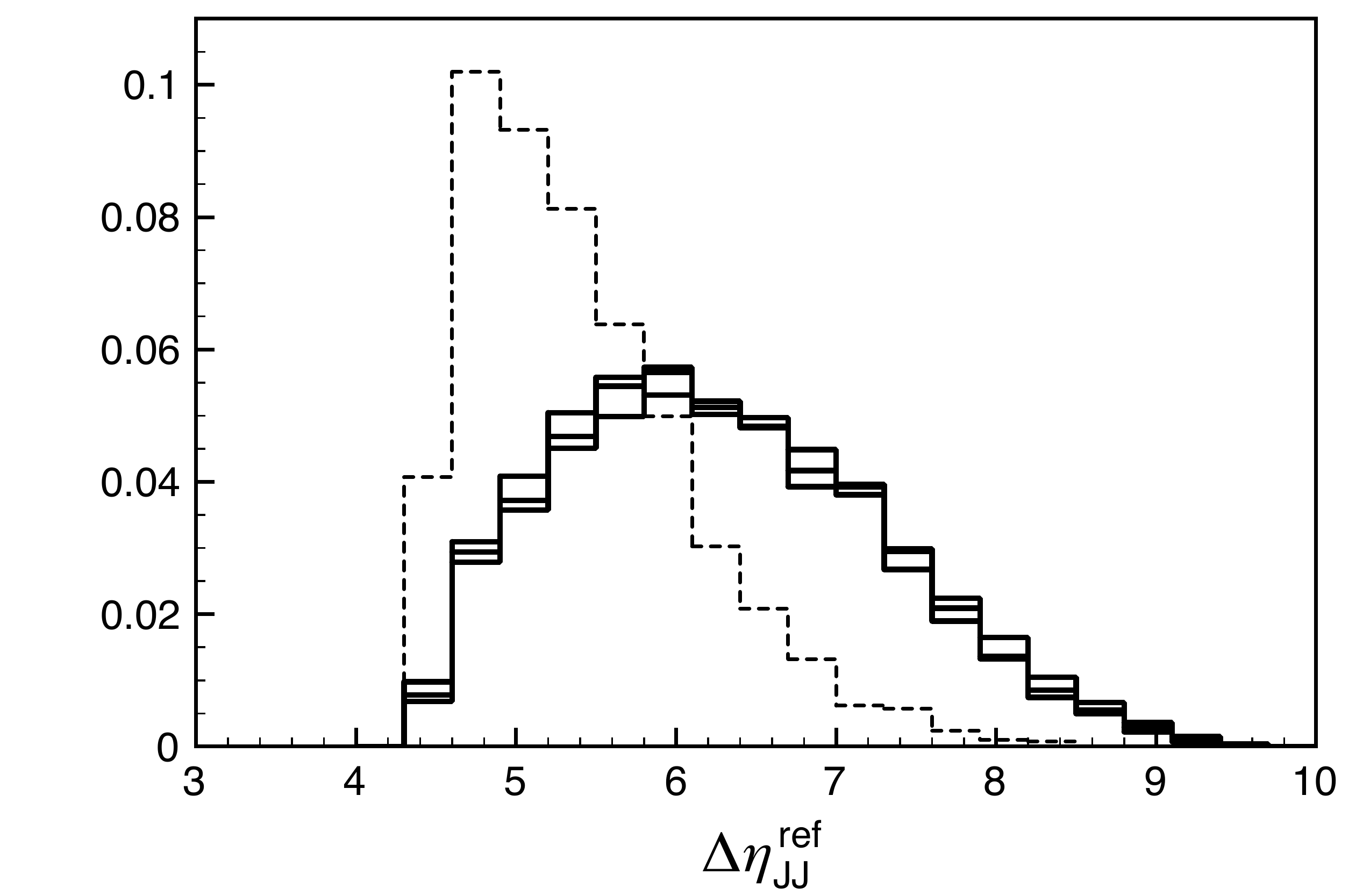}
\caption[]{
\label{fig:OPT2norm}
\small
Differential cross sections  for the three-lepton channel  after imposing  the optimized cuts of 
Eqs.~(\ref{eqn:acceptance})--(\ref{eqn:1levcut}) and (\ref{eqn:mwatlas}).
From up left to down right:   invariant mass of the system of the two Higgs candidates, excluding the neutrinos, $m_{hh}^{vis}$;
scalar sum of the $p_T$'s of the jets and leptons forming the two Higgs candidates, $H_T$;  invariant mass, $m_{JJ}^{ref}$, 
and separation, $\Delta \eta_{JJ}^{ref}$, of the two reference jets.
Continuous curves: signal ${\cal S}_3$  in the MCHM4 for $\xi=1,0.5,0$. Dashed curve: total background.
All curves have been normalized to unit area.
}
\end{center}
\end{figure}

The discussion   insofar focused on the possibility to detect the signal over a relatively large background.
This was done as a counting experiment. The very  limited number of events left  no other possibility
open. Assuming a much larger statistics one can try  to establish the distinguished features of  strong double Higgs production. 
These are basically two.  The first and most important one  is the hardness of the $W_LW_L\to hh$  subprocess
in  the SILH scenario, corresponding to an s-wave dominated cross section growing with  the invariant mass squared 
$m_{hh}^2=(p_h^{(1)}+p_h^{(2)})^2$ of the two Higgs system: $\sigma(WW\to hh)\approx m_{hh}^2/(32\pi f^4)$. 
In spite of this obvious property of the signal, as we will discuss below, a harder cut on $m_{hh}$ would not help our analysis.
A second feature is  the presence of two energetic forward jets with a transverse momentum $p_T$ peaked at $p_T\sim m_W$, 
independently of the jet energy. The absence of a typical  scale in the collinear momentum of the virtual $W_L$ emitted from the quark lines 
implies that also the partonic cross section  $\hat\sigma(q q' \to hh jj)$  grows with the square of the center of mass energy 
$\hat s$ of the $hh jj$ system. 
For the same reason, the quantities $m_{hh}$, $H_T$  (where $H_T$ is defined as the scalar sum of the transverse momenta of all the 
jets and charged leptons forming the two Higgs candidates) and $m^{ref}_{JJ}$ will all be distributed, for fixed $\hat s$, with a typical value of  
order $\sqrt{ \hat s}$. Given that the partonic cross section of the signal grows with $\hat s$, one would naively 
expect the distribution of these variables to be  harder for the signal than for the background. 
Similarly,  the rapidity separation of the reference jets $\Delta \eta_{JJ}^{ref}$, which for the signal is directly correlated with $\ln m^{ref}_{JJ}$ 
(given that the $p_T$ of the jets is peaked at $\sim m_W$) is expected to have a more significant tail at large values than for the 
background. In practice things are however more complicated. First of all, in order to realize the above expectations  it is essential to identify the 
Higgs decay products and to impose the optimized cuts of Eq.~(\ref{eqn:1levcut}).   The results are shown in Fig.~\ref{fig:OPT2norm},
where we plot the distributions for $m_{hh}^{vis}$ (the visible $m_{hh}$, defined as the invariant mass of the system of the two Higgs 
candidates, \textit{i.e.}, excluding the neutrinos), $H_T$, $m_{JJ}^{ref}$ and $\Delta \eta_{JJ}^{ref}$  for the three-lepton channel  after imposing all the 
optimized cuts of Eqs.~(\ref{eqn:acceptance})--(\ref{eqn:1levcut}) and (\ref{eqn:mwatlas}).
 Before the optimized cuts, the background distributions are actually harder than  the  signal ones,   and this is more so 
for $m_{hh}$ and $H_T$.
Secondly when devising optimized cuts not all variables work equally well. In particular the signal significance is better enhanced by cutting on 
$\Delta \eta_{JJ}$ as shown in the analysis.  This is  largely  due to  complex features of the background that are not immediately described analytically. 
There are however  features 
of the signal than can be easily understood analytically.  In particular the relative hardness in the distributions of $m_{JJ}$ and $m_{hh}$ is one such feature.
Indeed, working in first approximation with 
$\sqrt {\hat s}\gg, m_W, m_h$ and using $\sigma(WW\to hh)\propto m_{hh}^2$ and the splitting quark function 
$P_{q\to W_Lq}(x)\propto (1-x)/x$, we find the following partonic distributions  at fixed $\hat s$:
\begin{align}
\frac{d\hat\sigma^{\xi=1}}{ d (m_{hh}^2/\hat s)}   \propto
  &\, \frac{\hat s}{v^4} \left [ \left(1+\frac{m_{hh}^2}{\hat s} \right)\ln\left(\frac{\hat s}{m_{hh}^2}\right) 
      -2\left(1-\frac{m_{hh}^2}{\hat s} \right)\right ]\\[0.3cm]
\frac{d\hat\sigma^{\xi=1}}{ d (m_{JJ}^{ref\, 2}/\hat s)} \propto  
  &\,  \frac{\hat s}{v^4} \; \frac{m_{JJ}^{ref\, 2}}{\hat s} \ln\bigg(\frac{\hat s}{m_{JJ}^{ref\, 2}}\bigg)\,.
 \end{align}
This result shows that, for the signal, $m_{hh}$ is distributed with lower values than $m_{JJ}^{ref}$. This is a consequence of the
soft $1/x$ singularity in the splitting function that favors softer $hh$ invariant masses. This property is clearly shown 
by  Fig.~\ref{fig:OPT2norm}:  One can see that
$m_{JJ}^{ref}$ has a significant tail up to $3.5$ TeV while $m_{hh}^{vis}$ dies off already above $1$ TeV
(the total $m_{hh}$ dies off above $1.5$ TeV). Notice that, after optimized cuts, also for the background the distribution of $m_{JJ}$ is  harder than that of $m_{hh}$. 
%
%
%
\begin{figure}[tp]
\begin{center}
\includegraphics[width=0.485\textwidth,clip,angle=0]{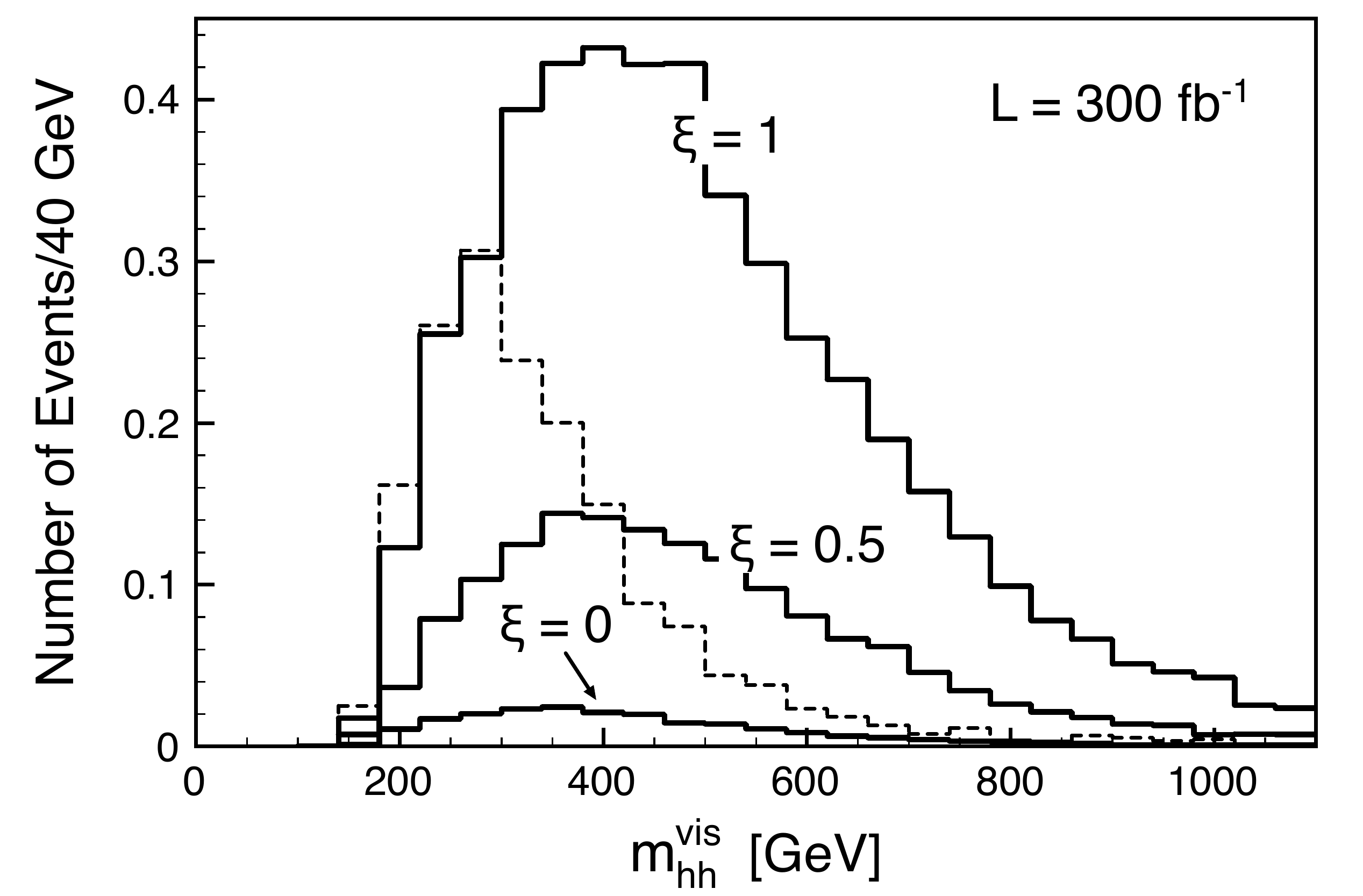}
\hspace{0.2cm}
\includegraphics[width=0.485\textwidth,clip,angle=0]{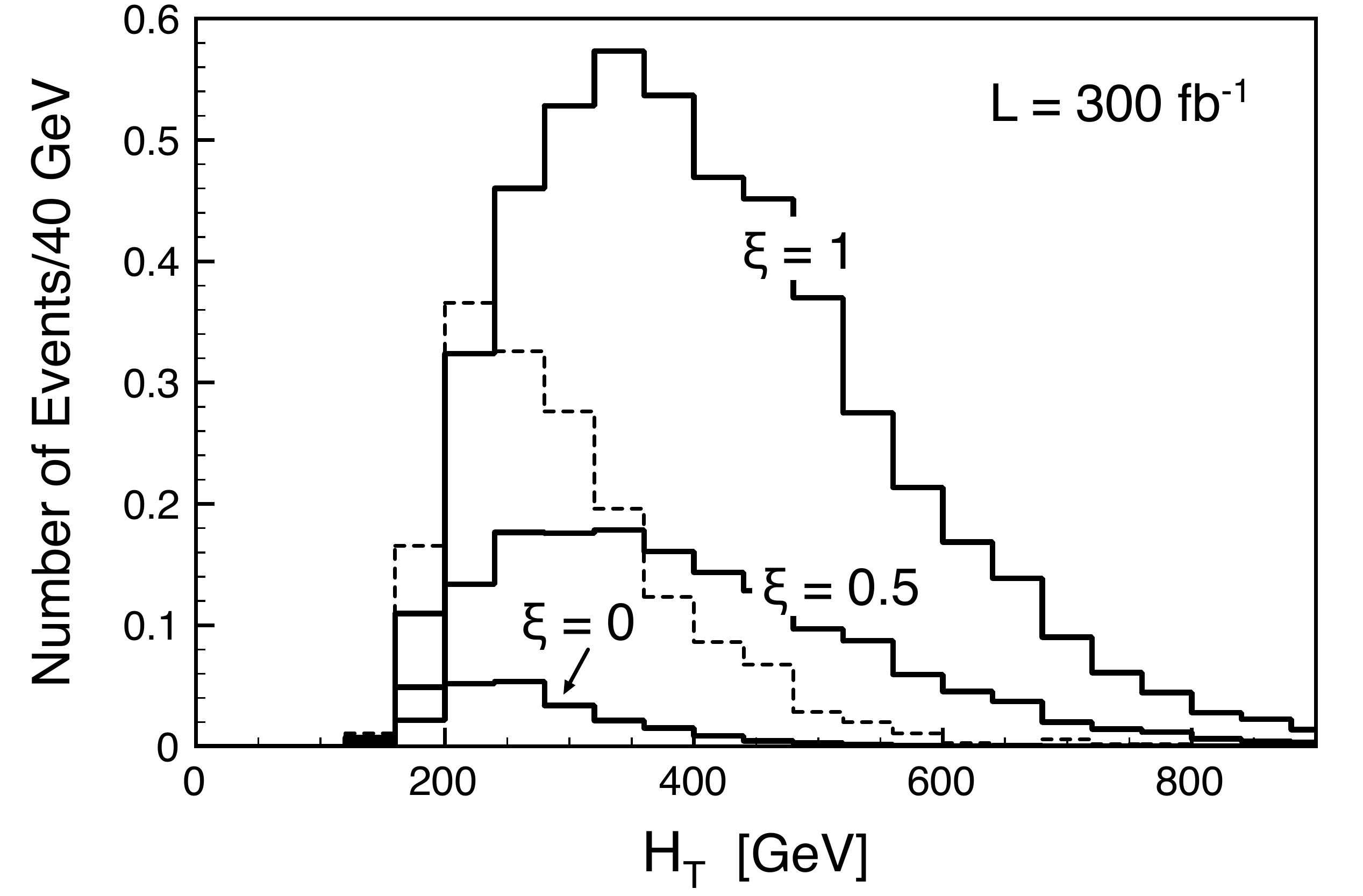}
\caption[]{
\label{fig:OPT2nevts}
\small
Number of three-lepton events with 300 fb$^{-1}$ after imposing  the optimized cuts of 
Eqs.~(\ref{eqn:acceptance})--(\ref{eqn:1levcut}) and (\ref{eqn:mwatlas}) as a function of
the  invariant mass of the system of the two Higgs candidates, excluding the neutrinos, $m_{hh}^{vis}$ (left plot), and
the  scalar sum of the $p_T$'s of the jets and leptons forming the two Higgs candidates, $H_T$ (right plot). 
Continuous curves: signal ${\cal S}_3$  in the MCHM4 for $\xi=1,0.5,0$. Dashed curve: total background.
}
\end{center}
\end{figure}

We already explained that  at the stage of optimization cuts $\Delta\eta_{JJ}^{ref}$ is the best variable to cut on.
On observing  Fig.~\ref{fig:OPT2norm}, one may wonder if additional cuts on any of the above observables
could further enhance the signal. In practice we have checked that below the already optimistic luminosity of $3 \,{\rm ab}^{-1}$
this is not the case, due to the loss of statistics.  This is, for example, illustrated by Fig.~\ref{fig:OPT2nevts}, where we show the 
number of three-lepton events at the end of the analysis (\textit{i.e.}, after the optimized cuts) as a function of $m^{vis}_{hh}$ and $H_T$.
Additional cuts on $m^{vis}_{hh}$ or $H_T$ would always further reduce the significance. 
The only possible and marginal improvement would be obtained by a further cut on $\Delta\eta_{JJ}^{ref}$ in the case $\xi =0.5$. 
Of course, if an excess were to be discovered, the study of the distributions in the above variables would provide an essential 
handle to attribute the excess to strong double Higgs production. It turns out that  
the scalar $p_T$ sum, $H_T$, seems overall the best variable in this regard: 
its shape is  distinguished from both the SM background and from the $\xi=0$ 
limit of $pp\to hhjj$, and this is a simple consequence of the signal consisting of a pure s-wave amplitude. Notice that the 
normalized distributions in $m^{vis}_{hh}$, $m_{JJ}^{ref}$ and $\Delta\eta_{JJ}^{ref}$,while they significantly differ
from the background,  surprisingly depend very little on $\xi$. In particular they are basically the same as for $\xi=0$, where 
$\sigma(WW\to hh)$ is dominated by the forward $t-$channel vector boson exchange and goes to  a constant $\propto m_W^2$, 
rather than growing, at large energy. This flattening in the $\xi$ dependence is due to the rapidly decreasing quark PDFs that makes 
the cross section  dominated by events close to threshold, that is with $m_{hh}^2/{\hat s}$ fixed. On the other hand, even close to threshold 
the distribution in $H_T$ of the signal stands out, on both QCD background and on $\xi=0$.

\section {Higgs mass dependence}
\label{sec:Higgsmass}

\begin{figure}[tbp]
\begin{center}
\includegraphics[width=0.55\textwidth,clip,angle=0]{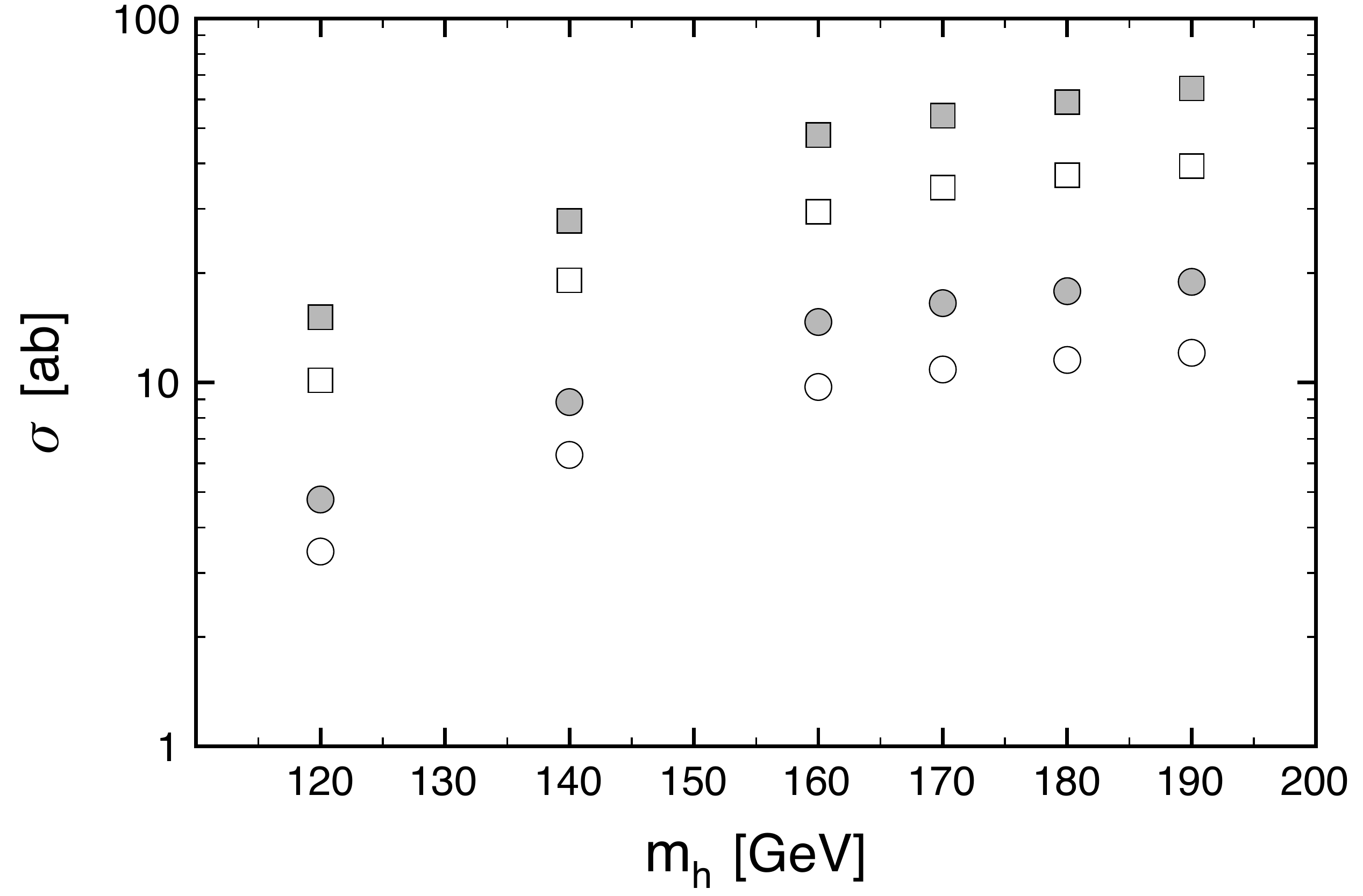}
\caption[]{
\label{fig:varyingmh}
\small
Cross section (in ab) for the signal ${\cal S}_3$ after the acceptance cuts of Eq.~(\ref{eqn:acceptance}) 
as a function of the Higgs mass.  The $BR(h\to WW)$ has been set to 1 (see text).
Filled (empty) squares and circles correspond  to the MCHM5 (MCHM4) with respectively $\xi=1$ and $\xi=0.5$.
}
\end{center}
\end{figure}

All the results presented so far were obtained by setting the Higgs mass to 180~GeV.
This choice was made to enhance the decay branching fraction to two $W$'s.
Varying the Higgs mass affects  the decay branching ratios and the signal cross section, as well as the kinematics of the events.
For example, Fig.~\ref{fig:varyingmh} shows how the value of the cross section of the three-lepton channel changes
after the acceptance cuts  when varying the Higgs mass.  In order to extract the different effects, we have set the $BR(h\to WW)$ to one.
%
%

The overall decrease of the signal for lighter Higgs masses is the result of two competing effects. 
On one hand,  due to the fast decrease of the quark PDFs at large energies, the cross section is dominated by events close to threshold, 
\textit{i.e.}, corresponding to quarks carrying away  a  fraction of the proton's momentum of  order $x_1 x_2 \sim 4m_h^2/s$.
The cross section is thus expected to increase for lighter Higgs masses, as smaller values of $x_{1,2}$ are probed.
This is indeed the case {\it before} the acceptance cuts, as shown for the MCHM4 and the MCHM5 by Table~\ref{tab:mh}.
\begin{table}[!h]
\begin{center}
\begin{tabular}{llcccccc}
			& & \multicolumn{6}{c}{$m_h$ [GeV]}\\ [.1cm]  \cline{3-8}
\multicolumn{2}{c}{$\sigma(pp\to hhjj)$ [fb]}\hspace{0.5cm} 	 & 120 & 140 & 160 & 170 & 180 & 190  \\[.1cm]
\hline
\multirow{2}{*}{$\xi=1$} 
                                  & MCHM4 &  12.4 & 11.3  & 10.3 & 9.77 & 9.31 & 8.88  \\[0.1cm]
                                  & MCHM5 &  17.4 & 16.3  & 15.3 & 14.6 & 14.2 & 13.8 \\[0.4cm]
\multirow{2}{*}{$\xi=0.5$} 
                                  & MCHM4 &  4.31 & 3.75  & 3.29 & 3.07 & 2.87 & 2.70  \\[0.1cm]
                                  & MCHM5 &  5.75 & 5.17  & 4.69 & 4.40 & 4.25 & 4.03 
\end{tabular}
\caption[]{
\label{tab:mh}
\small
Values of the $pp\to hhjj$ cross sections for undecayed Higgses with acceptance cuts applied only on the jets.
}
\end{center}
\end{table}
On the other hand, the lighter the Higgs is, the softer its decay products, and the less effective the acceptance cuts. 
In fact, this second effect dominates and leads to the overall decrease of the signal cross section when the Higgs mass
diminishes. 
We have checked that, as expected, the bulk of the reduction comes from the $p_T$ cut on the softest jet and lepton.

As already noticed, the value of the Higgs mass also affects the final number of signal events through the  decay branching ratios. 
In models like the MCHM4, where the Higgs couplings are shifted by a common factor, the branching ratio to two $W$'s is the
same as in the SM, and thus rapidly falls to zero below the $WW$ threshold. 
In general, however,  the branching ratios can be significantly modified compared to the SM prediction, and the branching 
ratio $BR(h\to WW)$ can be still sizable even for very light Higgses. This is for example the case of the MCHM5 with
 $\xi \sim 0.5$,  as illustrated by  Fig.~\ref{fig:BRs}.

\section {Luminosity vs energy upgrade}
\label{sec:varyingE}

The key feature of the composite Higgs  scenario is the partonic cross section growing with~$\hat s$. 
This behaviour persists until the strong coupling scale is reached where new states are expected to 
come in and  the growth in the cross section  saturates. 
Of course, with a sufficiently high beam energy, it is the direct study of the new, possibly narrow, resonances
that conveys most information on the compositeness dynamics.~\footnote{In the simplest models based on 5-dimensional constructions 
there exists no spin-0 resonance that could provide an s-channel enhancement of $WW\to hh$. Such a resonance instead exists in a 
4-dimensional example  based on a linear $SO(5)/SO(4)$ $\sigma$-model \cite{Barbieri:2007bh}. 
For recent studies on the detection of vector and scalar heavy resonances at the LHC see for example~\cite{resonances}.
}
Still, it is fair to ask how better a higher beam energy would allow one to ascertain the growth in the partonic cross section
below the resonance threshold. Unfortunately, since after the acceptance cuts the signal is still largely dominated by the background, 
it turns out that it is not possible to properly answer that question without 
a dedicated study, in particular  without cut optimization, a task  that exceeds the purpose of this paper. Here we limit ourselves
to a qualitative discussion based on ``standard''
(at the LHC) acceptance cuts and on a few additional cuts which 
seem the most obvious to enhance the signal to background ratio. Since the most promising channel is 
the one with three leptons
and the respective background is dominantly $W l^+l^-4j$,
we restrict our discussion to this channel and we examine
the behaviour of this background only.

A reasonable expectation is  that, as the centre of mass energy grows, 
the signal features become more prominent  over the background. 
In the upper panel of Table~\ref{tab:3Lenscan}, we report the cross section, with the same acceptance cuts as for 14 TeV, 
as a function of the collider energy $\sqrt{s}$ for both the signal and the background.
%
\begin{table}[!h]
\begin{center}
\begin{tabular}{|l||c|c|c|c|c|}
\cline{2-6}
   \multicolumn{1}{c|}{}   & 10 TeV & 14  TeV & 20 TeV  & 28 TeV & 40 TeV  \\ 
\cline{2-6}
\multicolumn{6}{c}{}  \\[-0.22cm]
\hline
${\cal S}_3$ (MCHM4 -- $\xi=1$)    & 12.1 & 30.4 & 70.0 & 135 & 252 \\
\hline
\hline
${\cal S}_3$ (SM -- $\xi=0$)  & 0.77 & 1.73 & 3.69 & 6.53 &  10.9 \\
\hline
\hline
$W l^+ l^- 4j$ & 4.75$\,\times 10^3$ & 12.0$\,\times 10^3$ & 28.6$\,\times 10^3$ & 59.7$\,\times 10^3$ & 122$\,\times 10^3$ \\
\hline
\end{tabular}

\vspace*{0.4cm}

\begin{tabular}{|l||c|c|c|c|c|}
\hline
${\cal S}_3$ (MCHM4 -- $\xi=1$)   & 11.1 & 24.5 & 45.4 & 66.3 &  81.0 \\
\hline
\hline
${\cal S}_3$ (SM -- $\xi=0$)  & 0.59 & 1.17 & 1.99 & 2.62 &  2.88 \\ 
\hline
\hline
$W l^+ l^- 4j$ & 3.44$\,\times 10^3$ & 6.54$\,\times 10^3$ & 10.9$\,\times 10^3$ & 15.0$\,\times 10^3$ & 17.2$\,\times 10^3$ \\
\hline
\end{tabular}
\caption[]{
\label{tab:3Lenscan} 
\small
Cross sections, in  ab, as a function of the collider energy $\sqrt{s}$,  
for signal and main background  in the three-lepton channel.
Upper panel:  values after imposing the acceptance cuts of Eq.~(\ref{eqn:acceptance});
Lower panel: values after imposing the acceptance cuts and the rescaled cut 
$\hat s >  0.01\, s$.
}
\end{center}
\end{table}
%
It is manifest that contrary to  naive expectations
the signal to background ratio is insensitive (if not degrading)
to the rising collider energy. As a matter of fact, this result is easily understood as follows. 

\begin{figure}[tbp]
\begin{center}
\includegraphics[width=0.55\textwidth,clip,angle=0]{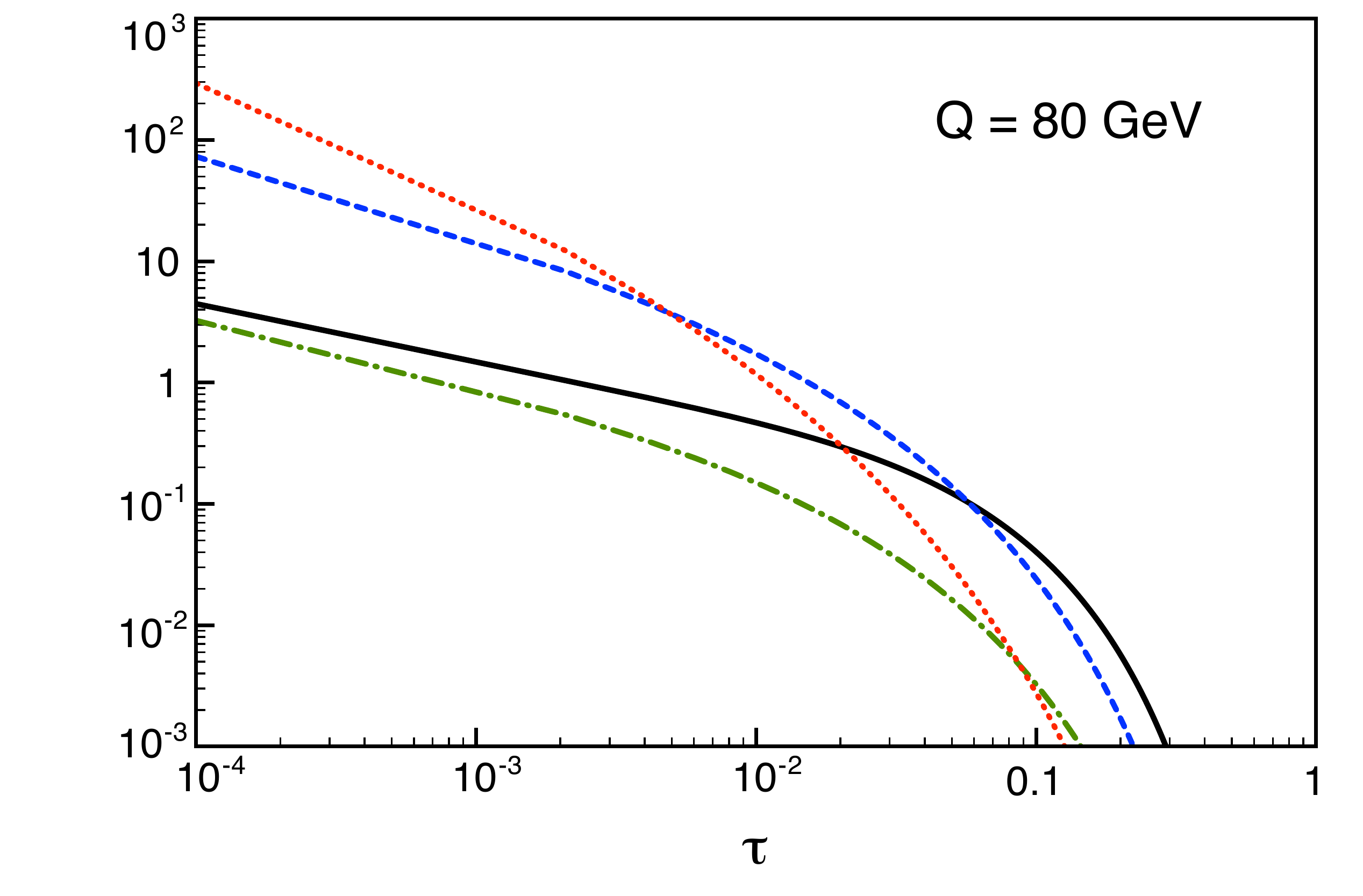}
\caption[]{
\label{fig:partonlum}
\small
Partonic luminosities $\rho_{AB}(\tau,Q^2)$ as a function of $\tau$ for partons $q_Aq_B=ud$ (black continuous curve); $gg$ (red dotted curve); 
$ug$ (blue dashed curve); $u\bar u$ (green dot-dashed curve). The factorization scale has been set to $Q=80\,$GeV.
}
\end{center}
\end{figure}

\begin{figure}[tbp]
\begin{center}
\includegraphics[width=0.485\textwidth,clip,angle=0]{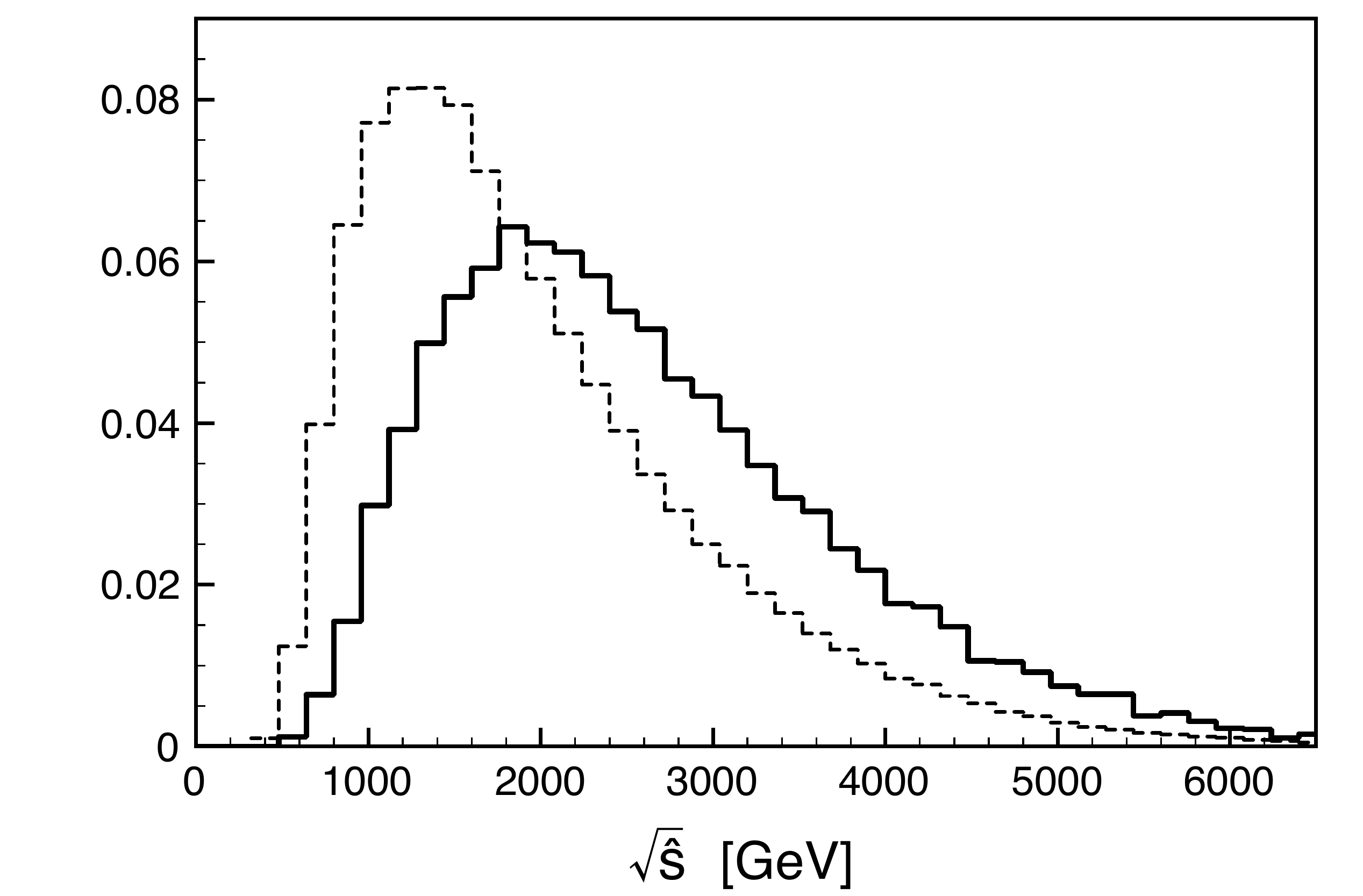}
\hspace{0.2cm}
\includegraphics[width=0.485\textwidth,clip,angle=0]{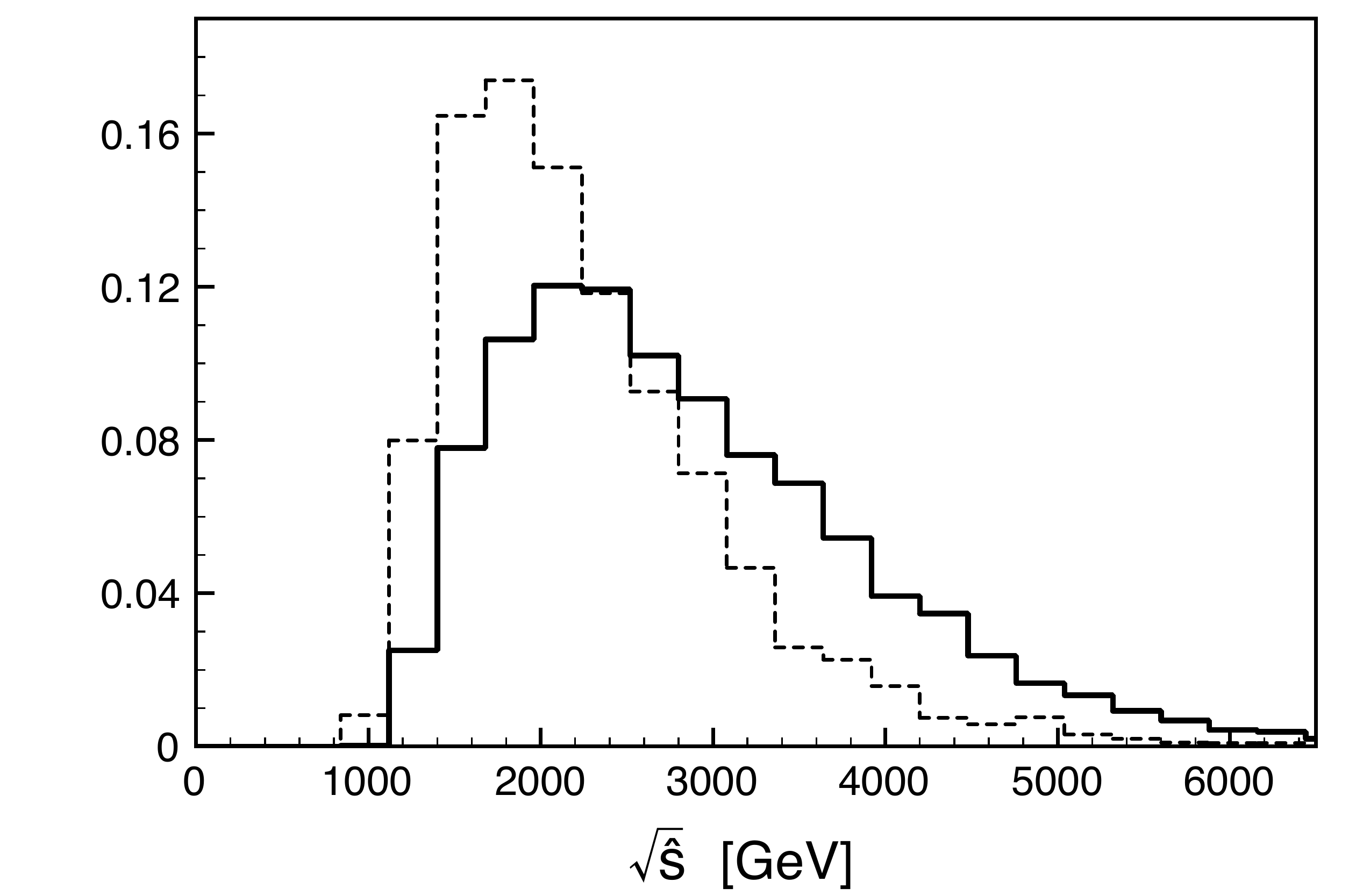}
\caption[]{
\label{fig:sqrtshat}
\small
Differential cross section for three-lepton events as a function of the total invariant mass $\sqrt{\hat s}$ (including the neutrinos) 
after the acceptance cuts of Eq.~(\ref{eqn:acceptance}) (left plot) and the optimized cuts of
Eqs.~(\ref{eqn:acceptance})--(\ref{eqn:1levcut}) and (\ref{eqn:mwatlas}) (right plot).
Continuous curve: signal ${\cal S}_3$ in the MCHM4 at $\xi=1$; Dashed curve: total background.
All curves have been normalized to unit area.
}
\end{center}
\end{figure}

In general, at fixed $\hat s$ the differential cross section to some final state $X$
can be written as the product of a partonic cross section $\hat \sigma(q_Aq_B\to X)$
times a partonic luminosity factor~$\rho_{AB}$:
\begin{equation}
\label{eq:factorization}
\begin{split}
& \frac{d\sigma}{d\hat s} = \frac{1}{\hat s}\,  \hat\sigma(q_Aq_B\to X)\, \rho_{AB}(\hat s/s, Q^2)   \\[0.2cm]
& \rho_{AB}(\tau, Q^2) = \tau \int_0^1 \!\! dx_1 \!\int_0^1 \!\! dx_2 \; f_{q_A}(x_1,Q^2) f_{q_B}(x_2,Q^2) \, \delta( x_1 x_2 - \tau)\, ,
\end{split}
\end{equation}
where $f_{q}(x,Q^2)$ denotes the PDF for the parton $q$, and $Q$ is the factorization scale. An implicit sum over all possible partons $q_A,q_B$
is understood.  The dependence on the collider energy $s$ only enters through the  luminosity factors $\rho_{AB}(\hat s/s, Q^2)$, which rapidly 
fall off when $\hat s/s$ increases, see Fig.~\ref{fig:partonlum}.
In fact, as a consequence of the luminosity fall off,  
at the LHC with $\sqrt{s}=14\,$TeV both the signal and background  cross section are  saturated near threshold, 
see Fig.~\ref{fig:sqrtshat}.
%
The total cross section can be written as 
\begin{equation}
\sigma = {\hat \sigma}(s_0) \times F(s,s_0),\ \ \textrm{with} \ \    
F(s,s_0) \equiv \int_{s_0}  \frac{d{\hat s}}{{\hat s}}\, \frac{{\hat \sigma}(\hat s)}{\hat \sigma(s_0)}   \rho(\hat s/s)
\end{equation}
where $s_0$ denotes the minimum value of $\hat s$ implied by the 
threshold constraint or by the cuts imposed. $F(s,s_0)$ is an effective luminosity function  depending on the form of  $\rho$ and $\hat \sigma$.
When the collider energy $s$ is increased, the growth of the total cross section is controlled by the factor $F(s,s_0)$, as a result of the
change in the parton luminosities. Consider for instance  a simple form $\rho(\tau,Q^2) = 1/\tau^q$, which gives a good fit of the
$ud$ ($gg$) parton luminosity for $\tau \lesssim 0.01$ with $q\simeq 0.5$ ($q\simeq 1.35$), see Fig.~\ref{fig:partonlum}. 
With that simple scaling, for all processes where the integral defining $F$ is saturated at the lower end ($\hat s\sim s_0$) one has that under 
$s\to \alpha \cdot s$ the integrated cross sections rescale universally as
$\sigma \to \alpha^q \cdot \sigma$. Even though this idealized situation is not exactly realized for our processes, we believe it largely explains  
the `universal' growth in the cross sections shown in Table~\ref{tab:3Lenscan}. That is  a simple reflection of the growth of
the PDFs at small $x$. This phenomenon is typical when considering rather  inclusive quantities, as it is the case for the total cross section after simple 
acceptance cuts. To the extreme case, with suitable hard and exclusive cuts, one should be able to contrast the $\propto \hat s$ growth of the  partonic 
signal cross section on the $\propto {1 / {\hat s}}$ decay of the background. 

The first obvious thing to do in order to put the underlying partonic dynamics in evidence is to rescale the lower cut as $ s_0=y\, s$, 
with fixed $y$.
Doing so, it is easy to see that, independent of the form of $\rho$, for a partonic cross section scaling like $\hat\sigma\propto \hat s^p$ one finds an 
integrated hadronic cross section  scaling in the same way: $\sigma\propto s^p$. 
The lower panel in Table~\ref{tab:3Lenscan} shows the signal and background cross sections as a function of $\sqrt{s}$ 
after imposing $\hat s >  0.01\, s$ 
in addition to the acceptance cuts.
One notices immediately that the background cross section still grows with $\sqrt{s}$, although with a much slower rate.
In fact,  this is not surprising, since in absence of more exclusive cuts the $t$-channel  singularities of the background $Wl^+l^- \!\!+jets$ 
imply a constant  cross section even at the partonic level,  $\hat \sigma \propto 1/M_W^2$, with a possible residual logarithmic growth
due to the soft and collinear singularities. Imposing more aggressive cuts can further uncover the  $1/\hat s$ behavior of the background at 
high energies, but  the efficiency on the signal would likely be too small,  and assessing the effectiveness of this strategy to enhance
the signal significance requires a dedicated  study.

\begin{figure}[tbp]
\begin{center}
\includegraphics[width=0.55\textwidth,clip,angle=0]{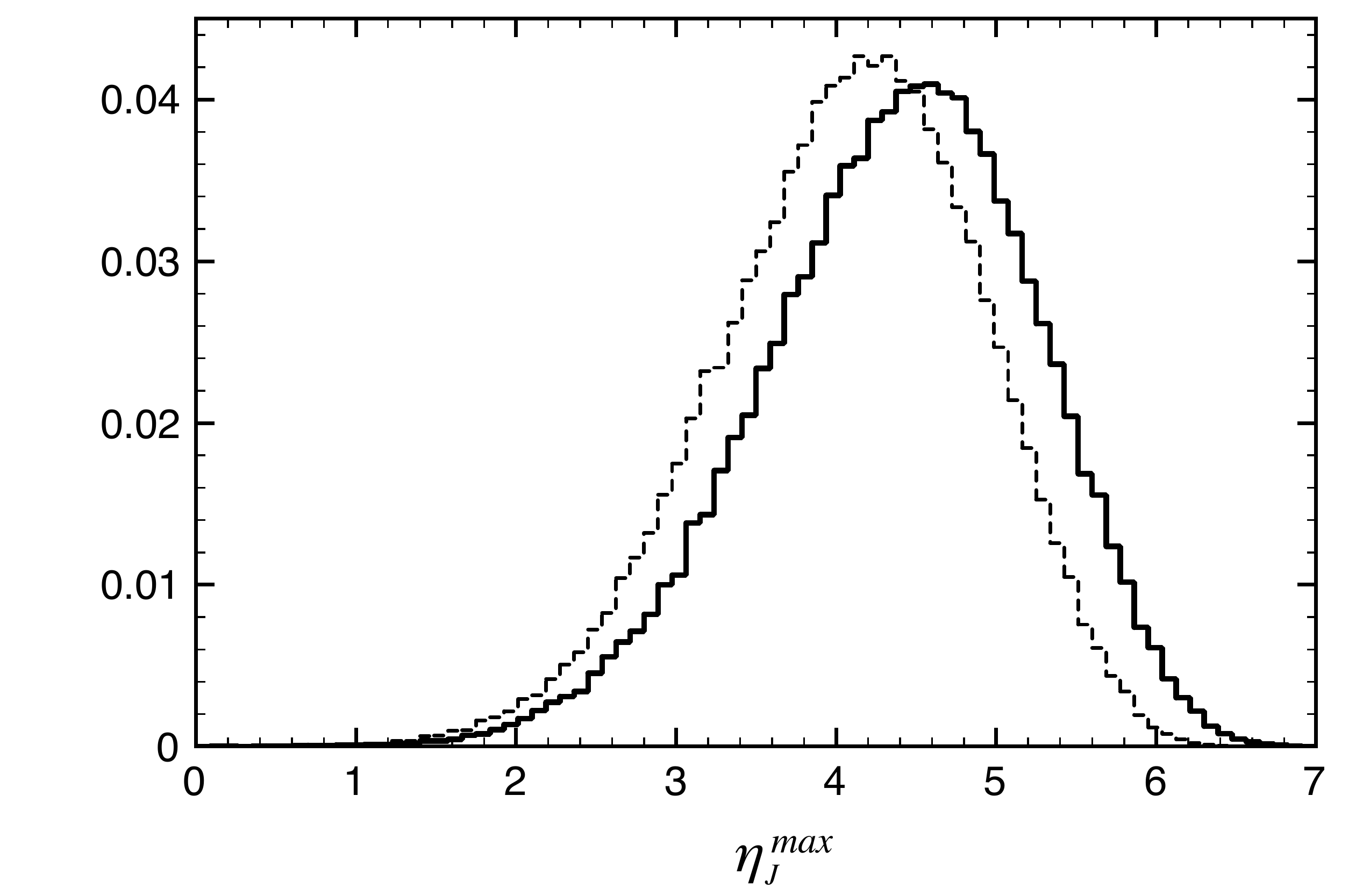}
\caption[]{
\label{fig:etamax2840TeV}
\small
Rapidity of the most forward jet (in absolute value) in the signal ${\cal S}_3$ at $\xi=1$ for $\sqrt{s} = 40\,$TeV (continuous curve)
and $\sqrt{s} = 28\,$TeV (dashed curve). The curves have been normalized to unit area.
}
\end{center}
\end{figure}

A more  surprising result is  the behavior of the signal in Table~\ref{tab:3Lenscan}: after the rescaled cut, one would expect the 
signal cross section  at $\xi=1$ to grow like $s$, modulo a mild logarithmic evolution of the PDFs. We do observe such a growth between 
10 and 20 TeV, but the growth saturates towards 40 TeV.
On inspection, this is a simple consequence of the acceptance cut we have imposed. A first effect comes from the constraint on the
rapidity of the forward jets: $|\eta _j|< 5$. Since the $p_T$ of the forward jets is $\sim m_W$, their rapidity will scale like $\ln \sqrt s/m_W$. 
Our Montecarlo simulation shows that above 40 TeV the $\eta_j$ distribution peaks above 4.5, 
and thus the apparently reasonable acceptance cut eliminates 
a significant  portion  of the signal (approximately $20\%$ at 40 TeV, which increases when selecting events at large $\hat s$, or large $m_{hh}$),
see Fig.~\ref{fig:etamax2840TeV}.  
%
%
We do not know how realistic is to consider detectors with larger rapidity acceptance, but it seems that one lesson to be drawn
is that forward jet tagging is a potential obstacle towards the exploitation of very high beam energies. 

A second and more dramatic effect comes from our request of having highly separated jets and leptons.
Quite intuitively, the more energetic the event is, the more boosted the Higgses, and the more collimated their decay products.
This implies that the efficiency of the ``standard'' isolation cuts in Eq.~(\ref{eqn:acceptance}) drastically decreases at large energies.
Rather than~$\hat s$, the best variable to look at in this case is $m_{hh}$, which is the real indicator of the strength
of the hard scattering in the signal and consequently of the boost of the Higgs decay products.
At 14 TeV,  in the MCHM4 at $\xi=1$,  the total fraction of three-lepton events where the two quarks from the decay of the hadronic $W$ 
are reconstructed as a single jet, so that the event has three hard and isolated jets, is 0.17. 
This has to be confronted with the fraction of events with four hard isolated jets, \textit{i.e.}, those selected
for the analysis of Section~\ref{sec:sig3L}, which is equal to 0.4.
If one requires $m_{hh} > 750\,$GeV,  the fraction of events where the hadronic $W$ is reconstructed as a single `fat' jet 
grows to 0.32, while the fraction of four jet events decreases to 0.36.  
For $m_{hh} > 1500\,$GeV, these  fractions become respectively 0.59 and 0.18.
It is thus clear that a different cut and event selection strategy has to be searched for if one wants to study the signal at very large energies.
Certainly, events with three jets will have to be included in the analysis, and jet substructure techniques~\cite{jetsubstructure} can prove 
extremely useful to beat the larger background. Ultimately, the very  identification and reconstruction of the signal events will probably have 
to be reconsidered, trying to better exploit the peculiar topology of the signal events at large energy, a limit in which the two Higgses and the 
two reference jets form four collimated and energetic clusters.

\begin{figure}[tbp]
\begin{center}
\includegraphics[width=0.485\textwidth,clip,angle=0]{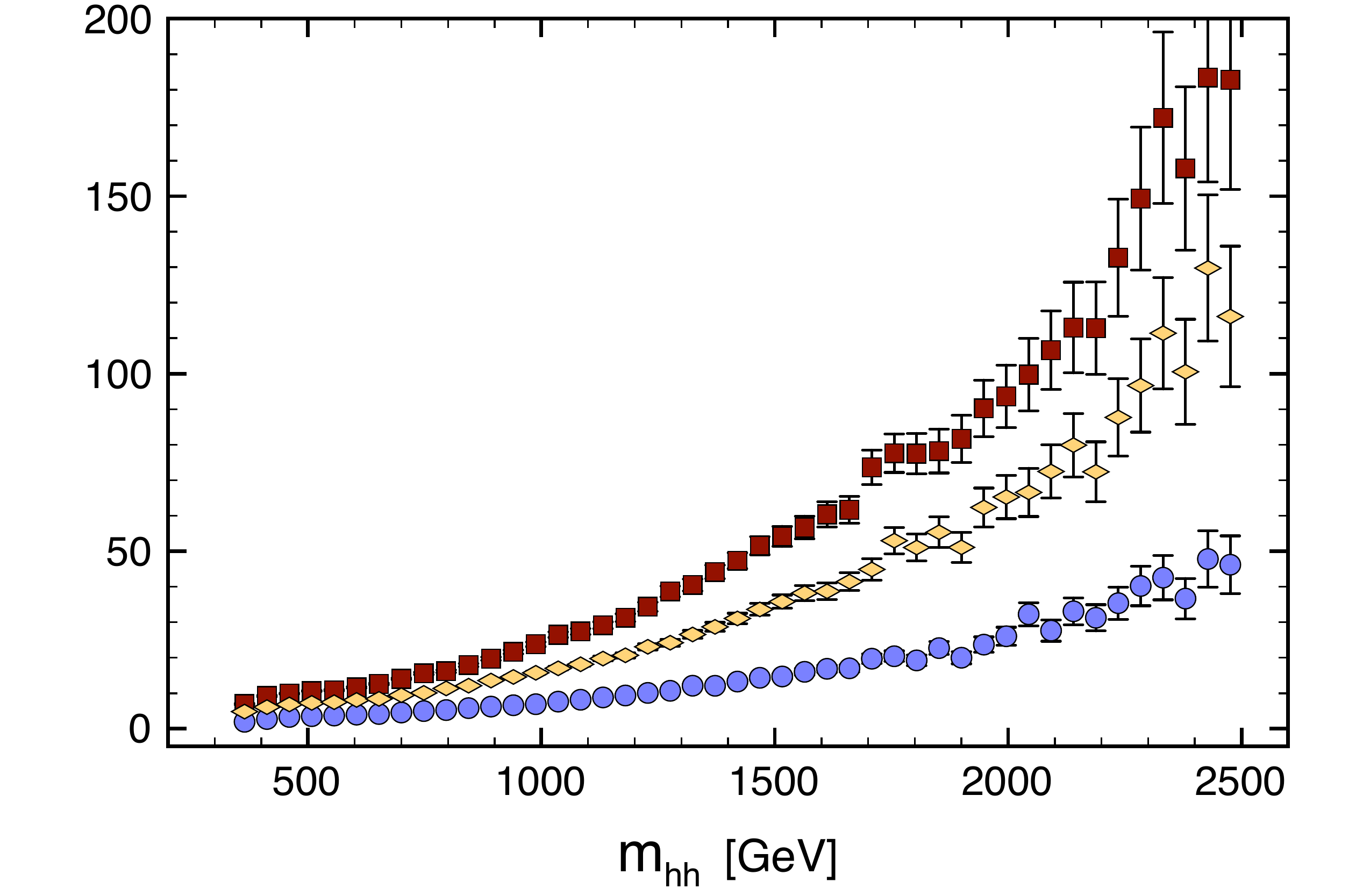}
\hspace{0.2cm}
\includegraphics[width=0.485\textwidth,clip,angle=0]{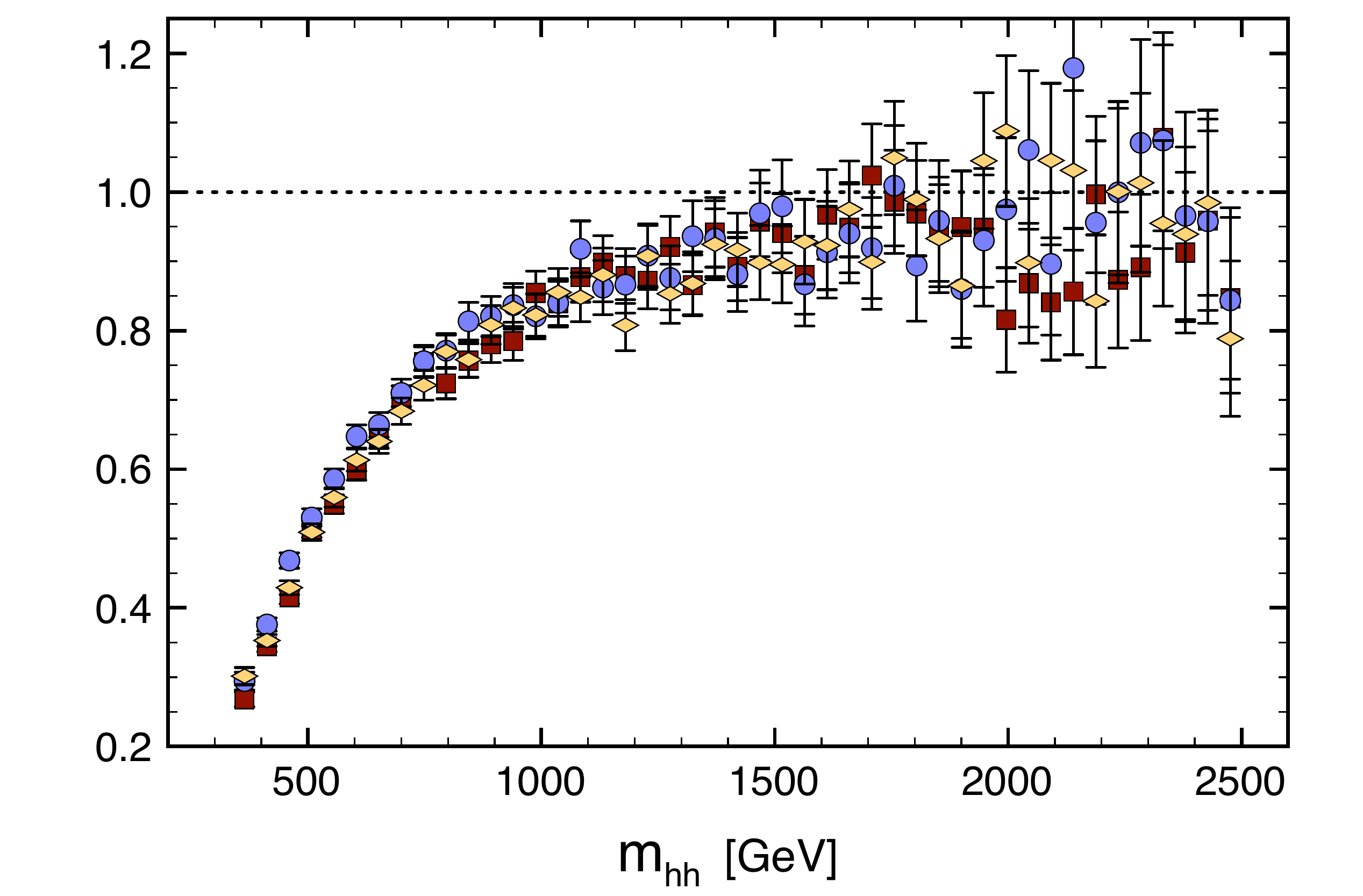}
\caption[]{
\label{fig:ratiosat14TeVvsmhh}
\small
Left plot: ratio of the differential cross sections for $pp\to hhjj$
in the MCHM4 and in the SM as a function of $m_{hh}$,
$(d\sigma/dm_{hh})|_\text{MCHM4}/(d\sigma/dm_{hh})|_\text{SM}$. Right plot: ratio of the differential cross sections 
for $pp\to hhjj$
in the MCHM4 and in the MCHM5 as a function of $m_{hh}$, $(d\sigma/dm_{hh})|_\text{MCHM4}/(d\sigma/dm_{hh})|_\text{MCHM5}$.
Red squares, yellow diamonds and blue circles  correspond respectively to $\xi = 1,0.8,0.5$. The vertical bars report the
statistical error on the ratio.
}
\end{center}
\end{figure}

Other than to beat the background,  studying the signal at large energies is crucial to disentangle its model dependency and extract $(a^2-b)$.
If the subdominant $ZZ\to hh$ contribution is neglected,  the signal cross section at fixed $m_{hh}$ can be written as
the product of a  $WW\to hh$ hard  cross section times a $W$ luminosity factor $\rho_W$:
\begin{equation}
\begin{split}
\label{eq:WWfactorization}
& \frac{d\sigma}{dm_{hh}^2} = \frac{1}{m_{hh}^2}\,  \hat\sigma(W_iW_j\to hh)\, \rho_W^{ij}(m_{hh}^2/s, Q^2)   \\[0.2cm]
&  \rho_W^{ij}(\tau, Q^2) = \tau \!\int_0^1 \!\!\! dx_1 \!\!\int_0^1 \!\!\! dx_2 \; f_{q_A}(x_1,Q^2) f_{q_B}(x_2,Q^2) 
    \!\int_0^1 \!\!\! dz_1 \!\!\int_0^1 \!\!\! dz_2 \; P^i_{A}(z_1) P^j_B(z_2)  \, \delta( x_1 x_2 z_1 z_2 - \tau)\, .
\end{split}
\end{equation}
An implicit sum over all partons $q_A,q_B$ and over transverse and longitudinal $W$ polarizations $i,j=T,L$ is understood.
$P^{T,L}_{A,B}(z)$ are the $W$ splitting functions  given in Eqs.~(\ref{eq:PTW}) and (\ref{eq:PLW}), 
which depend upon the parton flavor $A,B$ through the vectorial and axial couplings.
Unless a cut on the rapidity of the final Higgses is imposed (see Section~\ref{subsec:hhscatt}),  the contribution of the longitudinal $W$'s 
is by far dominating both at $\xi\not =0$ and at $\xi=0$. 
Hence, by taking the ratio of the observed number of signal events over the SM expectation at $\xi=0$,
the $W$ luminosity factors drop out, and the quadratic growth in $m_{hh}$ can be extracted.
The left plot of
Fig.~\ref{fig:ratiosat14TeVvsmhh} shows such ratio
for events with no cuts imposed.
After the cuts, one obtains a similar plot, although the range of accessible values of  $m_{hh}$ is reduced
as the consequence of the smaller efficiency 
at large $m_{hh}$ discussed above.
The  plot on the right in  Fig.~\ref{fig:ratiosat14TeVvsmhh} reports, instead, the ratio of the number of signal  events
predicted in two different models, respectively the MCHM4 and MCHM5, with $BR(h\to WW)$  set to one.
As expected, at large $m_{hh}$ the universal $\propto m_{hh}^2$ behavior dominates over the  model-dependent threshold effects 
controlled by the Higgs trilinear coupling,  and the ratio tends to 1.
These two plots show
that the strong scattering growth of the signal could be established, and $(a^2-b)$ be extracted, 
if one were able to study events with $m_{hh}$ up to $1.0-1.5\,$TeV,  corresponding to $m_{hh}^{vis}$ 
up to $\sim 0.7-1.0\,$TeV.
As Fig.~\ref{fig:OPT2nevts} clearly illustrates, at 14 TeV with 300 fb$^{-1}$ there are too few events at large $m_{hh}$
to perform such study.  It is thus necessary to have either a luminosity or an energy upgrade of the LHC.

\begin{figure}[tbp]
\begin{center}
\includegraphics[width=0.485\textwidth,clip,angle=0]{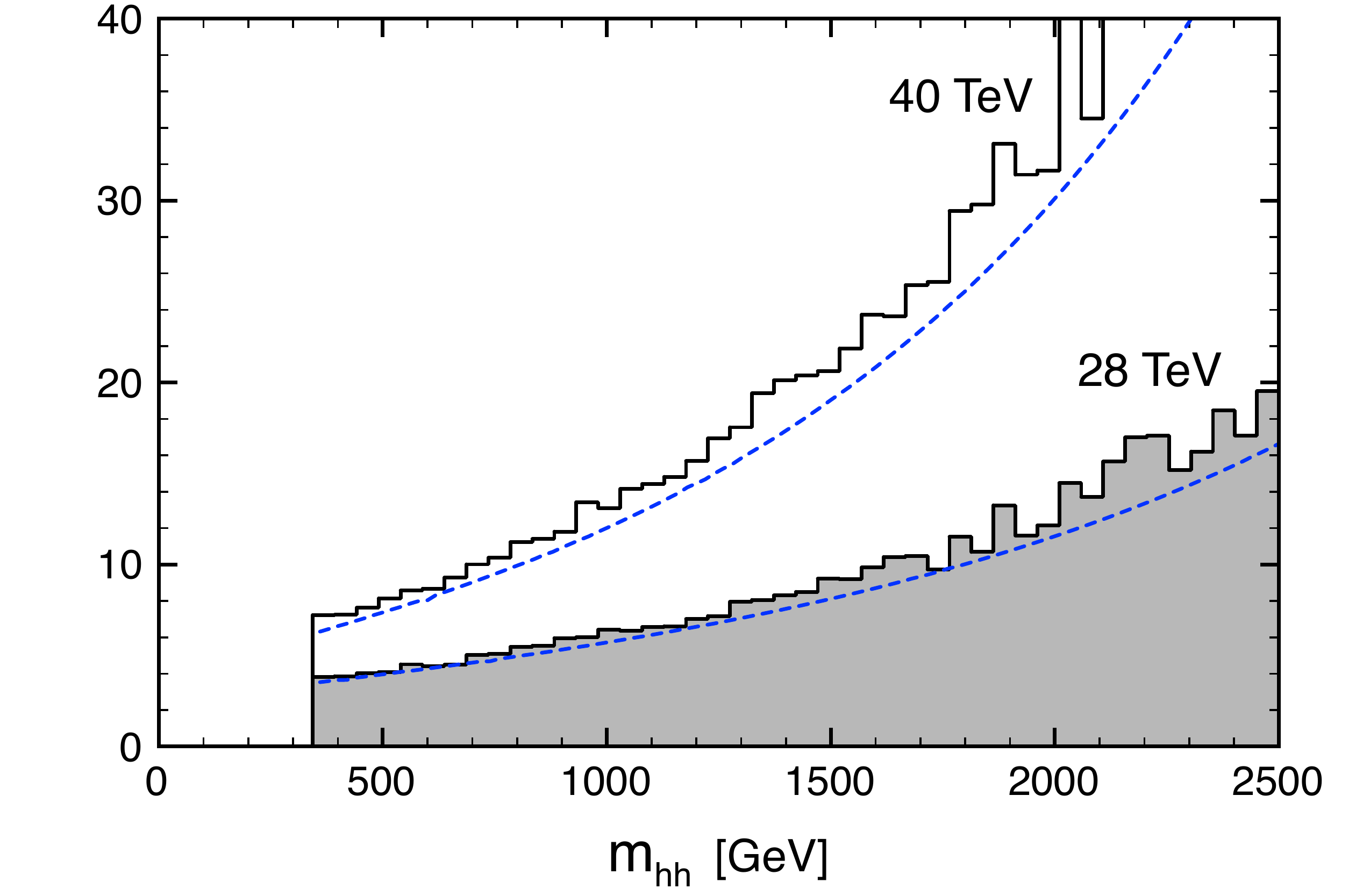}
\caption[]{
\label{fig:ratios28over14}
\small
Increase in the number of signal events at a given $m_{hh}$ when upgrading to 28 TeV (filled histogram)
and 40 TeV (empty histogram).  The two continuous blue curves correspond to the  ratio of $W$ luminosities
$r(m_{hh}^2/s) = \rho_W^{LL}(m_{hh}^2/s,Q^2)/\rho_W^{LL}((m_{hh}/14\,\text{TeV})^2,Q^2)$ respectively for $\sqrt{s} = 28\,$TeV
and $\sqrt{s} = 40\,$TeV.
}
\end{center}
\end{figure}

With 3 ab$^{-1}$ of integrated luminosity our analysis predicts approximately 50 three-lepton events and 150 two same-sign lepton
events in the MCHM4 at $\xi=1$, see Table~\ref{tab:finalnumbers}.  Although these are still small numbers,  this shows that 
even following a standard strategy  a tenfold luminosity upgrade of the LHC should be sufficient to extract the energy growing 
behavior of the signal.
The advantage of a higher-energy  collider compared to a luminosity upgrade 
is that,  for the same integrated luminosity, one can probe larger values of $m_{hh}$.
According to Eq.~(\ref{eq:WWfactorization}), when the collider energy is increased the differential cross section
gets rescaled due to the modified  luminosity factor $\rho_W$.
The plot of Fig.~\ref{fig:ratios28over14} shows  the  increase in the number of signal events 
at a given $m_{hh}$.  This is well approximated by the ratio of luminosity factors
$r(m_{hh}^2/s) = \rho_W^{LL}(m_{hh}^2/s,Q^2)/\rho_W^{LL}((m_{hh}/14\,\text{TeV})^2,Q^2)$ and is thus independent of the imposed cuts.
One can see that at 28 TeV the increase is larger than 10 only for events with $m_{hh}\gtrsim 1.6\,$TeV.
This suggests that in order to study the signal up to $m_{hh}\sim 1.5\,$TeV
a tenfold luminosity upgrade of the LHC would be as effective as, if not better than, a 28~TeV collider.
Of course a definitive conclusion on 
which of the two facilities is the most effective, whether a luminosity or an energy upgrade,
requires a precise estimate of the background, which scales differently in the two cases, and a more precise knowledge 
of how the various reconstruction  efficiencies are modified at the higher luminosity phase. 
We leave this to a future study.
%
%

\section{Conclusions and outlook}

In this paper we have considered the general scenario where a light composite CP even scalar $h$
with couplings similar, but different, to those of the Standard Model  Higgs arises from the electroweak symmetry breaking dynamics. 
We have simply called $h$ `the Higgs', although our parametrization also applies to situations where $h$ is quite distinguished from a Higgs, 
like for instance the case of a light dilaton.
We have noticed that besides deviations from the SM in single Higgs production and decay rates, this scenario is characterized by 
the growth  with the energy of the amplitudes for the processes 
$W_LW_L\to W_LW_L$, $W_LW_L\to hh$ and $W_LW_L \to t\bar t$.
In particular, the reaction of double Higgs production in vector boson fusion $W_LW_L\to hh$ emerges, along with the well studied process
of vector boson scattering $W_LW_L\to W_LW_L$, as a potentially interesting probe of strongly coupled electroweak dynamics. 
Specifically, the amplitude for $W_LW_L\to hh$ is predicted to grow with energy  at the same rate as $W_LW_L\to W_LW_L$ in models 
where $h$ is a pseudo-Goldstone boson, like those based on the $SO(5)/SO(4)$ coset of Refs.~\cite{Agashe:2004rs,Contino:2006qr}. 
On the other hand, when $h$ represents a dilaton, the amplitude for $W_LW_L\to hh$ does not grow at the leading linear order in the center of 
mass energy $s$.
 
Motivated by the above, we have performed a detailed analysis of the detectability of the process $W_LW_L\to hh$ at the LHC, more precisely
$pp\to hh jj $.
Our analysis focussed  for concreteness on the pseudo-Goldstone Higgs scenario, but our results have clearly a broader validity. 
Theoretically the physics of strong  $W_LW_L\to hh$ in hadron collisions resembles quite closely that of $W_LW_L\to W_LW_L$. 
In  practice there are important differences due to the different decay channels of  the final states and due to the different SM backgrounds. 
For instance, it is a known fact, which we reviewed in Section~\ref{sec:anatomy}, that the  cross section for the scattering of transversely polarized 
vector bosons $W_TW_T\to W_TW_T$ is numerically large in the SM, to the point that even in maximally coupled Higgsless models one must 
go to a center of mass energy around $700$ GeV in order for the signal $W_LW_L\to W_LW_L$ to win over. This `difficulty' is compensated by the 
availability of rather clean final states, in particular the purely leptonic gold-plated modes $W_LW_L\to \ell \ell +\etmiss$. 
The end result is that, at $14$ TeV  with 300 ${\rm fb}^{-1}$,  strong vector boson scattering should be detectable in  Higgsless models 
and in pseudo-Goldstone Higgs models with $v^2/f^2 \gtrsim 0.5$~\cite{Giudice:2007fh}.
In the case of $W_LW_L\to hh$ the situation is somewhat reversed. 
In realistic composite Higgs models the rate for $W_LW_L\to hh$ is significantly bigger than the one in the SM, already close to threshold. 
However, the final states from the decay of the Higgs pair most of the time involve QCD jets, thus making it more difficult to distinguish 
the signal from the background created by other SM processes. 
 
We  performed a partonic analysis of $pp\to hh jj $ using `standard' cuts as shown in Eq.~(\ref{eqn:acceptance}) to define jets. 
With that method we found that for the final state $hh\to \bar bb\bar b b$ the pure QCD background from $pp \to \bar bb\bar b bj j$ 
makes the signal undetectable. 
We have then focussed on the case where the Higgs decays dominantly to $W$'s, \textit{i.e.}, $hh \to W^+W^-W^+W^-$. While in the SM this 
requires $m_h\gsim 150$ GeV, it should be remarked that in the case of a composite Higgs the range can in principle extend to lower values 
of $m_h$ as the single Higgs couplings are also modified.  For example, in some interesting models like those based on the $SO(5)/SO(4)$ 
coset with matter transforming  as  the fundamental representation of $SO(5)$, the Higgs coupling to fermions is suppressed over a significant 
range of the parameter space, thus enhancing the relevance of the channel $h\to WW^*$ over $h\to bb$. 
We have made a detailed study of the detectability of the final states involving at least 2 leptons shown in Eq.~(\ref{eqn:channels}). 
One basic feature of the signal events that plays a crucial role in our analysis  is the presence of two very energetic forward jets
with large rapidity separation, large relative invariant mass and $p_T\lsim m_W$. Like in 
 $WW$-scattering, these jets originate from the collinear splitting $q\to q W_L^*$, where $W_L^*$ is a longitudinally polarized $W$ with 
virtuality $\sim p_T\sim m_W$. For each of the final states we have devised the optimal cuts by proceeding with a 3-step analysis. 
First we have performed standard acceptance cuts (Eq.~(\ref{eqn:acceptance})). In our simple partonic analysis those also provide our crude 
definition of jets. Secondly we have identified the relevant set of kinematical variables that characterize the signal against the background. 
These are the rapidity separation and invariant mass of the suitably identified forward jets $\Delta\eta_{JJ}^{ref}$, $m_{JJ}^{ref}$, 
and the mass shell conditions of the reconstructed candidate $h$'s and $W$'s. On those variables we then performed a
 set of master cuts defined in such way that cutting on each variable would not decrease the signal by more than 2\%. 
As a third final step we searched for the optimal set of cuts on the relevant kinematical variables
by following an iterative procedure: at each step we  cut over the observable providing the largest enhancement of the signal significance,
until no further improvement is possible. For instance, for the three lepton final state ${\cal S}_3$ the optimized cuts are shown in
Eqs.~(\ref{eqn:1levcut}),  (\ref{eqn:mwcms}) and (\ref{eqn:mwatlas}), where in the latter two equations we specialize the cut on the invariant 
mass on the candidate hadronic $W$'s to the energy resolution of respectively CMS and ATLAS.
In the case of two and four-lepton events we proceeded in a similar way.

The final results for the cross section of the signal and of the various backgrounds 
at different stages of the cut procedure are shown in Tables (\ref{tab:signbcks3L}), (\ref{tab:signbcks2L}) and
(\ref{tab:signbcks4L}), respectively for ${\cal S}_3,\,{\cal S}_2$ and ${\cal S}_4$. 
Some of the background processes needed in our study were not available in the literature, and we computed them by writing new routines 
in \texttt{ALPGEN}. We believe that the results of our simple partonic analysis are robust, and should remain stable  
when performing a more proper treatment of initial and final state radiation. We have not done a complete analysis, but only considered 
showering for the signal and the  leading sources of background. We found that the inclusion of  showering
enhances  the efficiency of our cuts. This is not surprising: while the energy scale  in the signal is large,
colored particles  have  a virtuality $\lsim m_W$ and  little QCD radiation is associated with them. 
This is not the case for the background:  extra radiation in this case increases the invariant masses of the Higgs and $W$ candidates 
and makes it more difficult for the background to pass our on-shell cuts.

The outcome of our analysis is synthesized in Tables (\ref{tab:finalnumbers}) and (\ref{tab:sigsig}).  With $300 \,{\rm fb}^{-1}$ 
only for very low compositeness scale $\xi=1$, basically the Technicolor limit, can one barely see the signal. 
A realistic viewpoint is therefore that the LHC luminosity upgrade
of 3 ${\rm ab}^{-1}$ will be needed to study strong double Higgs production. 
The three lepton final state would then provide a rather clean signal for $\xi >0.5$.
The two same-sign lepton final state is not as free from background, but yields a predicted number of  events a factor 3 larger.
Both channels would independently give a 
$9 \sigma$ signal in the limiting case $\xi=1$. It should be emphasized that for the case of the two  lepton signal 
a more careful estimate of the background, including correcting for detector efficiency, will be needed to reach the above mentioned significance, 
given that the background is more important for that channel. One should compare our results for strong double Higgs production to those 
of the more studied $WW$ scattering. In that case the final numbers are  significantly better. For instance, according to Ref.~\cite{bagger},
the reaction $W^+W^+\to W^+W^+$  in the purely leptonic final state would yield approximately 40 events of signal at $\xi =1$
 with $300 \,{\rm fb}^{-1}$,  with a background of about 10 events (mostly due to the scattering of transversely polarized $W$'s). 
It should however be emphasized that the $hh$ final state gives access to additional information on the independent parameters $b$ and $d_3$.
At large $m_{hh}$ the effect of $b$ dominates as it controls the energy growing part of the amplitude. In our analysis, we did not impose a lower 
cut on $m_{hh}$ and we thus collected also the events close to threshold,
which depend also on the Higgs cubic $d_3$. This parameter has a significant impact on the total cross section. 
For instance, this can be seen in Table~\ref{tab:finalnumbers} by inspecting  the two lepton channel  in the two different models 
MCHM4 and MCHM5 for the same value $\xi=0.8$, that is for coinciding $a$ and $b$: the $40\%$ mismatch in the number of events 
is a measure of the relevance of the cubic coupling $d_3$. In principle a scan of the dependence of the signal events on $m_{hh}$ should allow 
the extraction of both $b$ and $d_3$.  By putting together the information contained in Figs.~\ref{fig:OPT2nevts} and \ref{fig:ratiosat14TeVvsmhh}
one can deduce that with a tenfold luminosity upgrade of the LHC  it would be realistic to perform such a study, at least for models 
that  deviate sizably from the SM (\textit{i.e.}, with  $(a^2-b)=O(1)$).
We have not attempted to estimate how well we could extract $b$ and $d_3$, because
in order to do so in a general model-independent way, 
we would also need to study in more detail how accurately $a$ and $c$ can be extracted from single Higgs production. 
This is because these two parameters affect both the signal and the background cross sections.
On the other hand, if an excess in the total cross section is found,  it should be possible 
to decide whether it was a pseudo-Goldstone Higgs or a dilaton by considering the energy distribution of the events. 
In the case of a dilaton the dependence on $m_{hh}$ would be the same as in the SM, 
while for the pseudo-Goldstone Higgs  a characteristic growth $\propto m_{hh}^2$  as well as a harder distribution in $H_T$ would appear.

Our detailed study of the background was done assuming $14$ TeV collisions. We have not attempted  such an analysis at higher energies. 
We have however tried to assess how better an energy upgrade, as opposed to a luminosity upgrade, would improve things. 
We believe that the answer to the above dilemma is somewhat answered by Fig.~\ref{fig:ratios28over14}, where we show how the differential 
signal cross section rescales with the beam energy in the relevant region of $m_{hh}$. Assuming the same luminosity as the LHC, it seems that
an energy upgrade to 28 TeV would not do better than a tenfold luminosity upgrade at 14 TeV. Of course there are many other variables 
in such extrapolation, like for instance the issue of pile-up, which we cannot control. Our result should thus be taken as a hint. 
It should also not be forgotten that an increase in beam energy would increase the sensitivity to resonances.
In particular a scalar resonance in the s-channel could clearly enhance our signal.

\enlargethispage{0.5cm}
There are a few directions along which our analysis can be extended or improved. One source of limitation in our study was the small branching ratio to leptonic final states. A possible improvement could come from considering
$W$ decays to $\tau$. By a simple estimate, one concludes that by including 
events with three leptons of which one is a tau, the yield of this channel is almost doubled. 
A careful study of background, including consideration of the efficiency of $\tau$ tagging and $\tau$/jet mistagging, would however be in order.
Another limitation of our analysis is due to our `conservative' choice of acceptance cuts. The parton isolation criterion corresponding to these 
cuts  clearly disfavors the signal in the interesting energy range where the center of mass energy of the two Higgs system is large and the final 
decay  products are boosted. It would be interesting to explore another cut strategy where the jets
and leptons from each decaying Higgs are allowed to merge, and where the features of the signal are contrasted to those of the background by 
using jet substructure observables. On one hand this direction seems to make things worse by increasing the relevance of background events 
with fewer jets. On the other hand, it would allow  a more efficient collection of signal events in the region of large invariant $m_{hh}$ where the 
signal cross section becomes larger. Indeed with that more aggressive strategy one could in principle consider the possible relevance of the one 
lepton channel, where only one $W$ decays leptonically. One advantage of that channel is that one can reconstruct the momentum 
of the neutrino and  close the kinematics.
To the extreme one could even reconsider the 4$b$'s final state, which could well be the dominant one if the
Higgs is light.

\appendix

\section*{Appendix}

\section{Model parameters}
\label{app:models}

For convenience, we report here the values of the Lagrangian parameters for the two minimal $SO(5)/SO(4)$ models 
of Refs.~\cite{Agashe:2004rs, Contino:2006qr}, MCHM4 and MCHM5, with SM fermions transforming respectively as
spinorial and  fundamental representations of $SO(5)$ (the Higgs field is canonically normalized and $\xi=v^2/f^2$):

\begin{center}
\begin{tabular}[t]{|l|l|l|}
\hline
Coupling & \multicolumn{1}{c|}{MCHM4}   & \multicolumn{1}{c|}{MCHM5} \\
\hline
{\vrule height 14pt depth 0pt width 0pt}
$g_{hWW}^\xi=a\cdot  g_{hWW}^{SM}$ & $a=\sqrt{1-\xi}$ & $a=\sqrt{1-\xi}$ \\[.2cm]
$g_{hhWW}^\xi=b\cdot  g_{hhWW}^{SM}$ & $b=1-2\xi$ & $b=1-2\xi$ \\[.2cm]
$g_{hff}^\xi=c\cdot  g_{hff}^{SM}$ & $c=\sqrt{1-\xi}$ & $\displaystyle c=\frac{1-2\xi}{\sqrt{1-\xi}}$\\[.5cm]
$g_{hhh}^\xi=d_3\cdot  g_{hhh}^{SM}$ & $d_3=\sqrt{1-\xi}$ & $\displaystyle d_3=\frac{1-2\xi}{\sqrt{1-\xi}}$\\[.3cm]
\hline
\end{tabular}
\end{center}

\section{Montecarlo generation}
\label{app:MC}

We report here the details about the Montecarlo generation of background and signal.
In all the simulations we have used  the CTEQ6l1 PDF set.
All the signal samples have been generated with \texttt{MADGRAPH}, setting the factorization and normalization scale
to $Q=m_W$. For the simulation of the background samples we have used both \texttt{MADGRAPH}
and \texttt{ALPGEN}.
The following tables report the choice of the factorization and renormalization
scale $Q$ chosen for each sample (where $m_h=180\,$GeV, $m_t=171\,$GeV):

\begin{center}
\begin{minipage}[t]{0.45\linewidth}
\begin{tabular}[t]{|l|c|}
\multicolumn{2}{c}{Backgrounds generated} \\
\multicolumn{2}{c}{with \texttt{ALPGEN}} \\
\multicolumn{2}{c}{} \\[-0.2cm]
\hline
Sample & $Q$ \\
\hline
\hline
& \\[-0.35cm]
$Wl^+l^-4j$ & $\sqrt{m_W^2 + M_{ll}^2}$ \\[0.1cm]
$Wl^+l^-5j$ & $\sqrt{m_W^2 + M_{ll}^2}$ \\[0.1cm]
$W\tau^+\tau^-4j$ & $\sqrt{m_W^2 + M_{\tau\tau}^2}$ \\[0.1cm]
$W\tau^+\tau^-5j$ & $\sqrt{m_W^2 + M_{\tau\tau}^2}$ \\[0.1cm]
$W^{+(-)}W^{+(-)}5j$ & $2 m_W$ \\[0.1cm]
$WWWjjj$ & $3 m_W$ \\[0.1cm]
$WWW4j$ & $3 m_W$ \\[0.1cm]
$hWjjj$ & $m_h + m_W$ \\[0.1cm]
$WWWWj$  & $4 m_W$ \\[0.1cm]
$WWWWjj$ & $4 m_W$ \\[0.1cm]
$t\bar t hjjj$ & $2m_t+m_h$  \\[0.1cm]
$t\bar tjj$ & $\sqrt{2 m_t^2 +\sum_{i=t,j} (p_{Ti})^2}$  \\[0.1cm]
$t\bar t3j$ & $2m_t$ \\[0.1cm]
$t\bar t4j$ & $2m_t$ \\[0.1cm]
$W6j$ & $\sqrt{M_{l\nu}^2+\sum_j (p_{Tj})^2}$ \\[0.1cm]
$l^+l^-5j$ & $M_{ll}$ \\[0.1cm]
$l^+l^-l^+l^-jj$ & $M_{4l}$ \\[0.1cm]
$l^+l^-\tau^+ \tau^- jj$ & $M_{ll\tau\tau}$ \\[0.1cm]
$hjj$ & $m_h$ \\[0.1cm]
$WWl^+l^-jj$ & $\sqrt{M_{ll}^2 + 2 m_W^2}$ \\[0.1cm]
$WWhjj$ & $2 m_W + m_h$ \\[0.1cm]
\hline
\end{tabular}
\end{minipage}
\hspace{1cm}
\begin{minipage}[t]{0.3\linewidth}
\begin{tabular}[t]{|l|c|}
\multicolumn{2}{c}{Backgrounds generated} \\
\multicolumn{2}{c}{with \texttt{MADGRAPH}} \\
\multicolumn{2}{c}{} \\[-0.2cm]
\hline
Sample & $Q$ \\
\hline
\hline
& \\[-0.35cm]
$t\bar tWjj$ & $2m_t+m_W$  \\[0.1cm]
$t\bar t Wjjj$ & $2m_t+m_W$\\[0.1cm]
$t\bar tWW$ & $2m_t+m_h$ \\[0.1cm]
$t\bar tWWj$ & $2m_t+m_h$ \\[0.1cm]
$t\bar t hjj$ & $2m_t+m_h$ \\[0.1cm] 
$t\bar tl^+l^-$ & $2m_t+m_Z$\\[0.1cm]
$t\bar tl^+l^-j$ & $2m_t+m_Z$ \\[0.1cm]
$h l^+l^-jj$ & $m_h+m_Z$  \\[0.1cm]
\hline
\end{tabular}
\end{minipage}
\end{center}

\section*{Acknowledgments}

We would like to thank
M.~Chanowitz,
R.~Franceschini,
S.~Frixione,
A.~Giammanco,
G.~Giudice,
B.~Grinstein,
Z.~Kunszt,
T.~Lari,
M.~Mangano,
C.~Mariotti,
B.~Mele,
M.~Pierini,
S.~Pokorski,
V.~Rychkov,
G.~Salam,
R.~Tenchini,
J.~Wells and 
A.~Wulzer
for useful discussions and comments. We would also like to thank the Galileo Galilei Institute for Theoretical Physics and the  CERN TH Unit 
for hospitality and  support during various stages of this work.
The work of C.G. has been partly supported by European Commission under the contract  ERC advanced grant 226371 `MassTeV' and the 
contract PITN-GA-2009-237920 `UNILHC'. 
The work of R.R. is supported by the Swiss National Science Foundation
under contracts No. 200021-116372 and No. 200022-126941.


\end{document}